\documentclass[12pt]{article}
\usepackage{amsmath,amsfonts,amssymb,amsthm}
\usepackage{mathtools}
\usepackage[mathscr]{eucal}
\usepackage{epsfig}
\usepackage{caption}
\usepackage{subcaption}
\usepackage{slashed}
\usepackage{pbsi}
\usepackage[T1]{fontenc}
\usepackage{xcolor}
\usepackage{mathdots}
\usepackage{booktabs}
\usepackage{mathtools}
\usepackage{hyperref}
\theoremstyle{plain}
\newtheorem{theorem}{Theorem}

\renewcommand{\theequation}{\arabic{section}.\arabic{equation}}
\newcommand\scH{{\mathscr H}}
\newcommand\scI{{\mathscr I}}

\newcommand\scK{{\mathscr K}}

\newcommand\mvector{\boldsymbol}

\newcommand\vd{\mvector{d}}

\newcommand\vp{\mvector{p}}
\newcommand\vq{\mvector{q}}

\newcommand\field{\mathbb}

\newcommand\R{\field{R}}

\newcommand\C{\field{C}}
\newcommand\Z{\field{Z}}

\newcommand\Q{\field{Q}}

\newcommand\rmi{\mathrm{i}\mspace{1mu}}

\newcommand\Dt{\frac{\mathrm{d}\phantom{t} }{\mathrm{d}\mspace{1mu}
t}}

\DeclareMathOperator{\sgn}{sgn}

\newcommand\cn{\operatorname{cn}}

\newcommand\defset[2]{\left\{{#1}\;\vert \;\; {#2} \,\right\}}

\begin{document}
\title{\bf  Translational dynamics of diatomic molecule\\ in magnetic quadrupole trap}
\author{Yurij Yaremko*, Maria Przybylska\dag, Andrzej J. Maciejewski\ddag}
\date{\normalsize
*Yukhnovskii Institute for Condensed Matter Physics\\ of the National Academy of Sciences of Ukraine,\\
Svientsitskii St. 1, 79011 Lviv, Ukraine\\
e-mail: yar@icmp.lviv.ua\\
\dag  Institute of Physics, University of Zielona G\'ora,\\ Licealna St. 9, 65--417 Zielona G\'ora, Poland\\
e-mail: m.przybylska@if.uz.zgora.pl.\\
\ddag Janusz Gil Institute of Astronomy, University of Zielona G\'ora,\\ Licealna St. 9, 65--417 Zielona G\'ora, Poland\\
e-mail: a.maciejewski@ia.uz.zgora.pl\\
}
\maketitle

\begin{abstract}
 We study the translational motions of homonuclear diatomic molecules prepared in their ${}^3\Sigma$ electronic states, deeply bound vibrational states, and rotational states of well-defined parity. The trapping potential arises due to the interaction of the total spin of electrons and orbital angular momentum of nuclei with the trap's quadrupole magnetic field. The translational motion of a molecule is treated classically.  We examine the Hamilton equations that govern the center of mass dynamics numerically and analytically. Using data of a hydrogen molecule at the ground vibrational state, we present global dynamics using the  Poincar\'e section method and various types of trajectories: periodic, quasi-periodic and chaotic. We prove that the Hamiltonian system governing this motion is non-integrable. The particle's orbits are confined to a bound region of space that grows with energy, but for small energies (< 1.8 K), the motion is restricted to a processing chamber (a few centimetres).
 Solutions of equations of motion occurring on the symmetry axis and the horizontal plane are expressed in terms of Jacobi elliptic functions.
   \paragraph{Declaration} The article has been published in~\textit{J. Chem. Phys. 163, 154305~(2025)}, and the final version is available at: \textbf{\href{https://doi.org/10.1063/5.0291088}{https://doi.org/10.1063/5.0291088} }
\end{abstract}
\section{Introduction}\label{Intro}
\setcounter{equation}{0}

The effects of the electric and magnetic fields on the dynamical processes of atoms and molecules in the gas phase have been studied extensively in the last half a century. The cooling techniques, the deceleration methods and the trapping methods are the focus of the review article \cite{Soft2023} (see also references therein). They exploit spatially inhomogeneous and time-varying electric and magnetic fields to manipulate cold, dense samples of atoms and molecules in selected quantum states. The forces generated by inhomogeneous field distributions influence a neutral particle due to the interaction between the electromagnetic field and the particle's magnetic dipole moment. In particular, the Zeeman effect makes it possible to confine neutral particles by multipole magnetic fields. This happens because atoms and molecules that undergo a linear Zeeman effect can be thought of as having a magnetic dipole moment coupled to the applied external magnetic field.

At the beginning of atomic and molecular physics, electromagnetic forces were used to turn aside particles in well-collimated thermal beams. In their pioneering experiment, Stern and Gerlach \cite{Stern1922} applied the magnetic field to separate a beam of silver atoms into spin-up and spin-down magnetic components (see also \cite[\S 6.1]{Krems} and references therein). This idea was revived by Rabi and co-workers \cite{Rabi1939} who carried out the experiment which involved passing a molecular beam through regions with opposite-directed field gradients. This method provides accurate measurements of the magnetic dipole moments of atoms and molecules \cite{aquilanti1995}. Later, inhomogeneous fields have guaranteed control of the transverse motion of the supersonic beams of quantum-state-selected atoms and molecules \cite{Paul1978,hogan2011,narevicius2012,hogan2016}. Further development consisted in controlling also their longitudinal motion.

The present paper is devoted to the study of the dynamics of a diatomic molecule in a magnetic quadrupole trap. For this reason, we pay special attention to this device, which is widely used in experimental physics.
The research group of Migdall \cite{Migdal1985} carried out the first experiment on confining a neutral atom in a spherical quadrupole trap. Doyle and co-workers \cite{Doyle1985} reported the first experiment on trapping $\rm{CaH}$ molecules by a quadrupole magnetic field produced by two superconducting coils arranged in the anti-Helmholtz configuration. In Ref.~\cite{Riedel2011}, the experiment on confining of Stark-decelerated ground-state $\rm{NH}$ molecules by means of a quadrupole magnetic field is reported.

Magnetic trapping of methyl radicals ${\rm CH}_3$ was reported by Liu et al \cite{Liu2017}. The experimental setup involves an 85-stage Zeeman decelerator and two superconducting magnets in an anti-Helmholtz configuration. They produce a quadrupole magnetic field of maximum $0.75~\mathrm{T}$ in the $z$-direction and $0.45~\mathrm{T}$ in the radial plane.  McCarron et al. \cite{McCarron2018} carry out the two-stage experiment with a magnetic quadrupole trap. During the first stage, the laser-cooled beam of $\rm SrF$ molecules was captured in a magneto-optical trap. Then, in the second stage, forty percent of the molecular cloud was transferred into a single quantum state in the magnetic trap.

 Electromagnetic traps are still being developed in connection with quantum computer technology.
DeMille \cite{Mille2002} proposes a novel architecture of a quantum processor using ultra-cold diatomic molecules with electric dipole moments acted upon an external electric field. According to the author, ``this design can plausibly lead to a quantum computer with $\geq 10^4$ qubits, which can perform $\sim 10^5$ CNOT gates in the anticipated decoherence time of $\sim 5 s$.'' Many investigators have studied the possibility of realising quantum computations with polar molecules and have tried to extend the capabilities of this device.  We emphasise the series of papers \cite{Cald2018,Cald2019,Cald2020}. Caldwell {\em et al.} present experiments with ultra-cold polar molecules in a magnetic trap. So, in \cite{Cald2020}, the team studied rotational transitions in ${}^2\Sigma$ electronic state of molecules confined in a quadrupole magnetic trap. Magnetic sensitivities of transitions for the numerous diatomic molecules are calculated, suggesting that the long coherent times are realisable in a magnetic trap.
In \cite{Picard2025}, a two-qubit iSWAP gate is realised using $\rm{NaCs}$ molecules in $X^1\Sigma$ electronic states. The optical tweezer supplemented with the magnetic field $863.9~G$ is used. For recent progress in single quantum state preparation, rotational coherence of individual polar molecules, and intermolecular dipolar interactions leading to entanglement, see \cite{Picard2025} and references therein.

 The aim of this paper is to analyse the center-of-mass motion of a diatomic molecule in the quadrupole magnetic trap.
Since the de Broglie wavelength of the translational mode is very small, we adopt the semi-classical approach \cite{Hashemloo2015}. We treat the motion of a molecule as a whole classically, while its electronic, vibrational, and rotational motions are considered according to quantum theory.

The translational dynamics of ions and molecules influenced by a quadrupole magnetic field was announced by Bergeman \cite{metcalf1986c}. The results of numerical simulations are presented in \cite{metcalf1994}. Exemplary classical trajectories and Poincar\'e sections were given for the square root potential $\sqrt{z^2+\rho^2/4}$. In the present paper, we analyse the dynamics of polarised particles in a potential which also involves the additional term  $z^2+\rho^2/4$ originated from the second-order Zeeman shift. 
 
The plan of the article is as follows. In Section \ref{Zeeman}, we focus on the Zeeman shifts caused by the interaction of the total angular momentum of the molecule with the inhomogeneous field
\begin{equation}\label{B1}
\mathbf{B}(\mathbf{X})=\frac12 B_1\left(-X,-Y, 2Z\right),
\end{equation}
generated by a pair of identical coaxial coils whose currents are equal and flow in
opposite directions. We assume that the molecule is prepared in the ${}^3\Sigma$ electronic state and in a deeply bound vibrational state. We adopt the rigid rotor approximation. The Zeeman shifts result in the trapping potential $V(X,Y,Z)\varpropto |\mathbf{B}(\mathbf{X})|$ \cite{metcalf1986,metcalf1987,metcalf1994}. It is zero in the center of the processing chamber and increases in all directions.

The obtained classical center-of mass Hamiltonian is analysed in the following two sections. First, in Section \ref{Num-simu}, we present the results of numerical simulations of the translational motion of a molecule in the processing chamber by means of Poincar\'e sections and exemplary trajectories representing different types of motion.  In Section \ref{Integrability}, we present analytical results; that is, we prove the non-integrability of the analysed Hamilton equations and give explicit form of solutions on invariant submanifolds of the phase space.
The submanifolds correspond to the line $X=Y=0$ and to the plane $Z=0$ where the returning force $\mathbf{F}=-\nabla V$ is central and the angular momentum is conserved. 
To illustrate the situation, in Appendix \ref{App-A} we consider a quasi-stable hydrogen molecule. We adopt the Heitler-London approximation and use the trial wave function, which requires two electrons in the triplet spin configuration \cite{Grif}. We calculate the linear and quadratic Zeeman shifts to the free molecule energy levels. In Appendix \ref{App-B}, we briefly describe the necessary integrability conditions in the Liouville sense for natural Hamiltonians with homogeneous potentials with special emphasis on algebraic potentials. Quoted here, Theorem~\ref{thm:MoRaalg}  is a crucial tool used for the non-integrability proof in Section \ref{Integrability}. The turning points of regular motions governed by the central returning force of the trapping potential are given explicitly in the Appendix~\ref{App_C}.

\section{Zeeman effect}\label{Zeeman}
\setcounter{equation}{0}

In this Section, we will show how a molecule-field interaction can be employed to trap it. In our previous article \cite{JPhysB2025}, we discussed trapping a polar diatomic molecule by the sextupole electrostatic field. In the present work, we study the confinement of a homonuclear diatomic molecule by the quadrupole magnetic field. The combination of such fields creates a so-called polar trap introduced in \cite{NJPhys2020}, where classical translational and rotational dynamics of a polarised nanoparticle were analysed. The perturbation due to the magnetic field yields the splitting of the spectral lines. The spectral problem is governed by the Hamiltonian of a molecule consisting of two nuclei and $n$ electrons expressed in terms of center-of-mass variables presented in Appendix A of~\cite{JPhysB2025}.

We base our consideration on the Lagrangian function \cite[Eq.~(A.1)]{JPhysB2025}. Transforming into the nuclear center-of-mass relative coordinate system \cite[Eqs.~(A.2), (A.3)]{JPhysB2025}, we obtain the center of mass Hamiltonian \cite[Eq.~(A.13)]{JPhysB2025} after standard Legendre transformation. The kinetic energy part consists of the center-of-mass kinetic energy,  inner energies of constituents, and the mass-polarisation correction, which appears because of the transformation from laboratory to nuclear center-of-mass coordinates. The usual electrostatic potential describes the interaction of nuclei and electrons. The free molecule's terms are supplemented with the first-order \cite[Eq.~(A.19)]{JPhysB2025} and the second-order \cite[Eq.~(A.20)]{JPhysB2025} perturbations due to the external magnetic field.

In Appendix \ref{App-A} of the present paper, we adopt the multielectron Hamiltonian \cite[Eq.~(A.13)]{JPhysB2025} to the hydrogen molecule. Its behaviour will serve as a guideline. The influence of the magnetic field (\ref{B1}) on the electronic cloud of the molecule is analysed in detail. We ``sandwich'' the perturbations \cite[Eq.~(A.19)]{JPhysB2025} and \cite[Eq.~(A.20)]{JPhysB2025} between the trial wave function (\ref{Psi_minTot}) and its complex conjugate and integrate over electrons' coordinates. Since the total orbital angular momentum of electrons is zero, the linear Zeeman effect is defined by the interaction of the magnetic field and the orbital angular momentum of nuclei. The quadratic Zeeman gives the expression (\ref{psi-Hz-psi}) depending on the relative distance $R$ between the nuclei. Interaction with vibrational modes ``removes'' $R$ from consideration (see Eqs.~(\ref{A1A2_gr}) and (\ref{A1A2_1st})). The rotation of the molecule as a whole only remains to be processed.

The resulting expectation values are functions of coordinate-dependent magnetic field components and spherical angles $\theta$ and $\varphi$ defining the orientation of the internuclear axis in the laboratory frame. At this point, we assume that the expressions are applicable to a multielectron molecule.  In this Section, we will consider the linear and the quadratic Zeeman effects caused by the rotation of a rigid molecule in a magnetic field (\ref{B1}). The rotation of a molecule in the inhomogeneous magnetic field results in the space-dependent trapping potential \cite{metcalf1987,metcalf1994} involving in the center of mass Hamiltonian.

We start our consideration with the spin Zeeman effect in ${}^3\Sigma$ molecule separated from the ground electronic state by a large energy gap. We assume that the wave function is the direct product
\begin{equation}\label{Psi-total}
\Psi(\theta,\varphi;M_s)=\chi(M_s)\Psi_0^{\rm rot}(\theta,\varphi),
\end{equation}
where the triplet state $\chi(M_s)$ is the linear combination of spinors, corresponding to spin-up $M_s=+1$ orientation, spin-zero $M_s=0$ orientation, and spin-down  $M_s=-1$ orientation, respectively. The function $\Psi_0^{\rm rot}(\theta,\varphi)$ is the eigenstate of the rigid rotor Hamiltonian
\begin{equation}\label{rotor}
\hat{H}_0=b_e\hat{J}^2.
\end{equation}
Here $b_e=B_e/{\cal U}$ is the dimensionless rotational constant, i.e. rotational constant $B_e$ measured in the atomic unit of energy ${\cal U}=\hbar^2/(m_ea_0^2)$. For hydrogen we have $B_e=29.3\,\mathrm{cm}^{-1}$ \cite{NIST}, or $b_e=1.335\cdot 10^{-4}$ if it is measured in Hartree units.

\subsection{Spin Zeeman effect}\label{Spin-1st}

A diatomic molecule in ${}^3\Sigma$ electronic state possesses the magnetic dipole moment $\boldsymbol{\mu}=-\mu_Bg_S\mathbf{S}$ where $\mu_B=e\hbar/2m_e$ is the Bohr magneton and $g_S$ is the electron $g$-factor. The spin operator $\mathbf{S}$ is presented by Hermitian matrices
\begin{equation}\label{S1-matr}
S_x=\frac{1}{\sqrt{2}}\left[
\begin{array}{ccc}
 0 & 1 & 0 \\
 1 & 0 & 1 \\
 0 & 1 & 0
\end{array}\right],
\,
S_y=\frac{1}{\sqrt{2}}\left[
\begin{array}{ccc}
 0 & -\rmi& 0 \\
 \rmi& 0 &-\rmi \\
 0 & \rmi & 0
\end{array}\right],\,
S_z=\left[
\begin{array}{ccc}
 1 & 0 & 0 \\
 0 & 0 & 0 \\
 0 & 0 & -1
\end{array}\right],
\end{equation}
in spatial dimensions. The molecule experiences torque $[\vec{\mu}\times\mathbf{B}]$ in the magnetic field of the trap (\ref{B1}). The interaction produces the Zeeman operator $H_s=-(\vec{\mu}\cdot\mathbf{B})$. We measure the energy in Hartree. We scale the space coordinate in the magnetic field (\ref{B1}) by the characteristic trap size $D$, i.e. $(X,Y,Z)$ are replaced by $(x,y,z)=(X/D,Y/D,Z/D)$. In terms of dimensionless
variables, the operator takes the form
\begin{equation}\label{HS1}
H_s=\frac{g_S}{2}\beta_L\left[
\begin{array}{ccc}
 z & -\frac{\displaystyle x-\rmi y}{\displaystyle 2\sqrt{2}} & 0 \\
-\frac{\displaystyle x+\rmi y}{\displaystyle 2\sqrt{2}} & 0 & -\frac{\displaystyle x-\rmi y}{\displaystyle 2\sqrt{2}} \\
 0 & -\frac{\displaystyle x+\rmi y}{\displaystyle 2\sqrt{2}} & -z
\end{array}\right].
\end{equation}
The dimensionless ruling parameter
\begin{equation}\label{beta_L}
\beta_L=\frac{eB_1Da_0^2}{\hbar}
\end{equation}
characterises the intensity of interaction of spin (\ref{S1-matr}) and the magnetic field (\ref{B1}). Recall that $B_1$ is the constant field gradient, $D$ is the size of the processing chamber, $\hbar$ is the reduced Planck constant, $e$ is the atomic unit of charge, and $a_0$ is the atomic unit of length (Bohr radius).

The triplet states are eigenvectors of the matrix (\ref{HS1}) with eigenvalues
\begin{equation}\label{lmb-S1}
\lambda_+=\sqrt{z^2+\frac{x^2+y^2}{4}},\quad \lambda_0=0,\quad \lambda_-=-\sqrt{z^2+\frac{x^2+y^2}{4}},
\end{equation}
corresponding to spin-up, spin-zero, and spin-down orientations of spinors, respectively. The spinors are as follows:
\begin{equation}
\label{chis}
\chi_+=\left[
\begin{array}{c}
\frac{\displaystyle (B_x-\rmi B_y)^2}{\displaystyle 2|B|(|B|-B_z)}\\[1em]
\frac{\displaystyle B_x-\rmi B_y}{\displaystyle \sqrt{2}|B|}\\[1em]
\frac{\displaystyle |B|-B_z}{\displaystyle 2|B|}
\end{array}\right],\,\,
\chi_0=\left[
\begin{array}{c}
\frac{-\displaystyle (B_x-\rmi B_y)^2}{\displaystyle |B|\sqrt{2(B_x^2+B_y^2)}}\\[1em]
\frac{\displaystyle (B_x-\rmi B_y)B_z}{\displaystyle |B|\sqrt{B_x^2+B_y^2}}\\[1em]
\frac{\displaystyle \sqrt{B_x^2+B_y^2}}{\displaystyle \sqrt{2}|B|}
\end{array}\right],\,\,
\chi_-=\left[
\begin{array}{c}
\frac{\displaystyle (B_x-\rmi B_y)^2}{\displaystyle 2|B|(|B|+B_z)}\\[1em]
-\frac{\displaystyle B_x-\rmi B_y}{\displaystyle \sqrt{2}|B|}\\[1em]
\frac{\displaystyle |B|+B_z}{\displaystyle 2|B|}
\end{array}\right],
\end{equation}
where $B_x=-x/2$, $B_y=-y/2$, and $B_z=z$ are dimensionless components of the magnetic field (\ref{B1}) and $|B|$ is the magnitude of the field. They span the linear space of solutions to the Schr\"{o}dinger equation
\begin{equation}\label{chi-H_s}
\rmi\frac{\partial\chi(t)}{\partial t}=H_s\chi(t).
\end{equation}
The general solution is
\begin{equation}\label{chiS1}
\chi(t)=a\chi_+{\rm e}^{-\rmi E_+t}+b\chi_0{\rm e}^{-\rmi E_0t}+c\chi_-{\rm e}^{-\rmi E_-t}
\end{equation}
where energies are
\begin{equation}\label{E_S1}
E_k=\frac{g_S}{2}\beta_L\lambda_k.
\end{equation}
The complex-valued constants $a$, $b$, and $c$ satisfy the
normalisation condition $a^*a+b^*b+c^*c=1$. They are defined by
initial conditions.

Exploiting the orthogonality of basic spinors (\ref{chis})
we derive the common energy $E_s=\langle\chi|H_s|\chi\rangle$ of this mixed state
\begin{equation}\label{Eshift-S1}
E_s=\frac{g_S}{2}\beta_L\left(a^*a-c^*c\right)\sqrt{z^2+\frac{x^2+y^2}{4}}.
\end{equation}
If the mixing parameter $\varpi=a^*a-c^*c$ is positive, the molecule experiences the positive Zeeman effect, i.e. it is a low-field seeker. If not, it becomes a high-field seeker and cannot be trapped.

\subsection{Linear Zeeman effect}\label{Zeeman-1st}

Let the initial quantum state (\ref{Psi-total}) be composed of the angular momentum states of the same length $\sqrt{J(J+1)}$ and all possible values of its projection $M$ on the space-fixed quantisation axis $z$:
\begin{equation}\label{Psi_00rot}
\Psi_0^{\rm rot}(\theta,\varphi)=\sum_{M=-J}^JD_{JM}Y_J^M(\theta,\varphi).
\end{equation}
The expansion coefficients $D_{JM}$ are complex values. We assume that the wave function is normalised. The linear combination (\ref{Psi_00rot}) is an eigenstate of the rigid rotor Hamiltonian (\ref{rotor}) with eigenvalue $J$. The unperturbed state is degenerate: $H_0\Psi_0^{\rm rot}(\theta,\varphi)=b_eJ(J+1)\Psi_0^{\rm rot}(\theta,\varphi)$.

We introduce dimensionless variables based on the atomic unit of energy ${\cal U}$. Careful analysis in Appendix A prompts that the linear Zeeman effect is defined by the perturbation
\begin{equation}\label{H_L-1st}
H_{\rm Zeeman}^{(1)}=-\alpha_L\beta_L(\mathbf{B}\cdot[\mathbf{R}\times\mathbf{\Pi}]).
\end{equation}
The dimensionless multiplier $\alpha_L$ depends on the particles' masses and charges
\begin{equation}\label{alpha_M}
\alpha_L=\frac{m_e}{2m}\left(Z_B\frac{m_A}{m_B}+Z_A\frac{m_B}{m_A}+n\frac{m_e}{m_B}\right).
\end{equation}
Here, $Z_A$ and $Z_B$ are the atomic numbers of the atoms of a diatomic molecule. The coefficient differs from the negative one-half $g_R$ factor \cite[eq.~(3.58)]{Krems} in an extremely small term $n m_e/m_B$ where $n=Z_A+Z_B$ is the number of electrons.

The term (\ref{H_L-1st}) describes the interaction of the dimensionless magnetic field $\mathbf{B}=(-x/2,-y/2,z)$ and the orbital angular momentum of the nuclei $\mathbf{L}=[\mathbf{R}\times\mathbf{\Pi}]$ with dimensionless components
\begin{eqnarray}
L^1&=&-\rmi \left(-\sin\varphi\frac{\partial}{\partial\theta}-\cos\varphi\cot\theta\frac{\partial}{\partial\varphi}\right),
\label{L1}\\
L^2&=&-\rmi \left(\cos\varphi\frac{\partial}{\partial\theta}-\sin\varphi\cot\theta\frac{\partial}{\partial\varphi}\right),
\label{L2}\\
L^3&=&-\rmi \frac{\partial}{\partial\varphi}.\label{L3}
\end{eqnarray}
Following \cite[\S 4.3.2]{Grif}, we introduce the raising $L_+=L^1+\mathrm{i}L^2$ and lowering operators $L_-=L^1-\mathrm{i}L^2$. Interaction with the magnetic field modifies the rotational modes. Using the relations \cite[eqs.~(4.120), (4.121)]{Grif} we derive the law
\begin{eqnarray}
(\mathbf{B}\cdot\mathbf{\hat{L}})Y_J^M(\theta,\varphi)&=&
-\frac14\left[(x-\rmi y)\sqrt{(J-M)(J+M+1)}Y_J^{M+1}(\theta,\varphi)\right.\nonumber\\
&+&\left.(x+\rmi y)\sqrt{(J+M)(J-M+1)}Y_J^{M-1}(\theta,\varphi)\right]\nonumber\\
&+&zMY_J^M(\theta,\varphi).
\end{eqnarray}
The perturbed wave function takes the form $\Psi^{\rm rot}=\Psi_0^{\rm rot}+\Psi_1^{\rm rot}$. The correction $\Psi_1^{\rm rot}$ to the wave function $\Psi_0^{\rm rot}$ satisfies the equation
\begin{equation}\label{E_1rot}
H_0\Psi_1^{\rm rot}+H_{\rm Zeeman}^{(1)}\Psi_0^{\rm rot}=b_eJ(J+1)\Psi_1^{\rm rot} + E_1\Psi_0^{\rm rot},
\end{equation}
where $E_1$ is the correction to the eigenvalue $b_eJ(J+1)$. Now we take the inner product of this equation with the spherical harmonic $Y_J^N(\theta,\varphi)$. But the operator $H_0$ is hermitian, so the first term in the left-hand side of the modified equation cancels the first term on the right. Exploiting the orthogonality of spherical harmonics, we arrive at the eigenvalue equation
\begin{equation}
H_{NM}D_{JM}=\varepsilon_1D_{JN},
\end{equation}
where $\varepsilon_1=-E_1/\alpha_L\beta_L$ and $H_{NM}=\langle Y_J^N|(\mathbf{B}\cdot\mathbf{\hat{L}})|Y_J^M\rangle$. The tridiagonal matrix $H_{NM}$ has the following non-trivial elements
\begin{eqnarray}
H_{M,M-1}&=&-\frac{x-\rmi y}{4}\sqrt{(J-M+1)(J+M)},\nonumber\\
H_{M,M}&=&z M,\nonumber\\
H_{M,M+1}&=&-\frac{x+\rmi y}{4}\sqrt{(J+M+1)(J-M)}.\nonumber
\end{eqnarray}
The eigenvalues of this tridiagonal matrix give the first-order corrections to the rotation energy $B_eJ(J+1)$:
\begin{equation}\label{E1_M}
\varepsilon_1^M=M\sqrt{z^2+\frac14(x^2+y^2)}.
\end{equation}
The perturbation $H_{\rm Zeeman}^{(1)}$ breaks the degeneracy. The correction of rotational energy $E_0=b_eJ(J+1)$ takes the form of coordinate-dependent function
\begin{equation}\label{E_1M}
E_1(x,y,z)=-\alpha_L\beta_LM\sqrt{z^2+\frac14(x^2+y^2)}.
\end{equation}
The result is in line with the Zeeman shift of the rotational states of molecules in a ${}^2\Pi$ electronic state \cite[\S~3.5.2]{Krems}. The state with rotational angular momentum $J$ splits into $2J+1$ Zeeman states corresponding to all possible values of its projection $M$ on the space-fixed quantisation axis $z$. The correction is analogous to that (\ref{Eshift-S1}). The difference is in the factors before the square root function. The spin Zeeman effect is much more intense than the linear Zeeman effect.

\subsection{Quadratic Zeeman effect}\label{Zeeman-2nd}

Let us look more closely at the product $\alpha_L\beta_L$of $g_R$ factor (\ref{alpha_M}) and ruling parameter (\ref{beta_L}) which define the intensity of interaction of trap's magnetic field and dipole momentum of nuclear orbital motion. The factor $\alpha_L$ is proportional to the ratio $m_e/m$ of the mass of the electron $m_e$ to the total mass $m$ of the diatomic molecule. It is significantly smaller than a unit. The second-order perturbation
is proportional to the factor $\beta_L^2$. This is approximately one-fifth of $\alpha_L\beta_L$. Therefore, the second-order Zeeman effect should be taken into account.

In Appendix \ref{App-A} we analyse the energy shifts caused by this perturbation for the hydrogen molecule in a ${}^3\Sigma$ electronic state. Summarising, the term
\begin{eqnarray}\label{Zee2R-H2}
(\mathbf{B}\cdot\mathbf{n}_R)^2&=&
\frac{x^2}{4}\cos^2\varphi\sin^2\theta+\frac{y^2}{4}\sin^2\varphi\sin^2\theta
+z^2\cos^2\theta\\
&+&\frac{xy}{2}\cos\varphi\sin\varphi\sin^2\theta-z\left(x\cos\varphi+y\sin\varphi\right)\sin\theta\cos\theta,\nonumber
\end{eqnarray}
defines the rotation energy shifts (see eq.~(\ref{psi-Hz-psi})). It is multiplied by a numerical factor depending on the particles' charges and masses. The dimensionless center-of-mass variables $(x,y,z)$ and polar angles survive in the space fixed frame.

Let us calculate the rotation energy shift corresponding to the initial quantum state (\ref{Psi_00rot}). The term (\ref{Zee2R-H2}) is the perturbation to the Hamiltonian $H_0=B_e\mathbf{J}^2$. Using Clebsch-Gordan coefficients, we find the expectation values of the combinations of trigonometric functions (\ref{Zee2R-H2}). Corrections to the rotational energy are defined by the eigenvalues of the pentadiagonal matrix $\hat{W}=\langle Y_J^N|(\mathbf{B}\cdot\mathbf{n}_R)^2|Y_J^M\rangle$ with elements
\begin{eqnarray}
W_{M,M-2}&=&-\frac{(x-\rmi y)^2}{8}\frac{\sqrt{(J-M+1)(J-M+2)(J+M-1)(J+M)}}{(2J-1)(2J+3)},\nonumber\\
W_{M,M-1}&=&\frac{(x-\rmi y)z}{2}\frac{\sqrt{(J-M+1)(J+M)}}{(2J-1)(2J+3)}(2M-1),\nonumber\\
W_{M,M}&=&\left(\frac{J^2+J-1+M^2}{(2J-1)(2J+3)}\right)\frac{x^2+y^2}{4}+\frac{2J^2+2J-1-2M^2}{(2J-1)(2J+3)}z^2,\\
W_{M,M+1}&=&\frac{(x+\rmi y)z}{2}\frac{\sqrt{(J-M)(J+M+1)}}{(2J-1)(2J+3)}(2M+1),\nonumber\\
W_{M,M+2}&=&-\frac{(x+\rmi y)^2}{8}\frac{\sqrt{(J-M-1)(J-M)(J+M+1)(J+M+2)}}{(2J-1)(2J+3)}.\nonumber
\end{eqnarray}
The eigenvalues are as follows
\begin{equation}\label{E_2M}
E_2^M=\frac{2J^2+2J-1-2M^2}{(2J-1)(2J+3)}\left(z^2+\frac{x^2+y^2}{4}\right).
\end{equation}
There are $J$ double degenerate eigenvalues corresponding to the pair $\pm M$ and non-degenerate ones for $M=0$. The correction is proportional to the square of the magnetic field magnitude.

 When we add all of the contributions to the energy -- the spin correction (\ref{Eshift-S1}), the linear (\ref{E_1M}), and the quadratic (\ref{E_2M}) Zeeman shifts -- we obtain the trapping potential as a function of space coordinates that span the cavity within the processing chamber:
\begin{eqnarray}\label{V_CM}
V_{\mathrm{cm}}(\mathbf{x})&=&\frac{g_S}{2}\beta_L\varpi\sqrt{z^2+\frac{x^2+y^2}{4}}\nonumber\\
&-&\alpha_L\beta_L M\sqrt{z^2+\frac{x^2+y^2}{4}}\nonumber\\
&+&\beta_L^2\left[A_1-A_2\frac{2J^2+2J-1-2M^2}{(2J-1)(2J+3)}\right]\left(z^2+\frac{x^2+y^2}{4}\right).
\end{eqnarray}
Recall that the parameter $\varpi=a^*a-c^*c$. Numerical coefficients $A_1$ and $A_2$ depend on electronic state and vibrational state matrix elements as well as on charges and masses of nuclei and electrons. In Appendix \ref{App-A} we calculate them for hydrogen molecule in  $f^3\Sigma_{u}^+4p\sigma$ electronic state (see eqs.~(\ref{A1A2_gr}) and (\ref{A1A2_1st})).

The kinetic part of the Hamiltonian \cite[eq.~(A.13)]{JPhysB2025} governing the evolution of a multi-electron diatomic molecule contains translational kinetic energy $\mathbf{P}^2/2m$, kinetic energies of constituents, and mass polarisation corrections. If one ``sandwiches'' $K$ between a particular electronic-vibrational-rotational state of a molecule, they obtain the term $\mathbf{P}^2/2m$ again, plus the expectation value of this part of the molecule's energy in this particular state. Since the electromagnetic Lagrangian \cite[eq.~(A.1)]{JPhysB2025} contains the vector potential coupled with particles' velocities, the magnetic field perturbation
\cite[eq.~(A.19)]{JPhysB2025} contains the translational momentum $\mathbf{P}$. In terms of dimensionless coordinates \cite[eq.~(A.25)]{JPhysB2025} and momenta \cite[eq.~(A.26)]{JPhysB2025} the scalar product $-(\mathbf{P}\cdot\mathbf{b}_X)/2m$ takes the form
\begin{equation}\label{weak}
-\frac{(\mathbf{P}\cdot\mathbf{b}_X)}{2m}\frac{1}{{\cal U}}=-\gamma_L p_i\varepsilon^i{}_{jk}B^jd^k+\frac12 \frac{a_0}{D}\gamma_Lp_i\varepsilon^i{}_{kr}\frac{\partial B^r}{\partial x^j}\Lambda^{kj},
\end{equation}
where factor $\gamma_L$ equals
\begin{equation}\label{gamma_L}
\gamma_L=\frac{m_e}{m}\beta_L \frac{|\mathbf{d}|}{ea_0}.
\end{equation}
Tensor $\Lambda^{kj}$ is given by \cite[eq.~(A.9)]{JPhysB2025}. By $\mathbf{d}$ we mean the electric dipole operator \cite[eq.~(A.18)]{JPhysB2025}. The terms (\ref{weak}) describe the relation between translational motion and intramolecular degrees of freedom \cite{fucuda19877}. If one ``sandwiches'' it between an electronic state and then a vibrational state, they derive the permanent electric dipole moment of a diatomic molecule in these particular states. The vector is directed along the internuclear axis. In the space-fixed frame, its components are
\begin{equation}\label{d123}
d^1=\mathrm{d}\cos\varphi\sin\theta,\quad d^2=\mathrm{d}\sin\varphi\sin\theta,\quad d^3=\mathrm{d}\cos\theta,
\end{equation}
where magnitude $d$ is measured in the atomic unit of electric dipole moment \cite[p.~298]{Krems}. The expectation value $\langle\Psi_0^{\rm rot}|\mathbf{d}|\Psi_0^{\rm rot}\rangle=0$ because the wave function (\ref{Psi_00rot}) has a
well-defined parity. The term is significant in classical dynamics of polarised nanoparticles moving in combined electric and magnetic fields of an electromagnetic trap for polar particles \cite{NJPhys2020}. The evolution of the electric dipole vector is analysed in detail, and the results of several numerical simulations are presented in this paper. 

Thanks to the ratio $a_0/D$, where $a_0$ is the Bohr radius and $D$ is the characteristic size of the trap, the second term in eq. (\ref{weak}) is extremely small. We neglect it in this paper. Neither the first term in  (\ref{weak})  nor the second one contributes to energy. Finally, the center of mass Hamiltonian is
\begin{equation}\label{Hvbr_rot_CM}
\overline{H}_{\mathrm{cm}}(\mathbf{x},\mathbf{p})=\frac12\left(p_x^2+p_y^2+p_z^2\right)+V_{\mathrm{cm}}(\mathbf{x}).
\end{equation}
The function is written in terms of dimensionless coordinates and momenta defined in eqs. (A.25) and (A.26) in \cite{JPhysB2025}; the parameter of evolution $\bar{\tau}$ is given by \cite[eq.~(A.27)]{JPhysB2025}. The Hamiltonian function (\ref{Hvbr_rot_CM}) governs the translational motion of the molecule as a whole. In subsequent sections, we use it for analysis of dynamics and numerical simulations.

\subsubsection{Depth of the trap}\label{Depth-trap}

 The depth of the confining potential depends on the properties of the molecule as well as on the magnitude of the magnetic field that can be experimentally realised. The field is produced by a pair of parallel circular superconducting coils running current in opposite directions. It is possible to generate a magnetic field $\sim5~\mathrm{T}$ \cite{Harris2004} using high-quality magnets. The diameter of the processing chamber is $8~\mathrm{cm}$ (see Fig.1 in this paper). Therefore, the gradient of the quadrupole magnetic field is $125\, {\rm T}\cdot {\rm m}^{-1}$. The state-of-the-art techniques can be used to produce a gradient of up to $250\, {\rm T}\cdot {\rm m}^{-1}$ \cite{Sarma2005}.

The depth of the trap is the difference in the potential energy (\ref{V_CM}) between the minimum point $x=y=z=0$ and the edge of the processing chamber $x=y=z=1$:
\begin{eqnarray}\label{Delta_V}
\triangle V&=&\frac{g_S}{2}\beta_L\varpi\sqrt{\frac{3}{2}}\nonumber\\
&-&\alpha_L\beta_L M\sqrt{\frac{3}{2}}\nonumber\\
&+&\beta_L^2\left[A_1-A_2\frac{2J^2+2J-1-2M^2}{(2J-1)(2J+3)}\right]\frac{3}{2}.
\end{eqnarray}
Let us analyse all three terms. For a rough estimate, it is enough to multiply the factors $\beta_L$, $\alpha_L\beta_L$, and $\beta_L^2$ in the atomic unit of energy measured in Kelvin: ${\cal U}_K=315775.23~\mathrm{K}$.
 The ruling parameter (\ref{beta_L}) contains the product $B_1D$. It is equal to the depth of the trap $B_{\rm trap}$ in \cite{Harris2004}. Further in this paragraph we take $B_{\rm trap}=5~{\rm T}$. So, the estimation of the first term in eq.~(\ref{Delta_V})  gives $\sim 6.7~\mathrm{K}$.

 The linear Zeeman effect is proportional to the factor $\alpha_L\beta_L$. It is less than the spin contribution in four orders. The quadratic Zeeman contribution in the potential energy balance is proportional to the factor $\beta_L^2$. It can reach values close to one-fifth of the linear contribution.

The values of $\alpha_L$ (\ref{alpha_M}) for homonuclear molecules ($g_R$ factor \cite[eq.~(3.58)]{Krems}) are given in Table~\ref{tab1}.
If the molecule jumps to the stable energy level with zero spin, the first term in the trapping potential (\ref{V_CM}) vanishes.
The trap depth decreases to hundreds of microkelvins (see Table 1, fifth column). If the spinless linear molecule is prepared in the state where the rotational angular momentum
is perpendicular to the quantisation axis, then the second term in the trapping potential (\ref{V_CM}) vanishes too. The molecule as a whole moves in a harmonic potential.

\begin{table}[h]
\begin{center}
\begin{tabular}{|c|c|c|c|c|}
\hline
No.&Molecule & Atomic number $Z$ & $\alpha_L\times 10^{-4}$ & Depth, $\mu\mathrm{K}$ \\
\hline
1.&Hydrogen $\mathrm{H}_2$&1& $2.72309$ & 1829.13 \\
\hline
2.&Nitrogen $\mathrm{N}_2$&7& $1.37085$ & 920.814 \\
\hline
3.& Oxygen $\mathrm{O}_2$&8& $1.37155$ &  921.285\\
\hline
4.& Fluorine $\mathrm{F}_2$&9& $1.28898$ & 865.826 \\
\hline
5.& Chlorine $\mathrm{Cl}_2$&17& $1.31526$ & 883.478 \\
\hline
6.& Bromine $\mathrm{Br}_2$&35& $1.20147$ & 807.041 \\
\hline
7.& Iodine $\mathrm{I}_2$&53& $1.14554$ & 769.474 \\
\hline
\end{tabular}
\end{center}
\caption{Estimation of values of the trap depth for homonuclear molecules. Magnetic field $B_{\rm trap}=5~{\rm T}$ produces trap depth $\sim 6.7~\mathrm{K}$ ($\beta_L=2.12718\cdot 10^{-5}$) due to spin Zeeman effect. Contributions caused by the linear Zeeman effect are presented in the fifth row. Quadratic Zeeman effect contributes a correction $\sim 142.8~\mu\mathrm{K}$.
\label{tab1}}
\end{table}

\section{Numerical simulations}\label{Num-simu}
\setcounter{equation}{0}

Hamiltonian \eqref{Hvbr_rot_CM} can be rewritten as
\begin{equation}
H_{\mathrm{cm}}(\mathbf{x},\mathbf{p})=\frac{1}{\beta_L}
\overline{H}_{\mathrm{cm}}(\mathbf{x},\mathbf{p})=
\frac12\left(p_x^2+p_y^2+p_z^2\right)+ V_{\mathrm{cm}}(\mathbf{x} ),
\label{Hcm_Edp}
\end{equation}
where 
\begin{equation}
  \label{eq:V_cart}
  V_{\mathrm{cm}}(\mathbf{x} ) = \sigma\sqrt{z^2+\frac{x^2+y^2}{4}}+
  2\delta\left(z^2+\frac{x^2+y^2}{4}\right),
\end{equation}
and 
\begin{equation}
\sigma= \frac{g_s}{2}\left(a^*a-c^*c\right)-\alpha_L M,\quad
\delta=\frac{\beta_L}{2}\left[A_1-A_2\frac{2J^2+2J-1-2M^2}{(2J-1)(2J+3)}\right].
\end{equation}
We approximate $g_s=2$.

The dimensionless parameter of evolution defined in \cite[eq.~(A.27)]{JPhysB2025} is replaced by a new time parameter 
\[
\tau=\sqrt{\beta_L}\overline{\tau}=\sqrt{\beta_L}\sqrt{\frac{M}{m_e}} \frac{\hbar}{a_0MD}t.
\]
Until now, the dot will denote the derivative with respect to $\tau$ and the momenta $p_i$ \cite[eq.~(A.26)]{JPhysB2025} change to $p_i/\sqrt{\beta_L}$ for $i=x,y,z$. 
 
 The numerical calculations in this Section were made for $\alpha_L=2.72309\cdot 10^{-4}$ and $\beta_L=2.12718\cdot 10^{-5}$, which corresponds to the hydrogen molecule. We also choose $a^*a-c^*c=1/2$ and values 
$A_1=0.5691906099701544$, $A_2=0.1665675408030196$, corresponding to the ground vibrational state of hydrogen, see eq.~\eqref{A1A2_gr} in Appendix~\ref{App-A}.
We present results of numerical simulations for values $J=10$ and $M=-10$. All these values give the coefficients
$\sigma=0.502723$ and $\delta=6.01911\cdot 10^{-6}$ and the time scaling $t=2.0812\cdot10^{-4} \tau\,\,(\mathrm{s})$.

First, we visualise the global dynamics analysing trajectories with various initial conditions by means of the Poincar\'e section method, see e.g. \cite[ch. 7]{Child}. To achieve this, it is convenient to use cylindrical coordinates $(r,z,\varphi)$,
$x=r\cos\varphi,$ $y=r\sin\varphi$ in which Hamiltonian \eqref{Hcm_Edp} has the form
\begin{equation}
\scH=\frac{1}{2}\left(p_r^2+p_z^2+\frac{p_{\varphi}^2}{r^2} \right)+
  \frac{\sigma}{2}\sqrt{r^2+4z^2}+
 \frac{\delta}{2}\left(r^2+4z^2\right).   
 \label{eq:Hcyl}
\end{equation}

By the axial symmetry of the magnetic field, $\varphi$ is a cyclic variable and the corresponding momentum $p_{\varphi}= r^2 \dot \varphi$ is a constant of motion. Thus, we reduce the system to one with two degrees of freedom with the parameter $p_{\varphi}$. 

For a fixed value of energy $h$, we consider the level
\[
M_{h}=\{(r,z,p_r,p_z)\in\R^4\ |\ \scH(r,z,p_r,p_z)= h \},
\]
where $H$ is given by~\eqref{eq:Hcyl} and $p_\varphi$ is considered as a parameter. 

\begin{figure}[h!tp]
  \centering
   \begin{subfigure}[b]{0.49\textwidth}
    \centering \includegraphics[width=\textwidth]{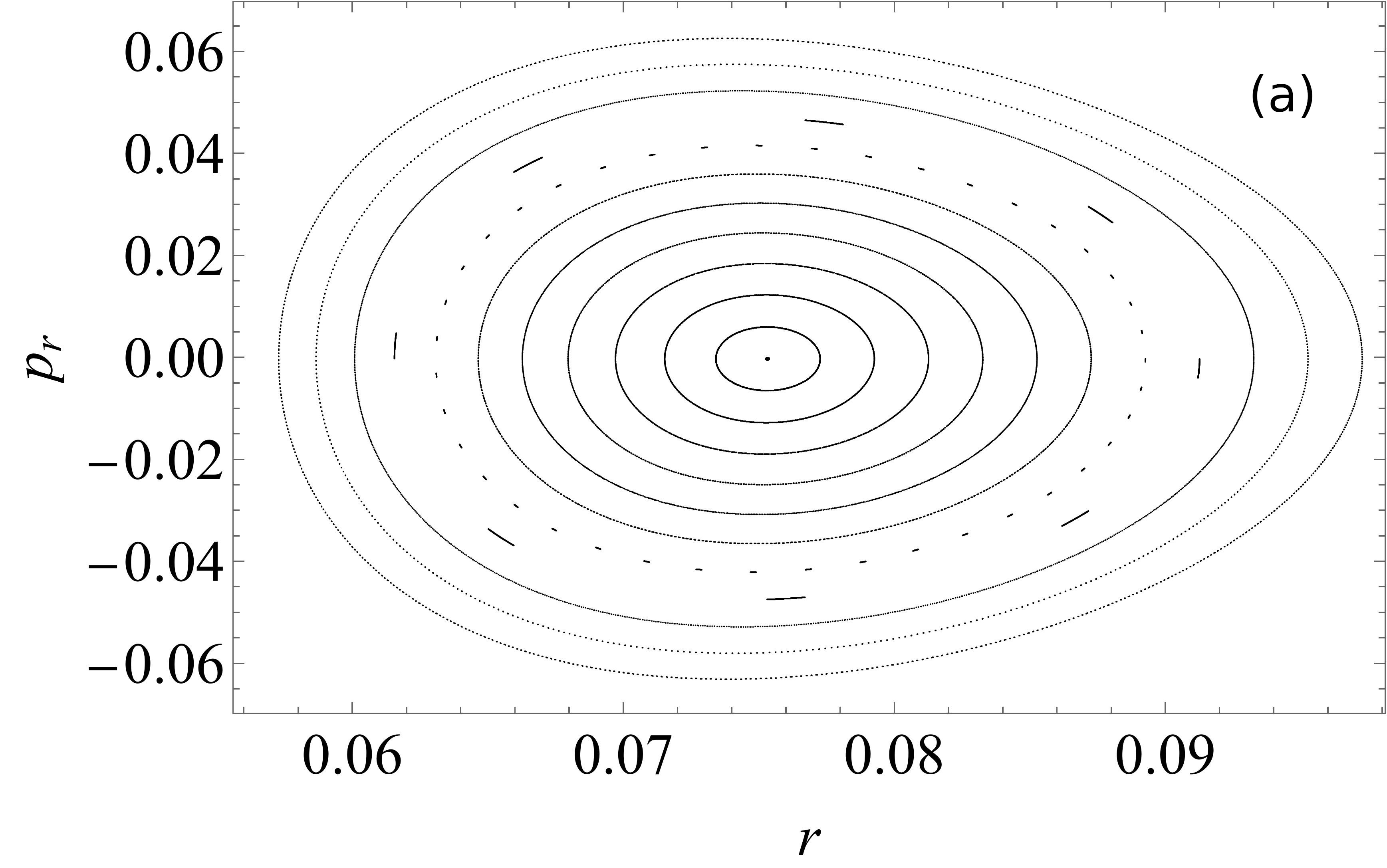}
  \end{subfigure}
   \begin{subfigure}[b]{0.49\textwidth}
    \centering \includegraphics[width=\textwidth]{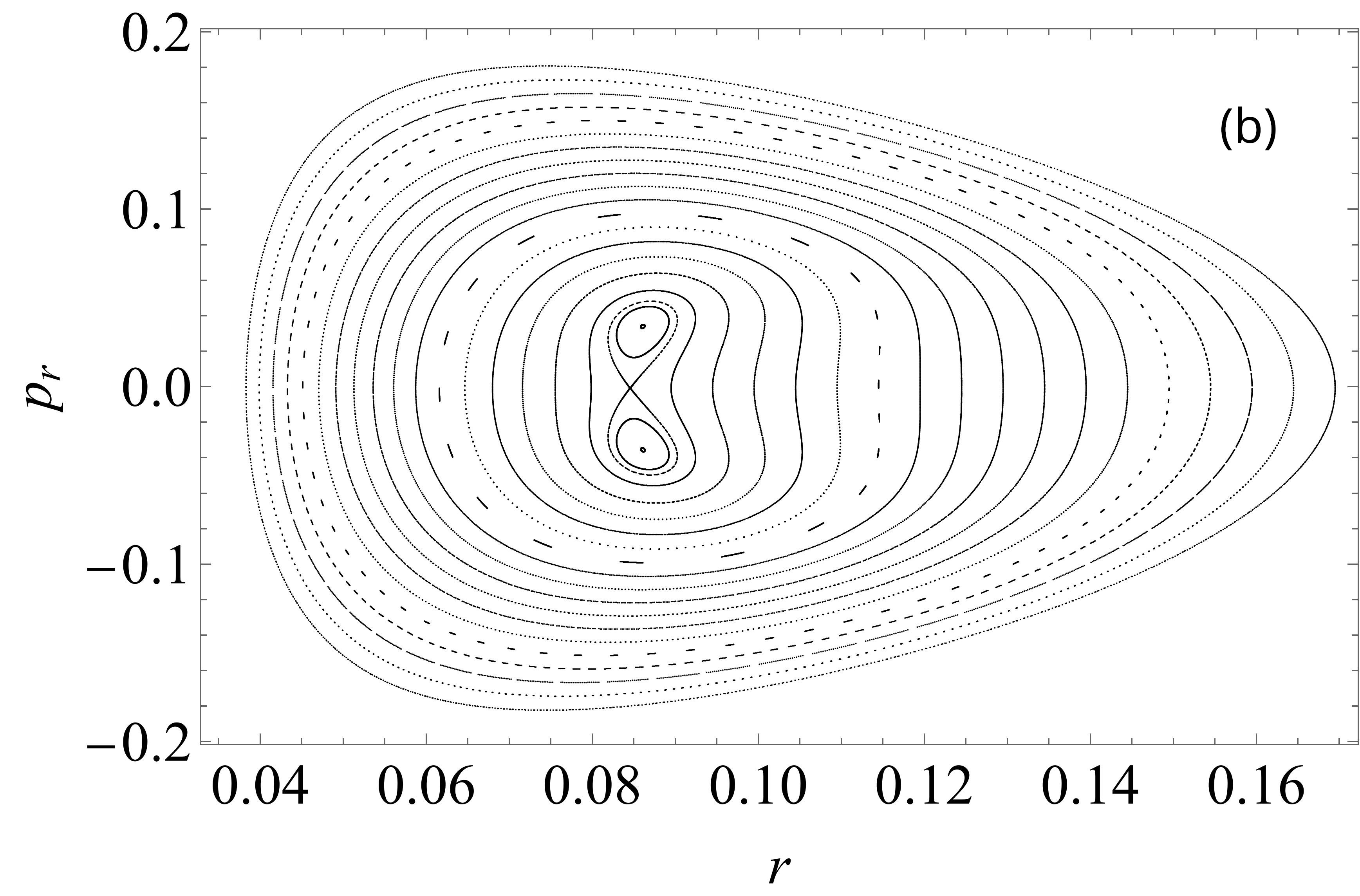}
  \end{subfigure}
    \begin{subfigure}[b]{0.49\textwidth}
    \centering \includegraphics[width=\textwidth]{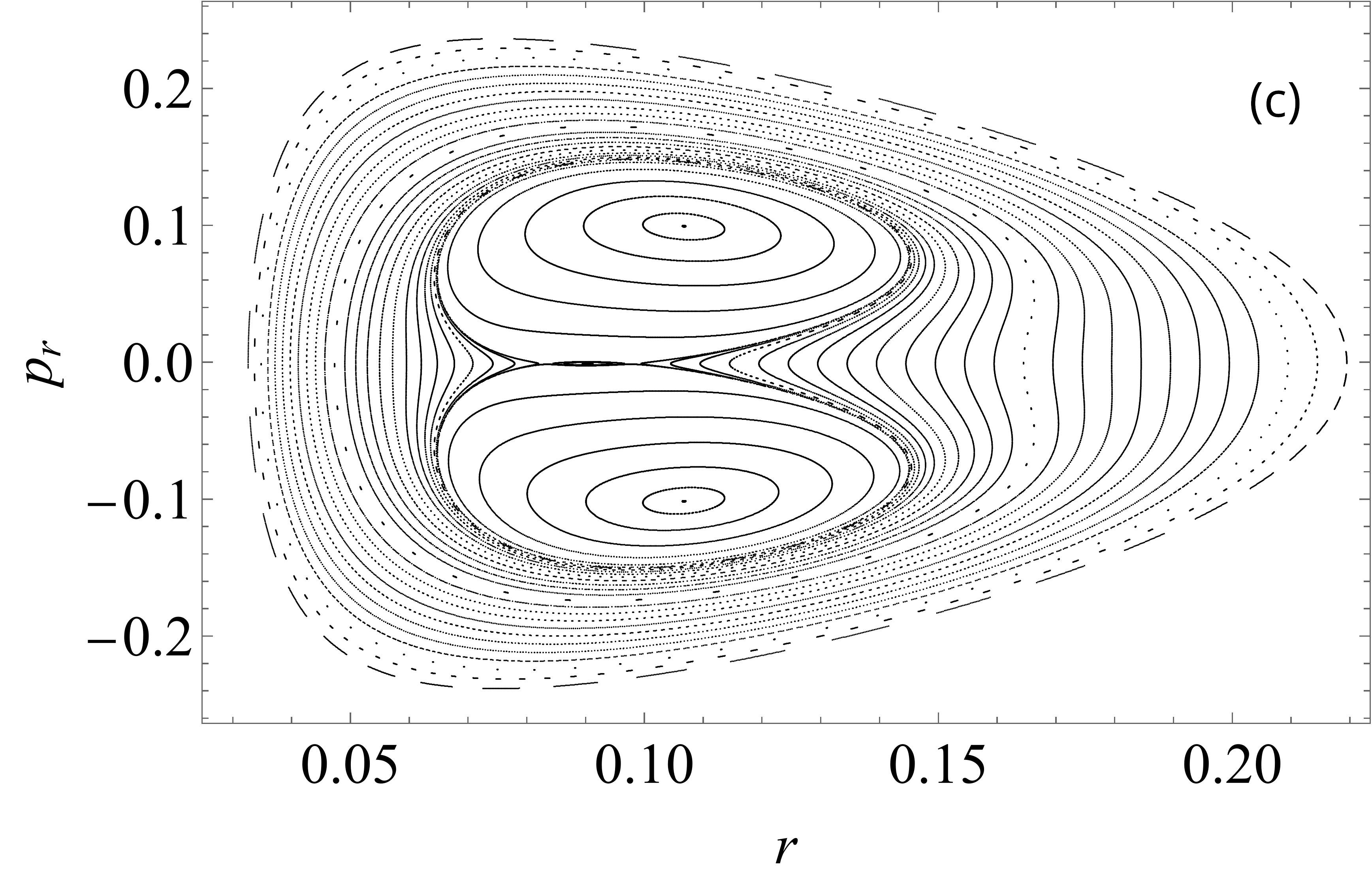}
  \end{subfigure}
   \begin{subfigure}[b]{0.49\textwidth}
    \centering \includegraphics[width=\textwidth]{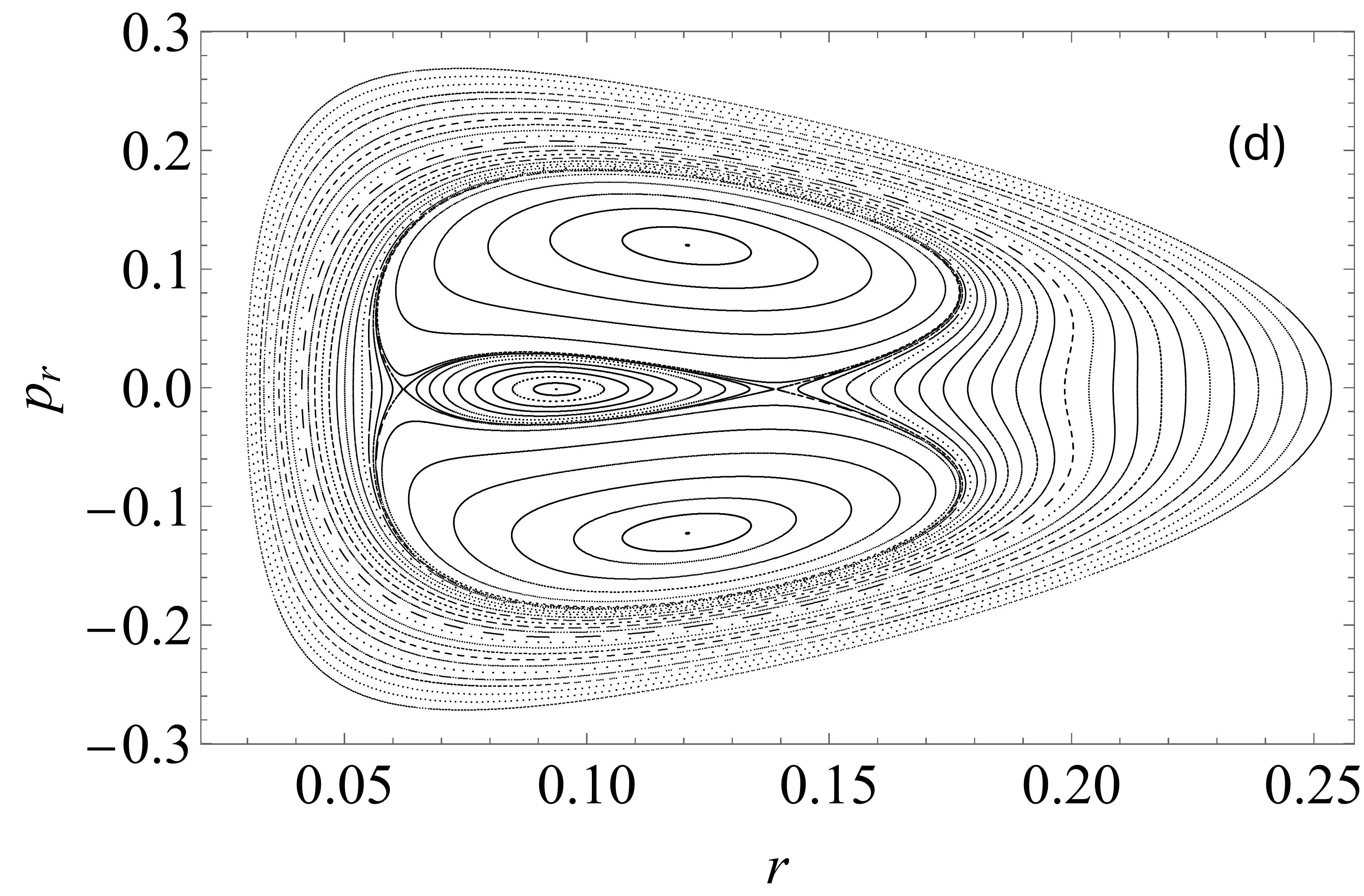}
  \end{subfigure}
   \caption{\small The Poincar\'e section for parameters $\sigma=0.502723$ and $\delta=0.0000179305$, $p_{\varphi}=\frac{1}{100}$ for different values of energy $h$. Cross-section plane $z=0$ and the orientation of the trajectories $p_z>0$. Energy and temperature values in the subfigures:  (a)  $h=0.03$ (0.4201~K);  (b) $h=0.0449$ (0.6288~K); (c) $h=0.057$ (0.7983~K);  (d) $h=0.065$ (0.9103~K).\label{fig:variousenergies}}
\end{figure}

Restricted to this level, our system becomes three-dimensional; thus, we can apply the Poincar\'e section method to investigate its dynamics. This method enables a qualitative study of the global properties of the system. 

As the cross-section plane, we choose $z=0$ with coordinates $(r,p_r)$. For a given initial condition, the trajectory is calculated numerically. We mark its intersections with this plane, but only points for which $p_z>0$ are displayed. The sections presented for $p_{\varphi}=\frac{1}{100}$ were calculated. It is important to notice that the levels $M_{h}$ are bounded, so the domain of motion of the particle is always bounded, and the domain size depends on the particle's energy.

A sequence of four cross sections is shown in Fig.~\ref{fig:variousenergies}. For small energy $h=0.03$ (0.4201~K), a stable periodic solution corresponds to a dot in the centre of the cross section shown in Fig.~\ref{fig:variousenergies}(a). The ovals surrounding this point correspond to quasiperiodic solutions. These ovals are intersections of invariant tori by the cross-section plane. In each torus, the motion is quasiperiodic and is characterised by two frequencies $\omega_1$ and $\omega_2$. If the ratio of these frequencies is irrational, then a solution fills the torus densely.  If $\omega_1/\omega_2 = l/k$ is a rational number, then the whole torus is filled by periodic solutions. Such a periodic solution is just a finite sequence of points in the cross-section. One can notice that the density of points changes with the oval size. This shows that $\omega_1/\omega_2$ is not constant.  

The dynamics of the system changes qualitatively as the energy increases. The central periodic solution mentioned above loses its stability and two new stable periodic solutions appear near it; see \ref{fig:variousenergies}(b). All of them are surrounded by invariant tori and form a characteristic "figure 8" shape.  

\begin{figure}[h!tp]
  \centering
   \includegraphics[scale=0.84]{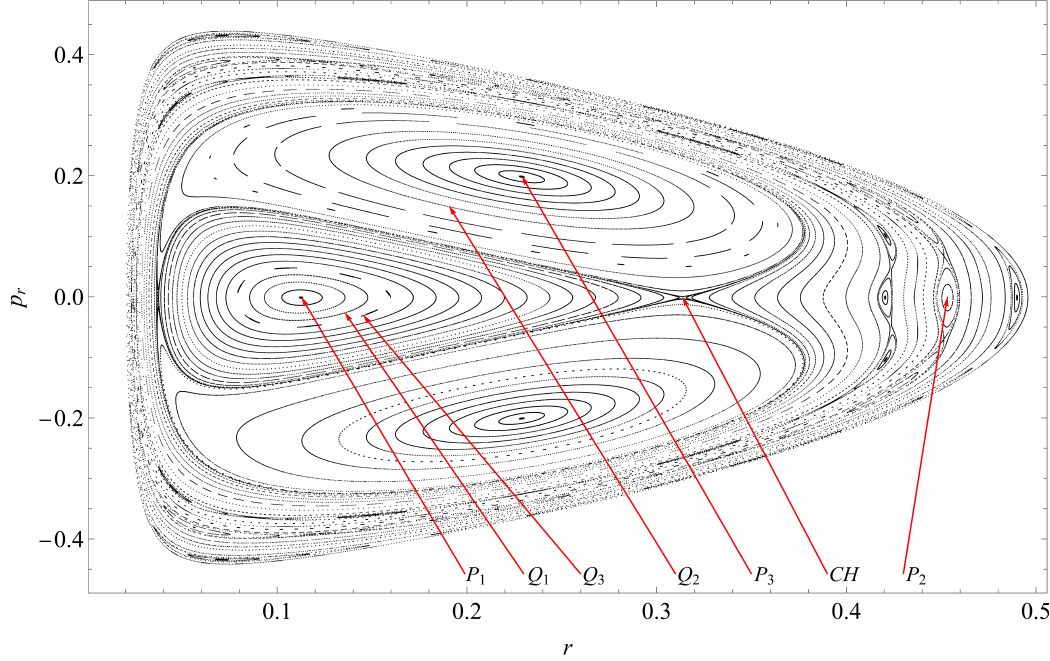}
\caption{\small The Poincar\'e section for parameters $\sigma=0.502723$ and $\delta=0.0000179305$,  $p_{\varphi}=\frac{1}{100}$ and the energy value $h=0.125$ (1.7507~K). Cross-section plane $z=0$ and the orientation of the trajectories $p_z>0$.\label{fig:J10Mm10_en_0d125_arrows}}
\end{figure}

A further increase in energy leads to the next bifurcation. The central unstable periodic solution now splits into two unstable and one stable periodic solution; see Fig.~\ref{fig:variousenergies}(c) and \ref{fig:variousenergies}(d), where the dominant central structure with three stable and two unstable periodic solutions is shown.

\begin{figure}[h!tp]
  \centering
   \includegraphics[scale=0.74]{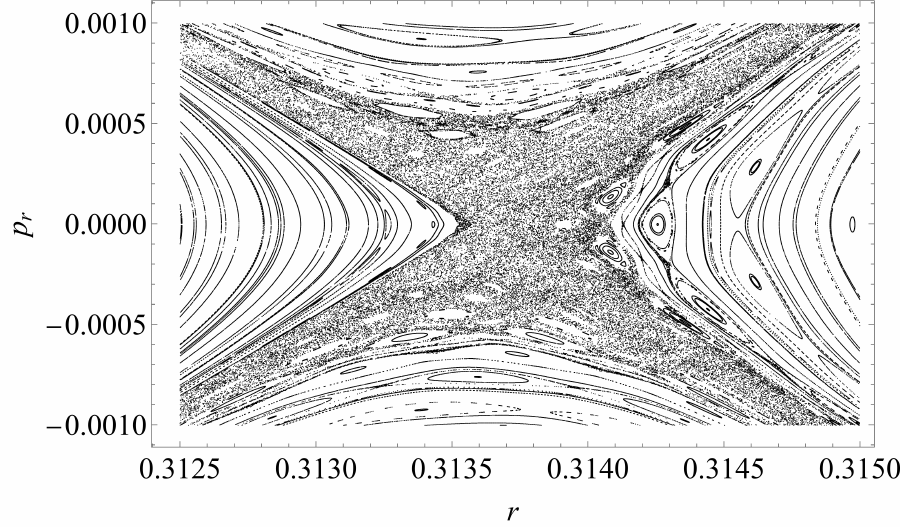}
\caption{\small Magnification of chaotic region near intersection point $CH$ in Fig.~\ref{fig:J10Mm10_en_0d125_arrows}. \label{fig:J10Mm10_cross_en_1over8_magnifi}}
\end{figure}

For $h=0.125$ (1.7507~K), the domain of motion has approximately the size of the trap. For this energy, the cross-section is shown in Figure~\ref{fig:J10Mm10_en_0d125_arrows}. 
Clearly, the central structure mentioned above dominates all the features. For this energy value, we can also find small regions where chaotic behaviour of the system is visible; see Fig.~\ref{fig:J10Mm10_cross_en_1over8_magnifi}. It is numerical evidence of non-integrability.

\begin{figure}[h!tp]
  \centering
   \begin{subfigure}[b]{0.46\textwidth}
    \centering \includegraphics[width=\textwidth]{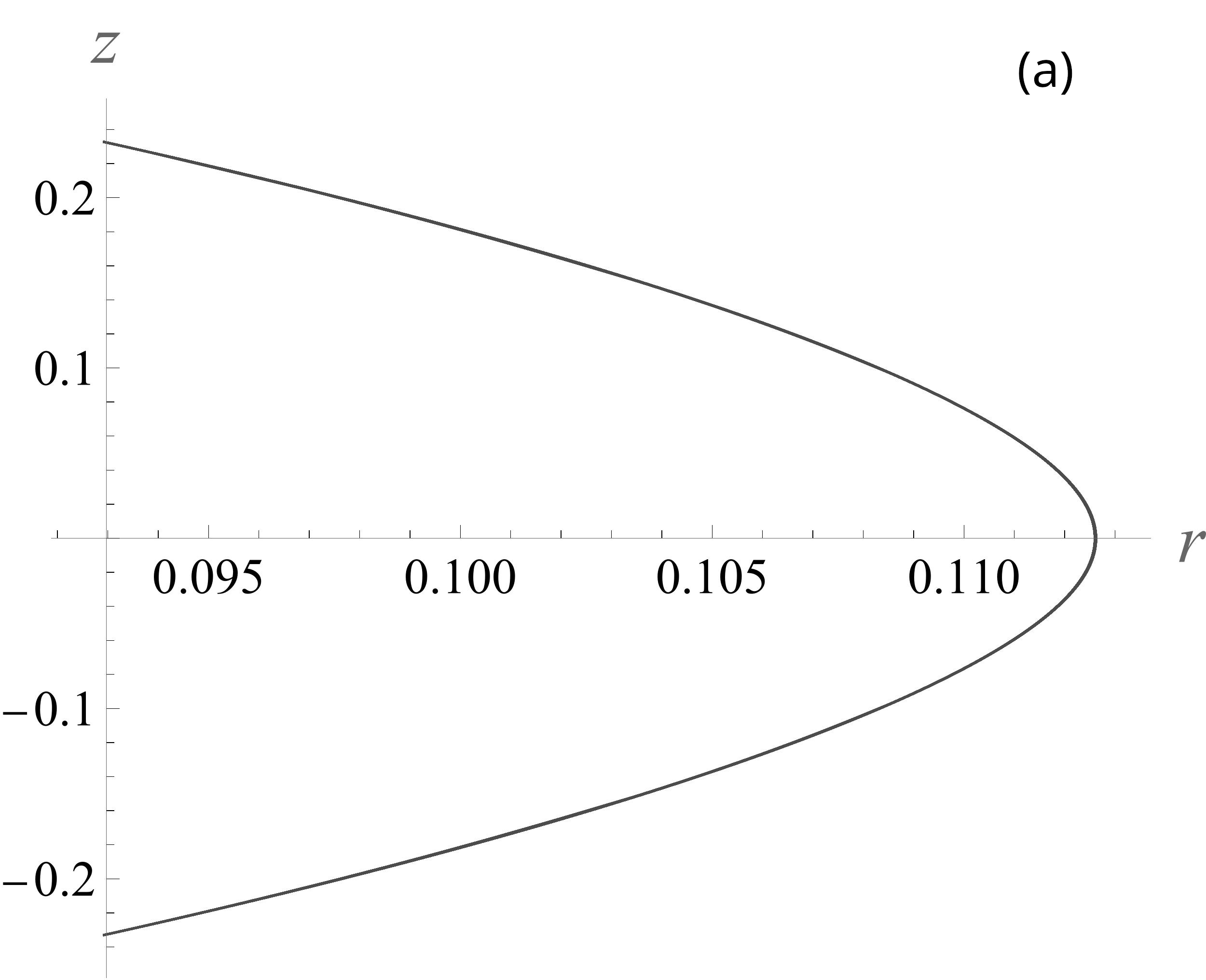}
  \end{subfigure}
   \begin{subfigure}[b]{0.43\textwidth}
    \centering \includegraphics[width=\textwidth]{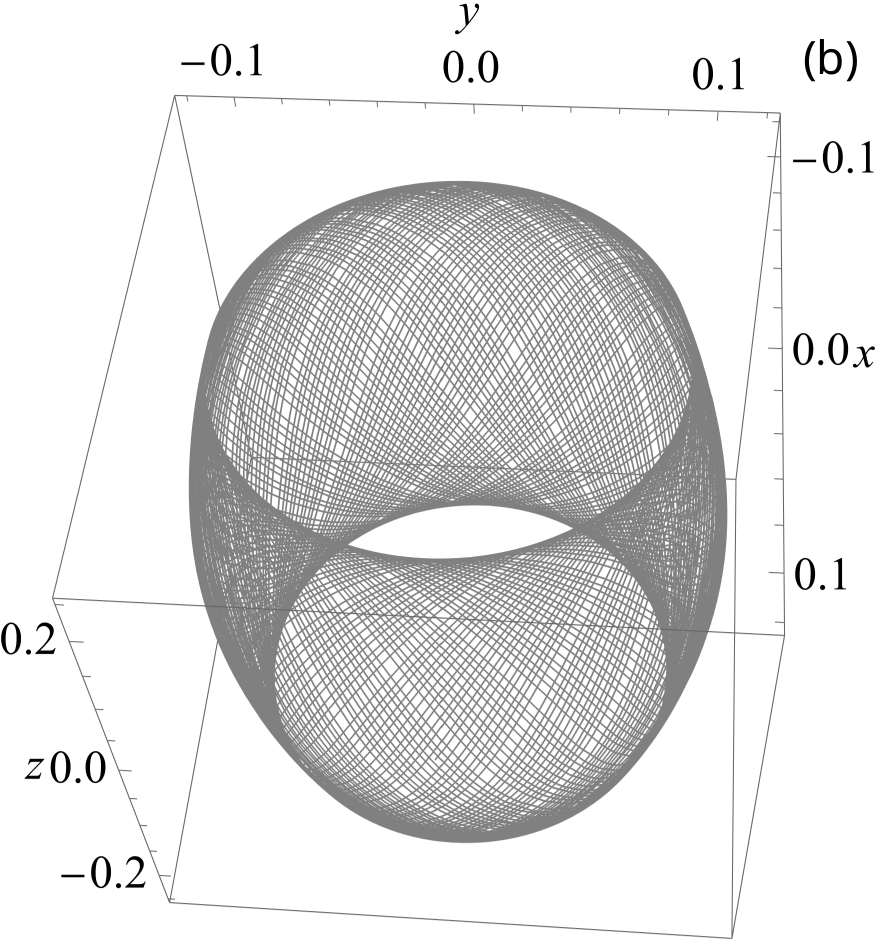}
  \end{subfigure}
    \begin{subfigure}[b]{0.46\textwidth}
    \centering \includegraphics[width=\textwidth]{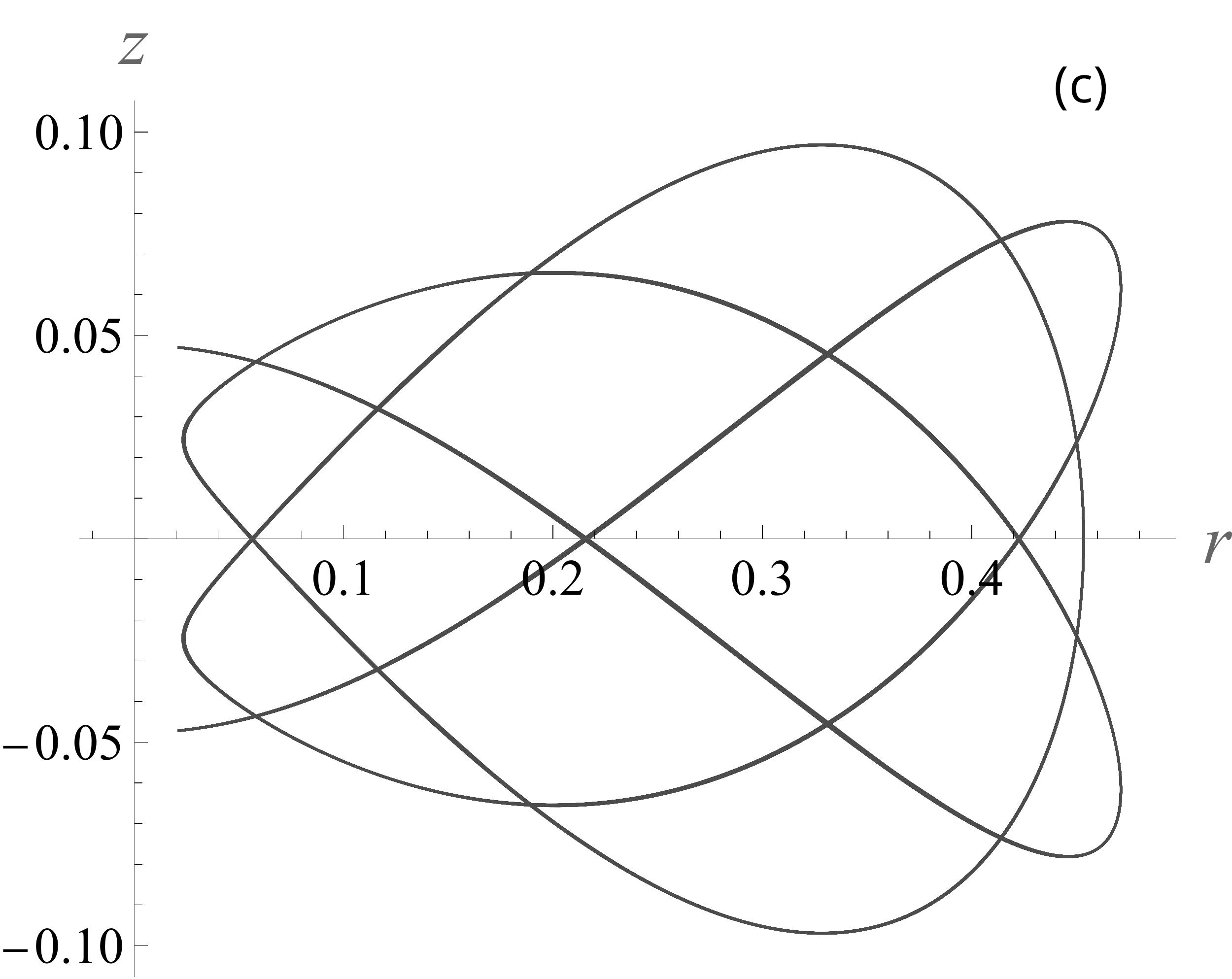}
  \end{subfigure}
   \begin{subfigure}[b]{0.43\textwidth}
    \centering \includegraphics[width=\textwidth]{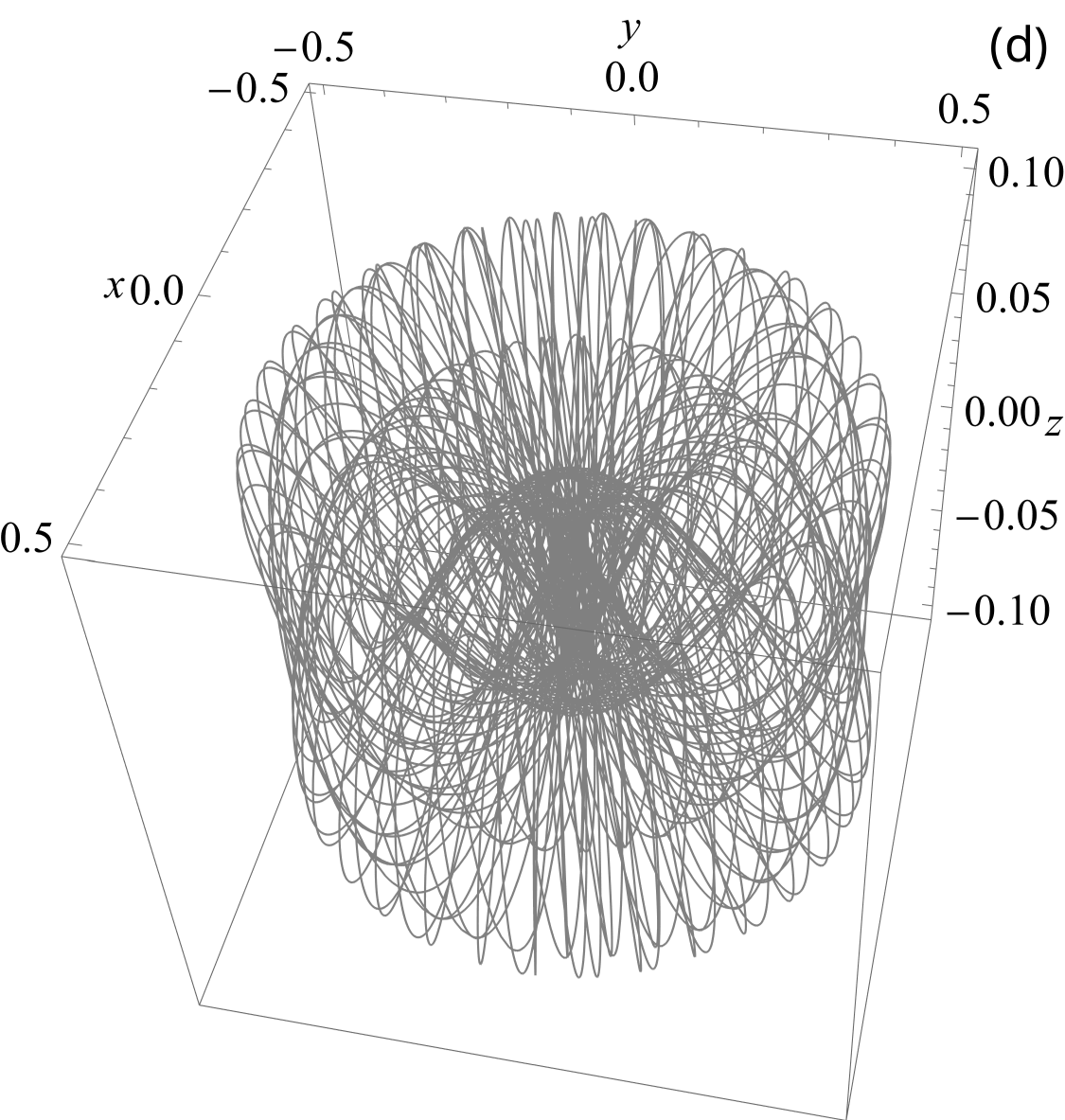}
  \end{subfigure}
   \begin{subfigure}[b]{0.46\textwidth}
    \centering \includegraphics[width=\textwidth]{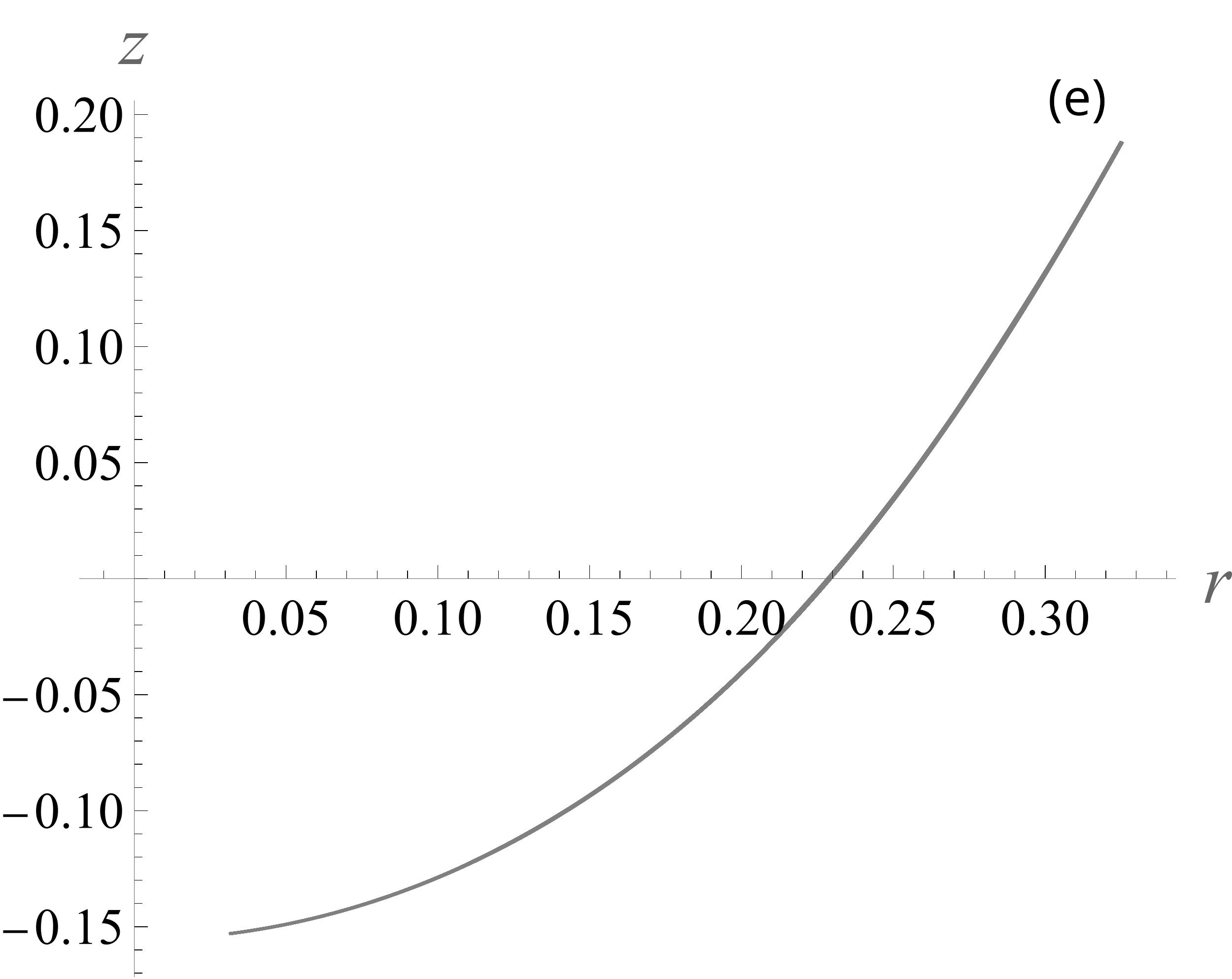}
  \end{subfigure}
   \begin{subfigure}[b]{0.43\textwidth}
    \centering \includegraphics[width=\textwidth]{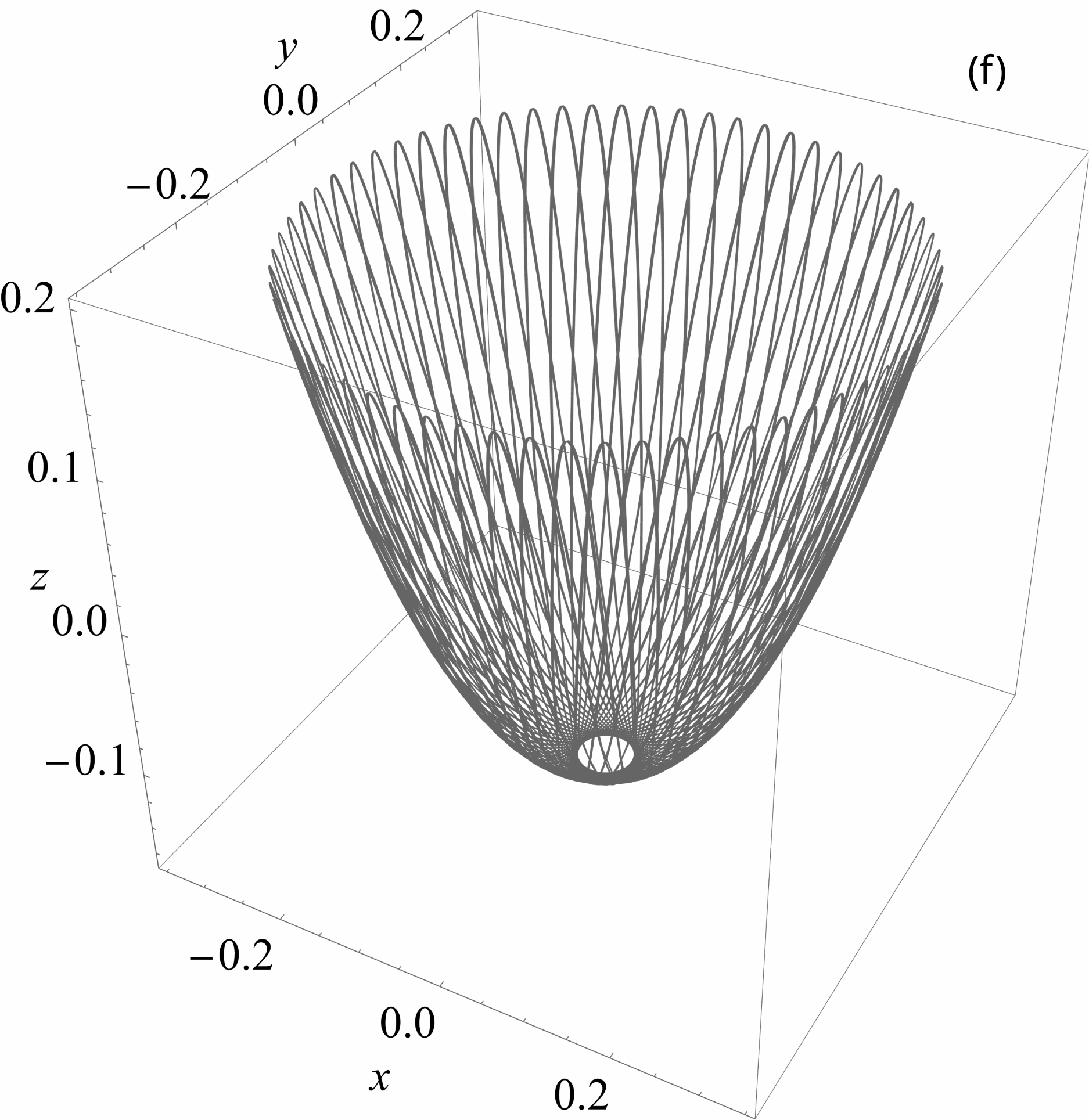}
  \end{subfigure}
   \caption{\small  Evolution of the center of mass vector $\mathbf{x}$ and the time of integration $\tau$ for selected periodic orbits $P_1,P_2$ and $P_3$ in Fig.\ref{fig:J10Mm10_en_0d125_arrows}. Subfigures: (a) orbit $P_1$ in plane $(r,z)$, $\tau=700$; (b)  orbit $P_1$ in space, $\tau=700$; (c)  orbit $P_2$ in plane $(r,z)$, $\tau=700$; (d) orbit $P_2$ in space, $\tau=700$; (e)  orbit $P_3$ in plane $(r,z)$, $\tau=700$; (f) orbit $P_3$ in space, $\tau=700$.  \label{fig:hydroperiodic}}
\end{figure}

\begin{figure}[h!tp]
  \centering
   \begin{subfigure}[b]{0.48\textwidth}
    \centering \includegraphics[width=\textwidth]{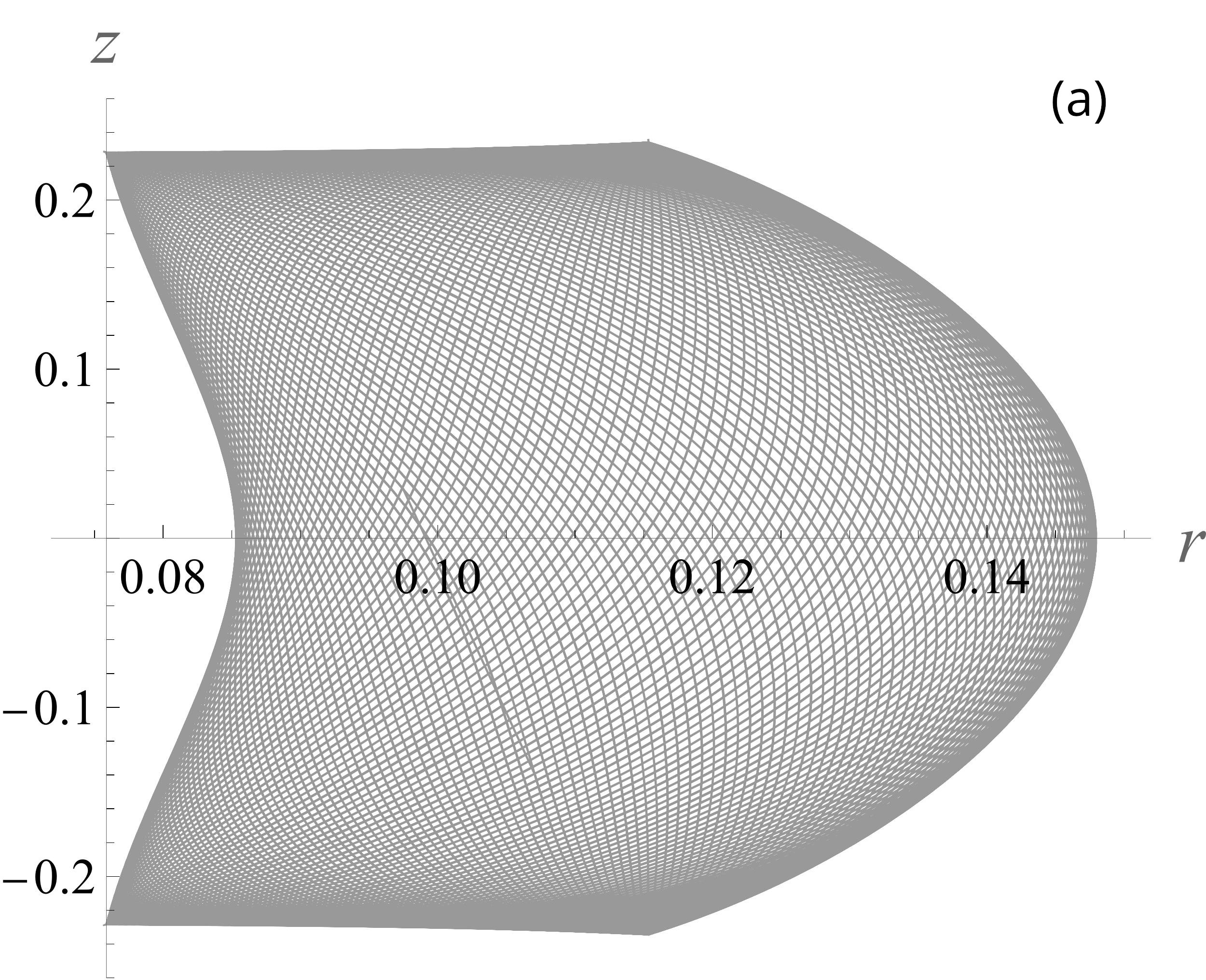}
  \end{subfigure}
   \begin{subfigure}[b]{0.44\textwidth}
    \centering \includegraphics[width=\textwidth]{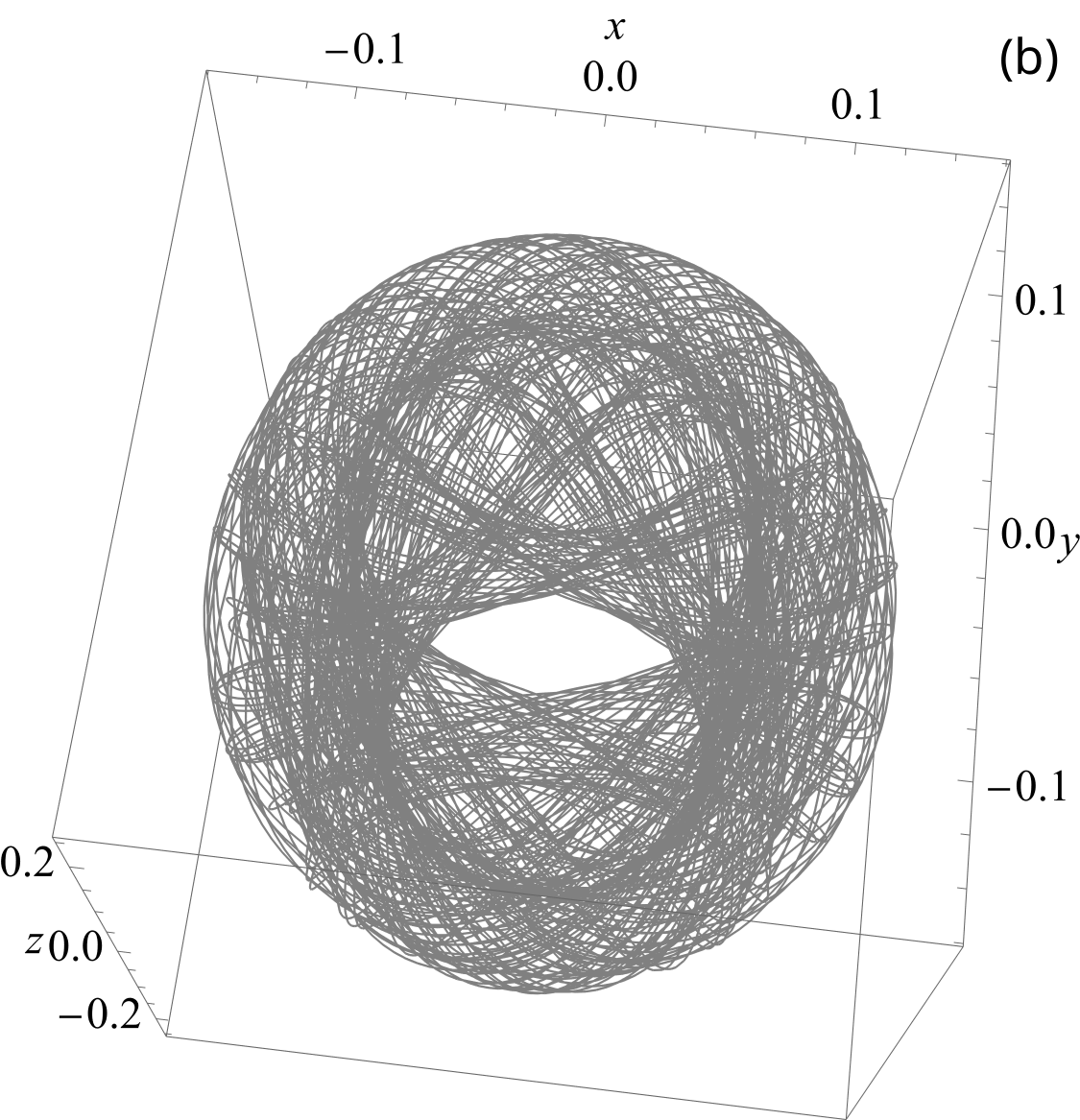}
  \end{subfigure}
    \begin{subfigure}[b]{0.48\textwidth}
    \centering \includegraphics[width=\textwidth]{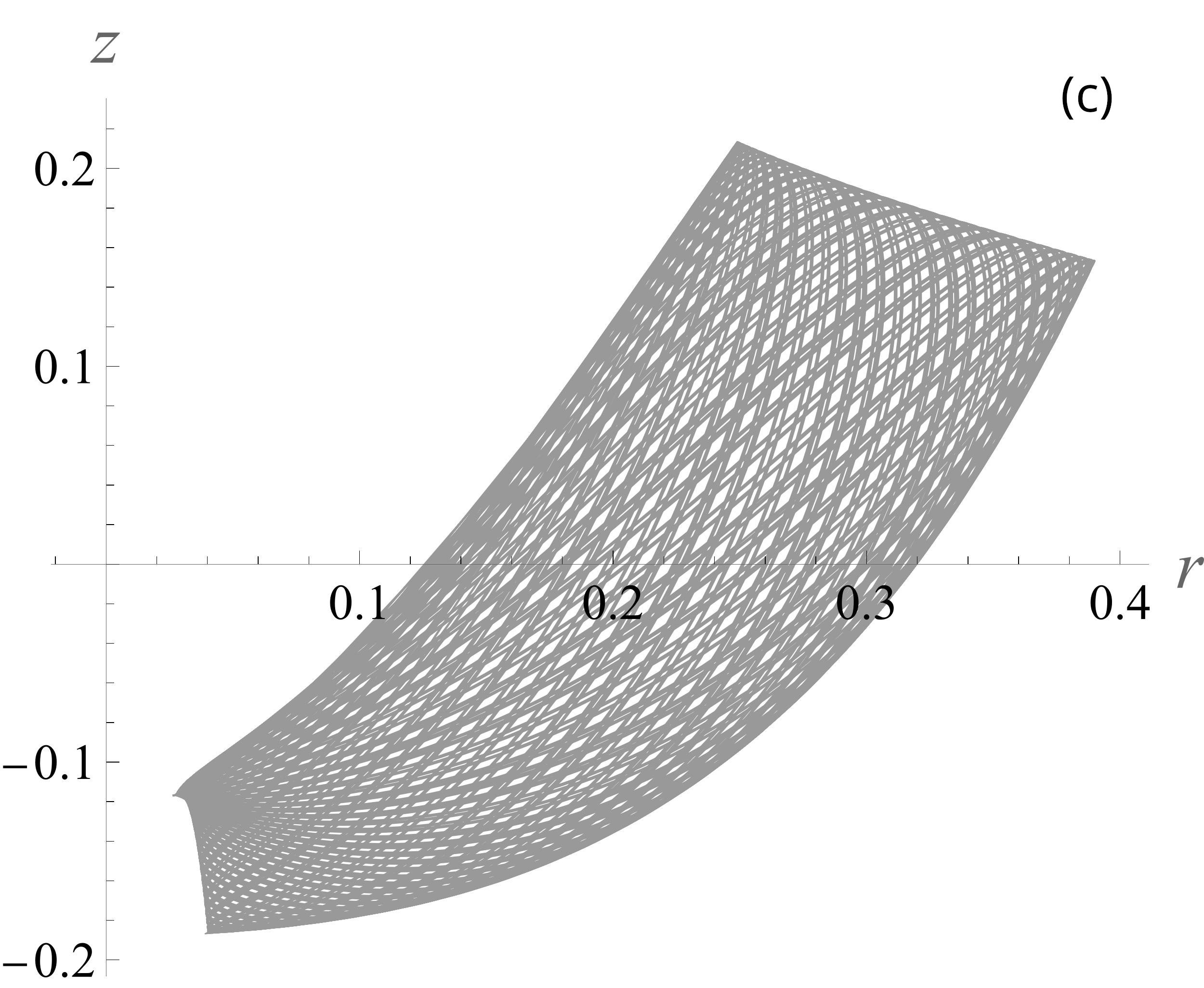}
  \end{subfigure}
   \begin{subfigure}[b]{0.43\textwidth}
    \centering \includegraphics[width=\textwidth]{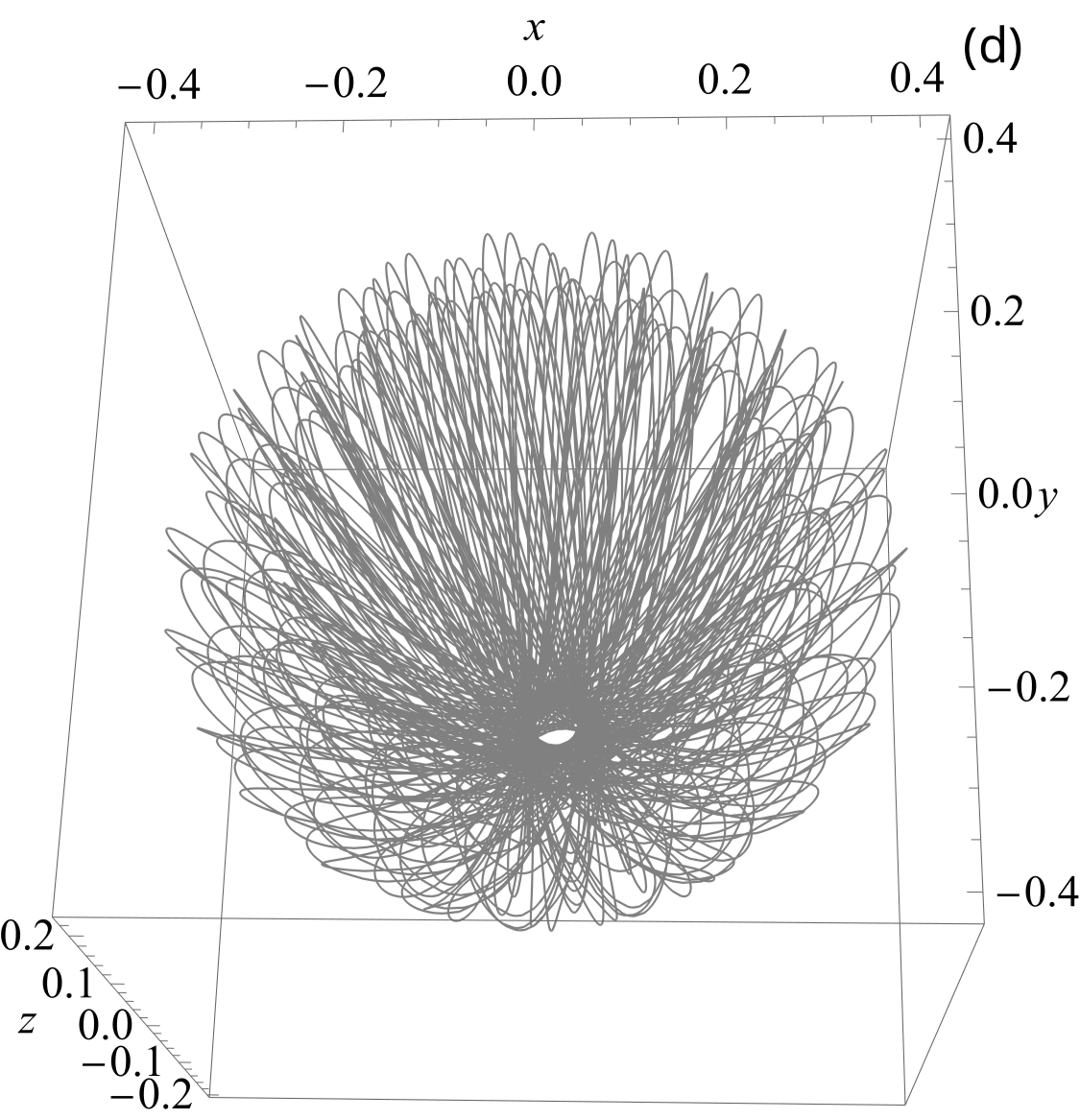}
  \end{subfigure}
   \begin{subfigure}[b]{0.48\textwidth}
    \centering \includegraphics[width=\textwidth]{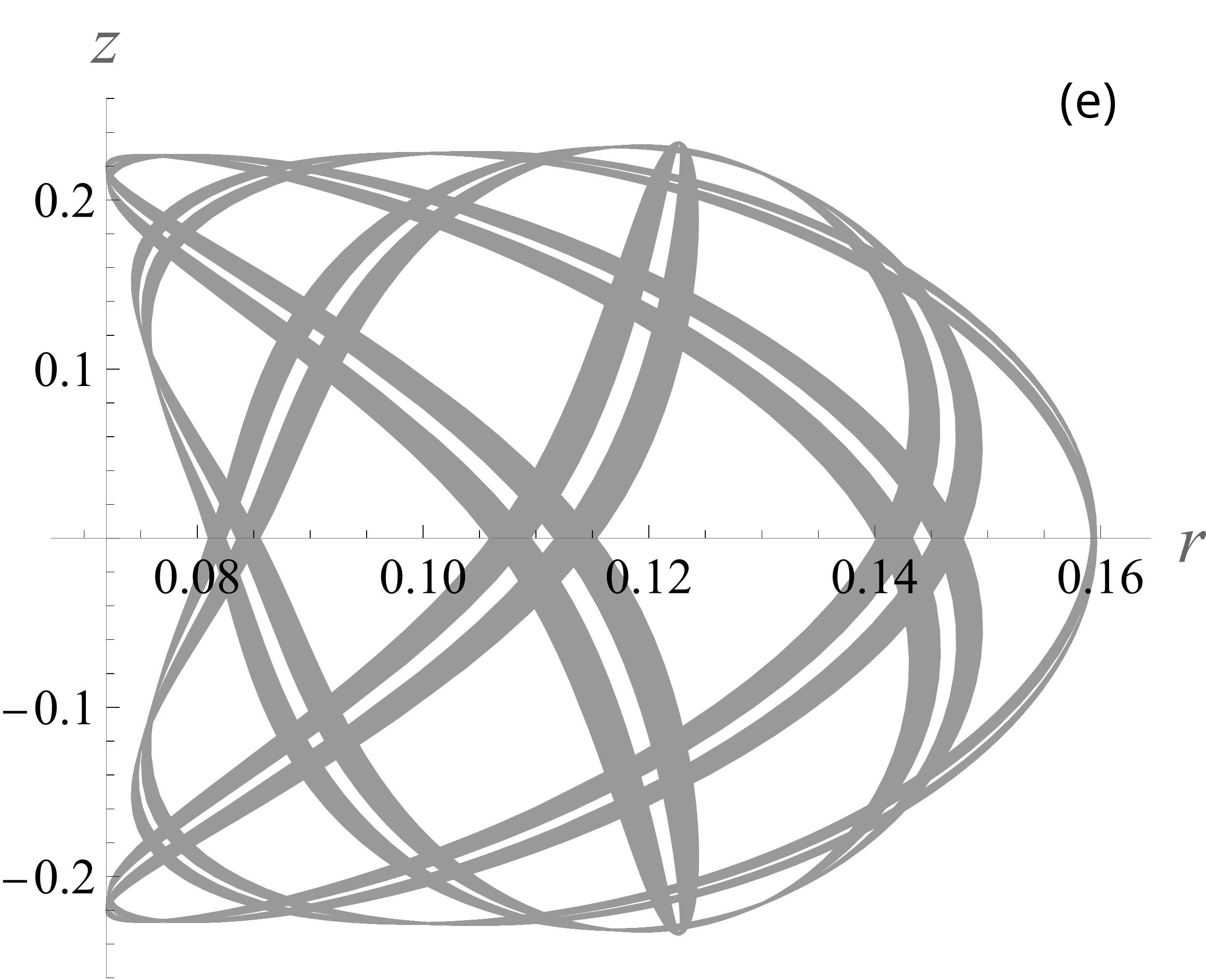}
  \end{subfigure}
   \begin{subfigure}[b]{0.43\textwidth}
    \centering \includegraphics[width=\textwidth]{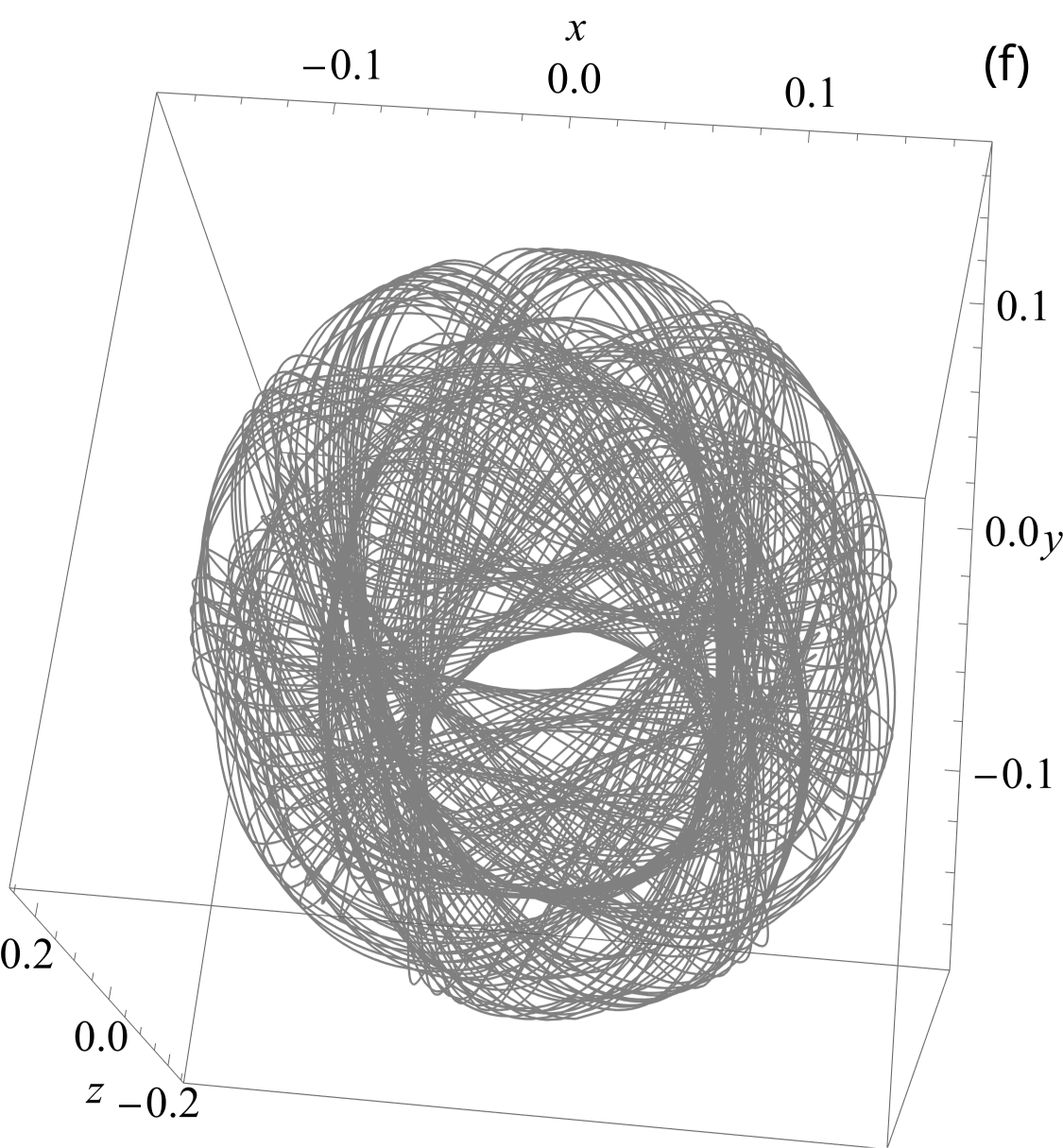}
  \end{subfigure}
  \caption{\small Evolution of the center of mass vector $\mathbf{x}$ and the time of integration $\tau$ for selected quasi-periodic orbits $Q_1,Q_2$ and $Q_3$ in Fig.~\ref{fig:J10Mm10_en_0d125_arrows}. Subfigures: (a) orbit $Q_1$ in plane $(r,z)$, $\tau=600$; (b)  orbit $Q_1$ in space, $\tau=1200$; (c) orbit $Q_2$ in plane $(r,z)$, $\tau=500$; (d) orbit $Q_2$ in space, $\tau=900$; (e) orbit $Q_3$ in plane $(r,z)$, $\tau=1000$; (f) orbit $Q_3$ in space, $\tau=1000$.
  \label{fig:hydroquasiperiodic}}
\end{figure}

In addition to the Poincar\'e sections showing the global qualitative properties of dynamics, it is convenient to present the particle trajectories $(x,y,z)$ in space. Let us choose some points from cross-section in Fig.~\ref{fig:J10Mm10_en_0d125_arrows} denoted by $P_1$, $P_2$, $Q_1$, $Q_2$, $Q_3$ and {\it CH} and plot the corresponding trajectories.  Thus, we numerically integrate Hamilton's equations of motion generated by Hamiltonian~\eqref{Hcm_Edp} with the following seven initial conditions:
\begin{itemize}
\item [$P_1$:] $x(0)=0.112615$, $y(0)=z(0)=0$, $p_x(0)=0$, $p_y(0)=0.0887981$ and $p_z(0)=0.430698$;
\item [$P_2$:] $x(0)=0.45325$, $y(0)=z(0)=0$, $p_x(0)=0$, $p_y(0)=0.0220629$ and $p_z(0)=0.14714$;
\item [$P_3$:] $x(0)=0.228784$, $y(0)=z(0)=0$, $p_x(0)=0.199993$, $p_y(0)=0.0437094$ and $p_z(0)=0.305084$;
\item [$Q_1$:]  $x(0)=0.13547$, $y(0)=z(0)=0$, $p_x(0)=-0.0254729$, $p_y(0)=0.0738171$ and $p_z(0)=0.419283$;
\item [$Q_2$:]  $x(0)=0.190487$, $y(0)=z(0)=0$, $p_x(0)=0.150348$, $p_y(0)=0.052497$ and $p_z(0)=0.358994$;
\item [$Q_3$:]  $x(0)=0.145072$, $y(0)=z(0)=0$, $p_x(0)=-0.0297181$, $p_y(0)=0.0689313$ and $p_z(0)=0.414046$;
\item [{\it CH:}]  $x(0)=0.313439$, $y(0)=z(0)=0$, $p_x(0)=0.000209503$, $p_y(0)=0.0319041$ and $p_z(0)=0.302336$.
\end{itemize}

Fig.~\ref{fig:hydroperiodic} presents orbits visible as elliptic points $P_1$, $P_2$ and $P_3$ on Fig.~\ref{fig:J10Mm10_en_0d125_arrows}, thus periodic in plane $(r,z)$, see the left column, but the corresponding spatial trajectories are not periodic but quasi-periodic as can be seen in the right column.

The next Fig.~\ref{fig:hydroquasiperiodic} presents solutions belonging to quasi-periodic orbits $Q_1$, $Q_2$ and $Q_3$ in Fig.~\ref{fig:J10Mm10_en_0d125_arrows}, thus they are quasi-periodic in the plane $(r,z)$, see the left column.  The corresponding spatial trajectories are also quasi-periodic; however, the patterns shown in the right column are quite complicated. Orbits $Q_1$ and $Q_2$ are simple quasi-period orbits around the elliptic periodic orbits $P_1$ and $P_2$. The trajectory for $Q_3$ lies on a torus surrounding a resonant periodic orbit (1:7 resonance). 

\begin{figure}[h!tp]
    \centering
     \begin{subfigure}[b]{0.49\textwidth}
    \centering \includegraphics[width=\textwidth]{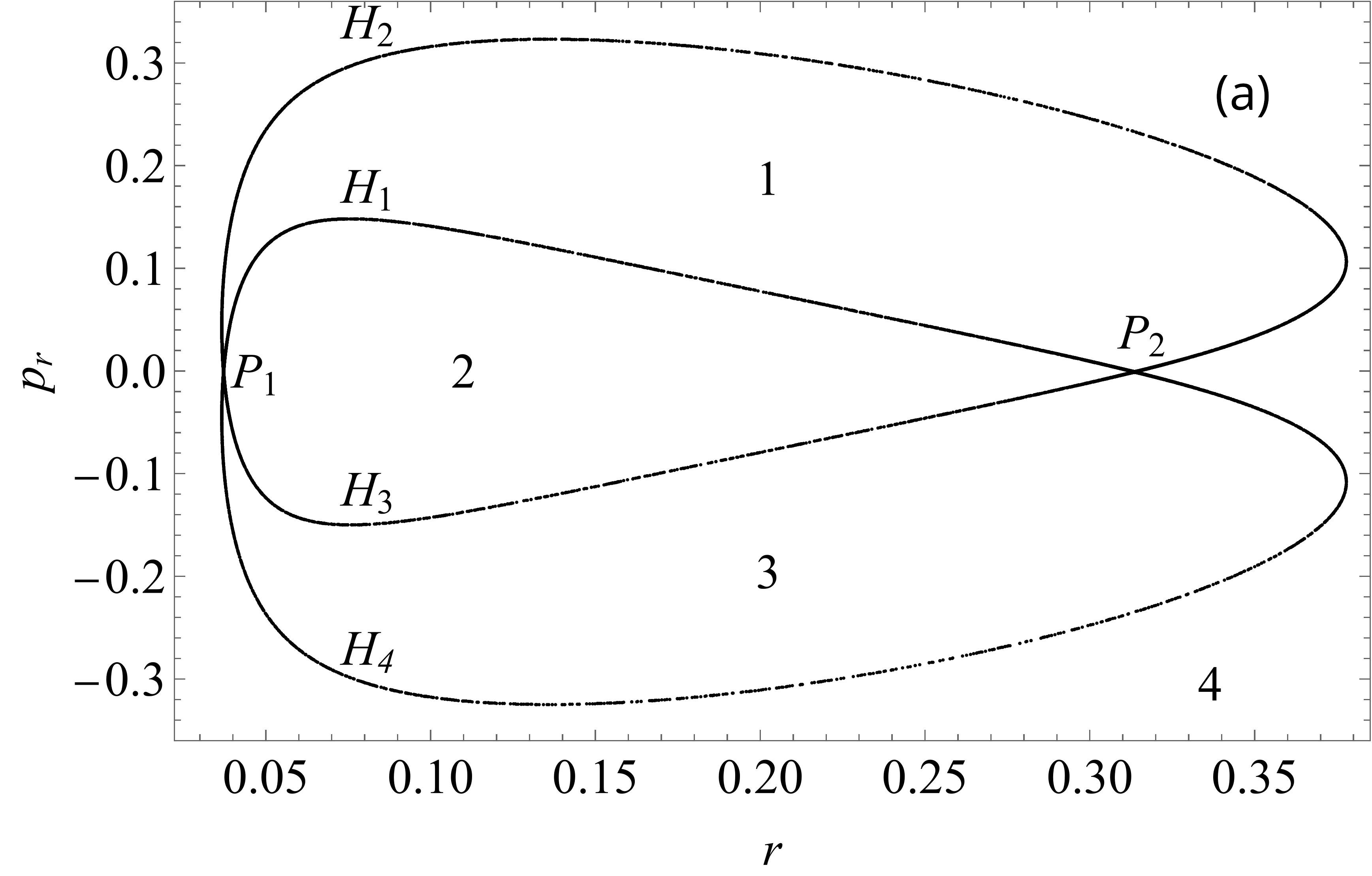}
  \end{subfigure}
    \begin{subfigure}[b]{0.49\textwidth}
    \centering \includegraphics[width=\textwidth]{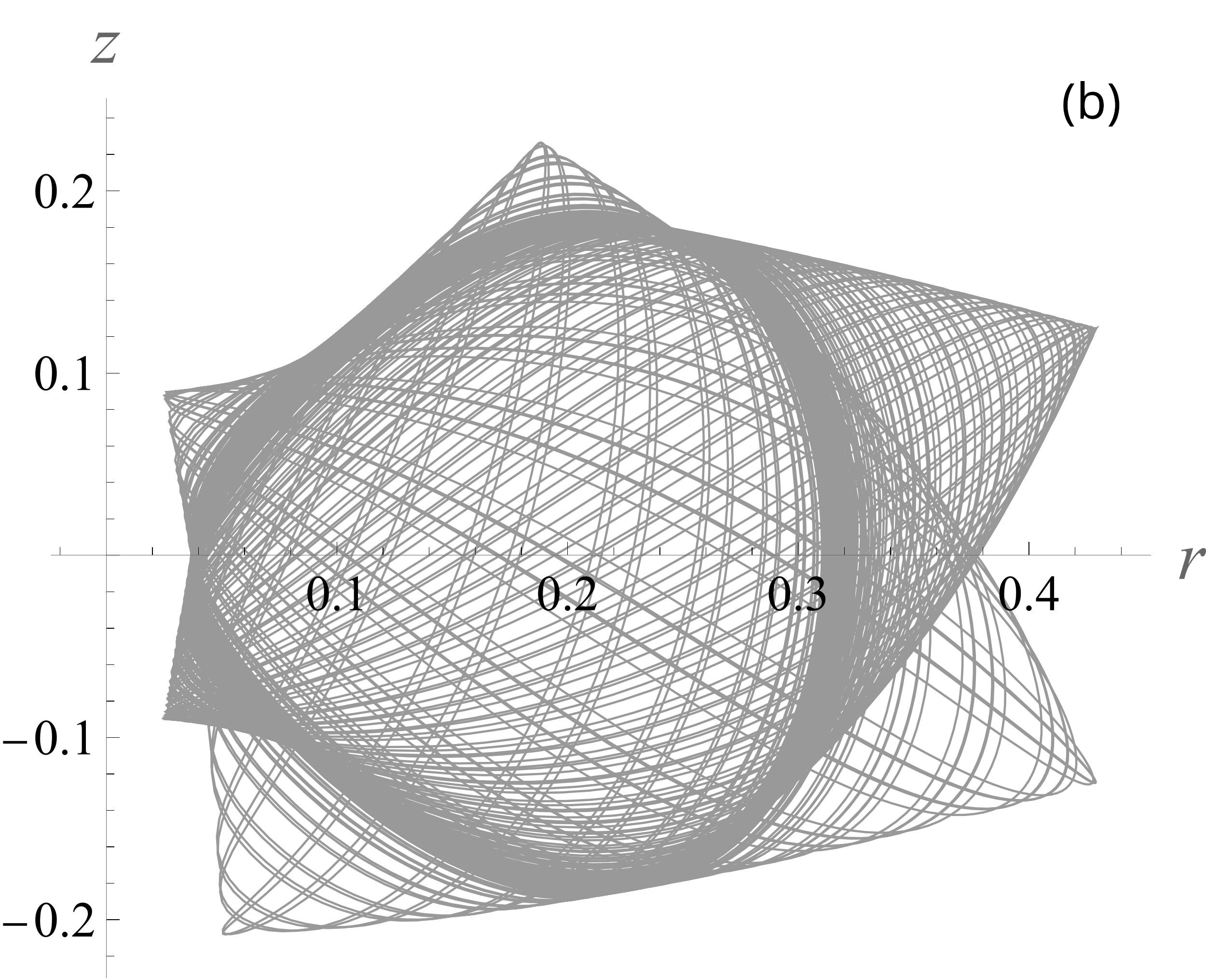}
  \end{subfigure}
   \begin{subfigure}[b]{0.50\textwidth}
    \centering \includegraphics[width=\textwidth]{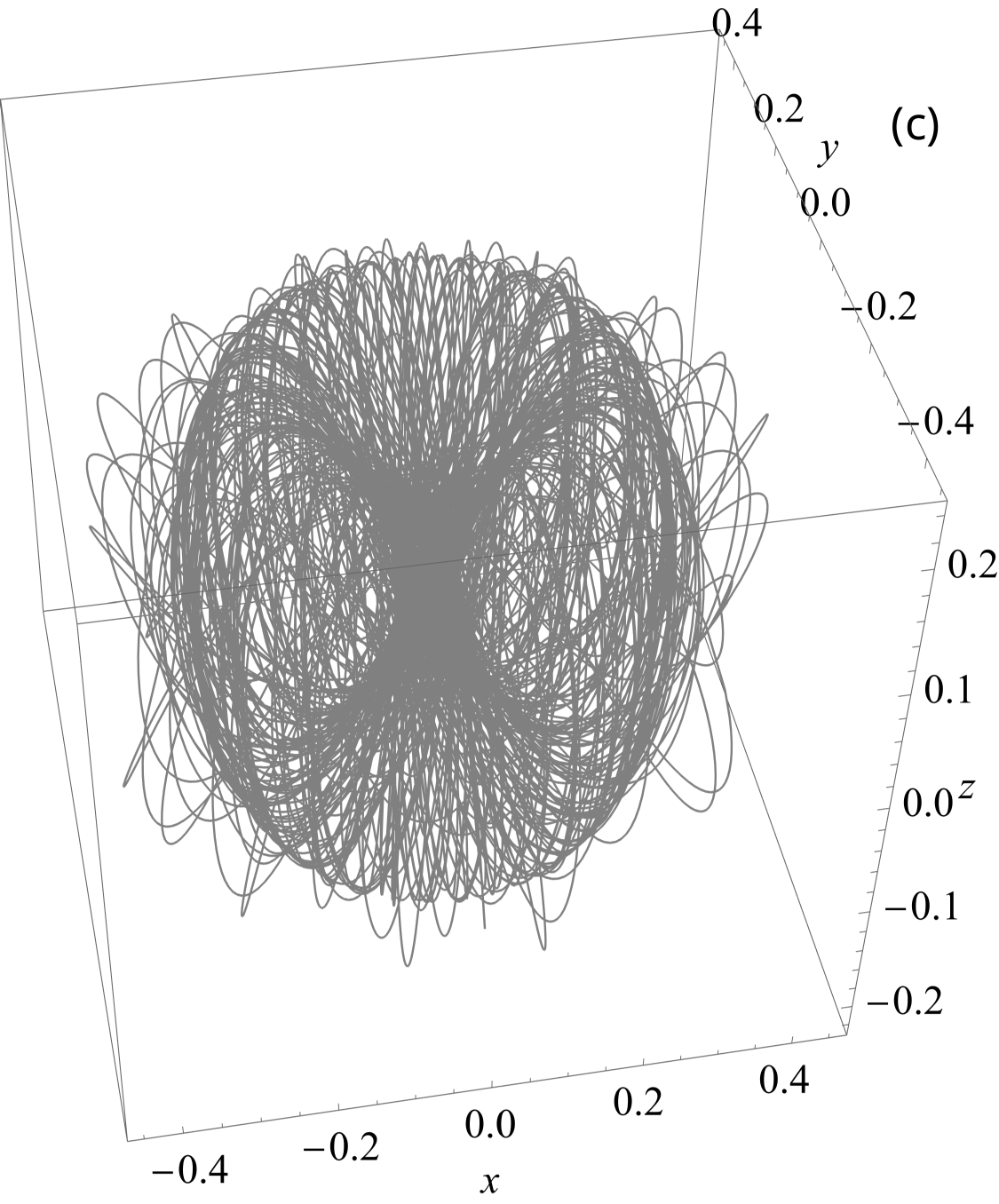}
  \end{subfigure}
   \begin{subfigure}[b]{0.49\textwidth}
    \centering \includegraphics[width=\textwidth]{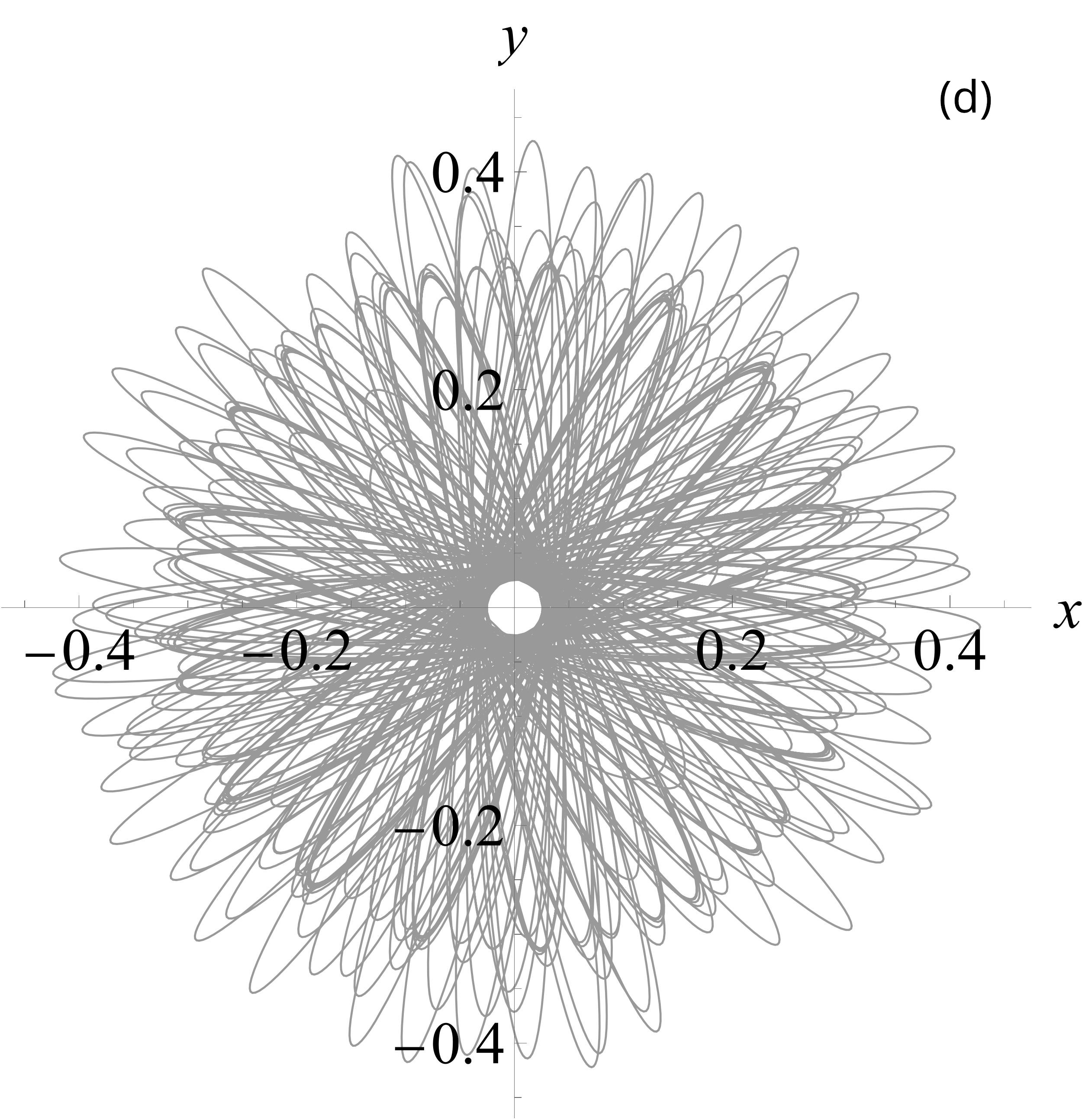}
  \end{subfigure}
  \caption{\small Time evolution of the center of mass vector $\mathbf{x}$ for  chaotic orbit $CH$ in Fig.~\ref{fig:J10Mm10_en_0d125_arrows}. Subfigures: (a)  Poincar\'e section of orbit {\it CH}; (b) orbit {\it CH} in plane $(r,z)$, $\tau=1000$; (c) orbit {\it CH} in space $(x,y,z)$, $\tau=1500$; (d)  orbit {\it CH} in plane $(x,y)$, $\tau=1500$.
  \label{fig:hydrochaotic}}
\end{figure}

Fig.~\ref{fig:hydrochaotic} shows a trajectory corresponding to an orbit {\it CH}  in Fig.~\ref{fig:J10Mm10_en_0d125_arrows} from the chaotic layer in the vicinity of a hyperbolic periodic solution. The visible scattered points belong to more complex trajectories; see an exemplary one in Fig.~\ref{fig:hydrochaotic} with the corresponding Poincar\'e section of this one orbit in Fig.~\ref{fig:hydrochaotic}(a). Let us notice that this chaotic layer is very narrow and is confined between the surrounding quasi-periodic orbits. In this figure, $P_1$ and $P_2$ mark unstable (hyperbolic) periodic solutions. Stable and unstable manifolds of these solutions divide the cross-section into four separate regions marked in the figure by 1, 2, 3 and 4. The points $P_1$ and $P_2$ are joined by heteroclinic solutions that cross the section plane in a close vicinity of curves $H_i$. A chaotic orbit can follow all of them.  The behaviour of this trajectory in $\R^3$ shows Fig.~\ref{fig:hydrochaotic}(c) and projections on $(r,z)$ and $(x,y)$ planes Figs.~\ref{fig:hydrochaotic}(b) and \ref{fig:hydrochaotic}(d), respectively.

\begin{figure}[h!tp]
  \centering
   \includegraphics[scale=0.75]{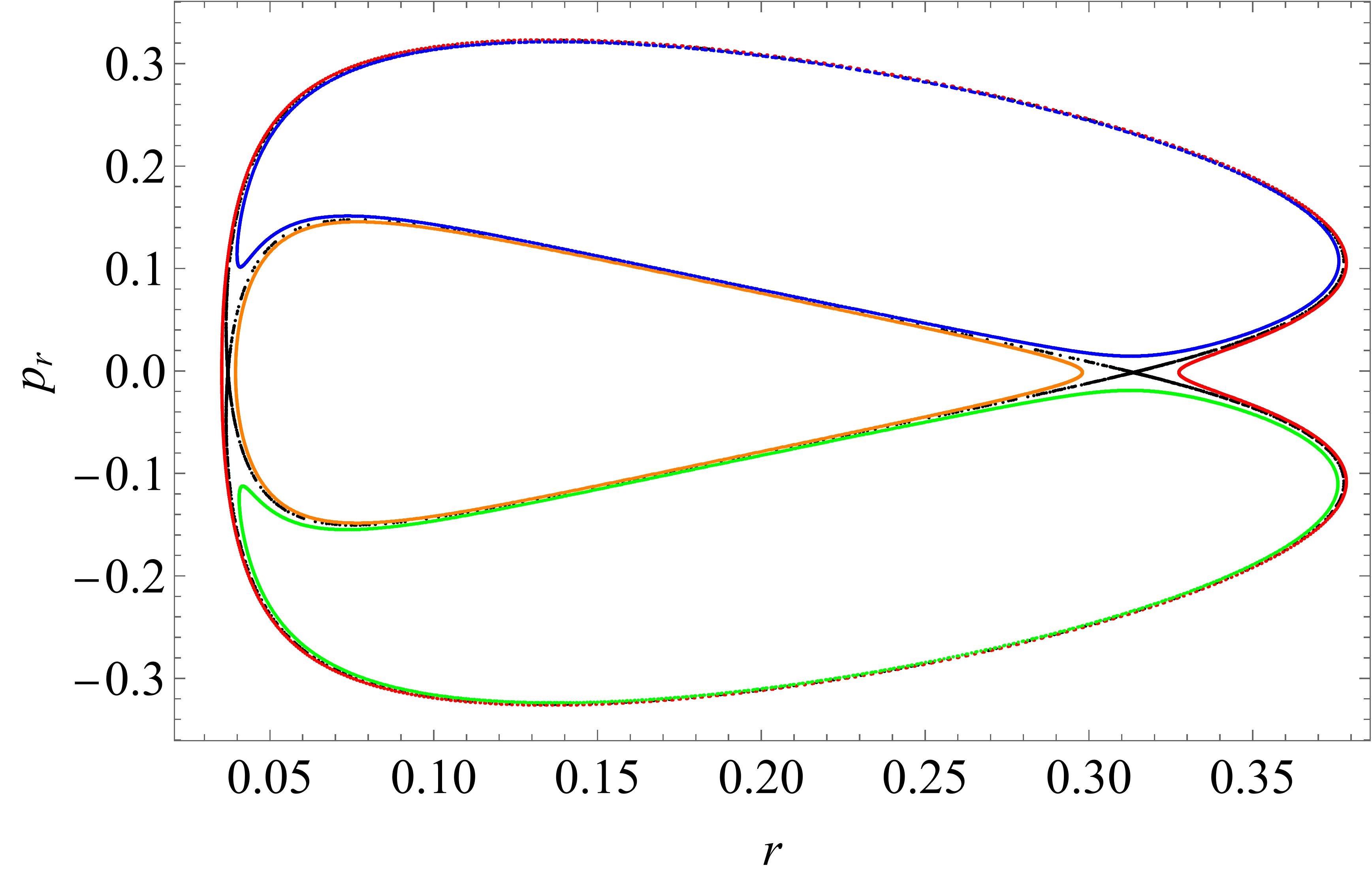}
\caption{\small Poincar\'e section of captured chaotic trajectory $CH$ (in black) given in Fig.~\ref{fig:hydrochaotic} between four neighbouring quasi-periodic trajectories (in blue, red, green and orange).\label{fig:J10Mm10_chaotic_captured}}
\end{figure}

\begin{figure}[h!tp]
    \centering
    \begin{subfigure}[b]{0.47\textwidth}
    \centering \includegraphics[width=\textwidth]{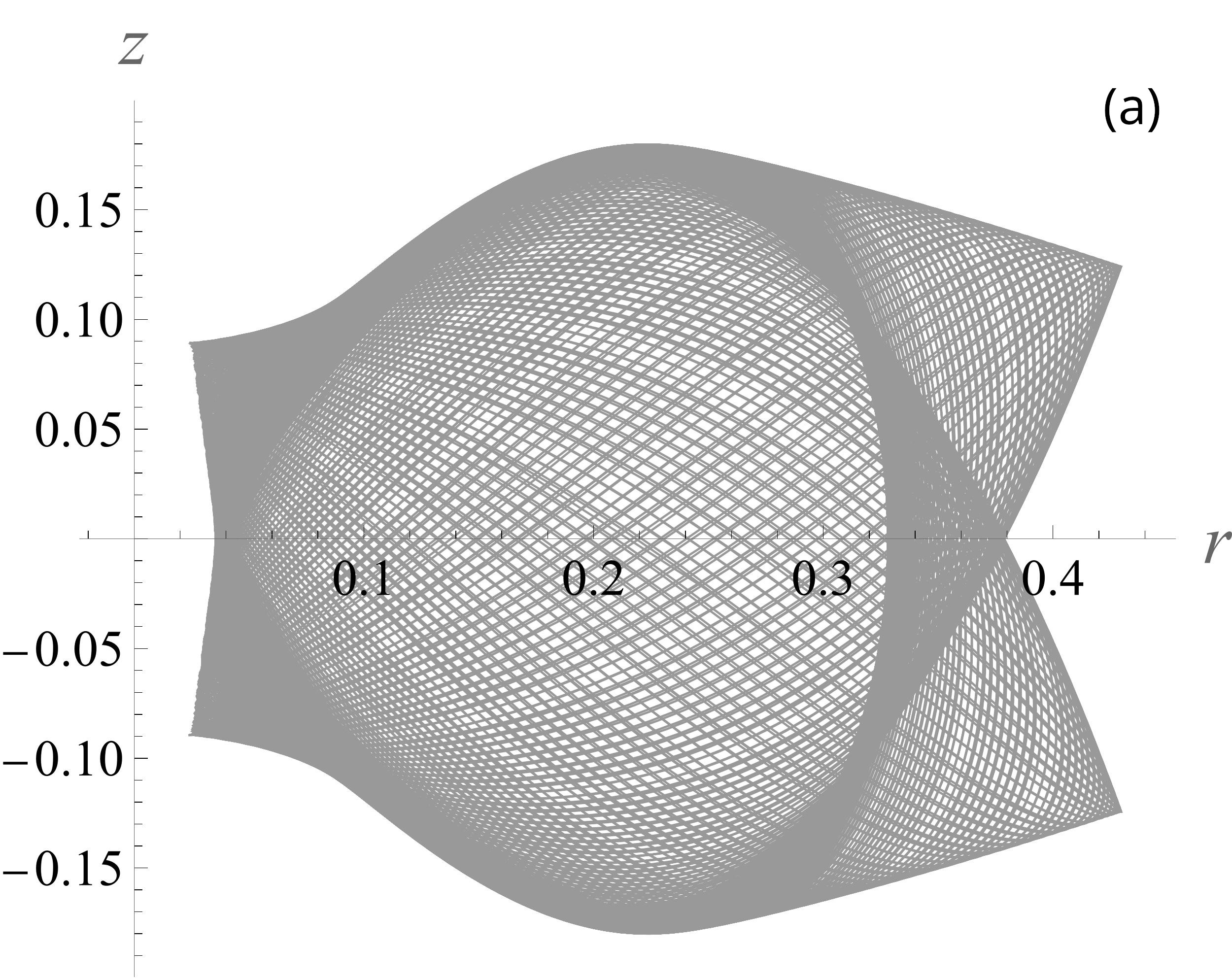}
  \end{subfigure}
   \begin{subfigure}[b]{0.39\textwidth}
    \centering \includegraphics[width=\textwidth]{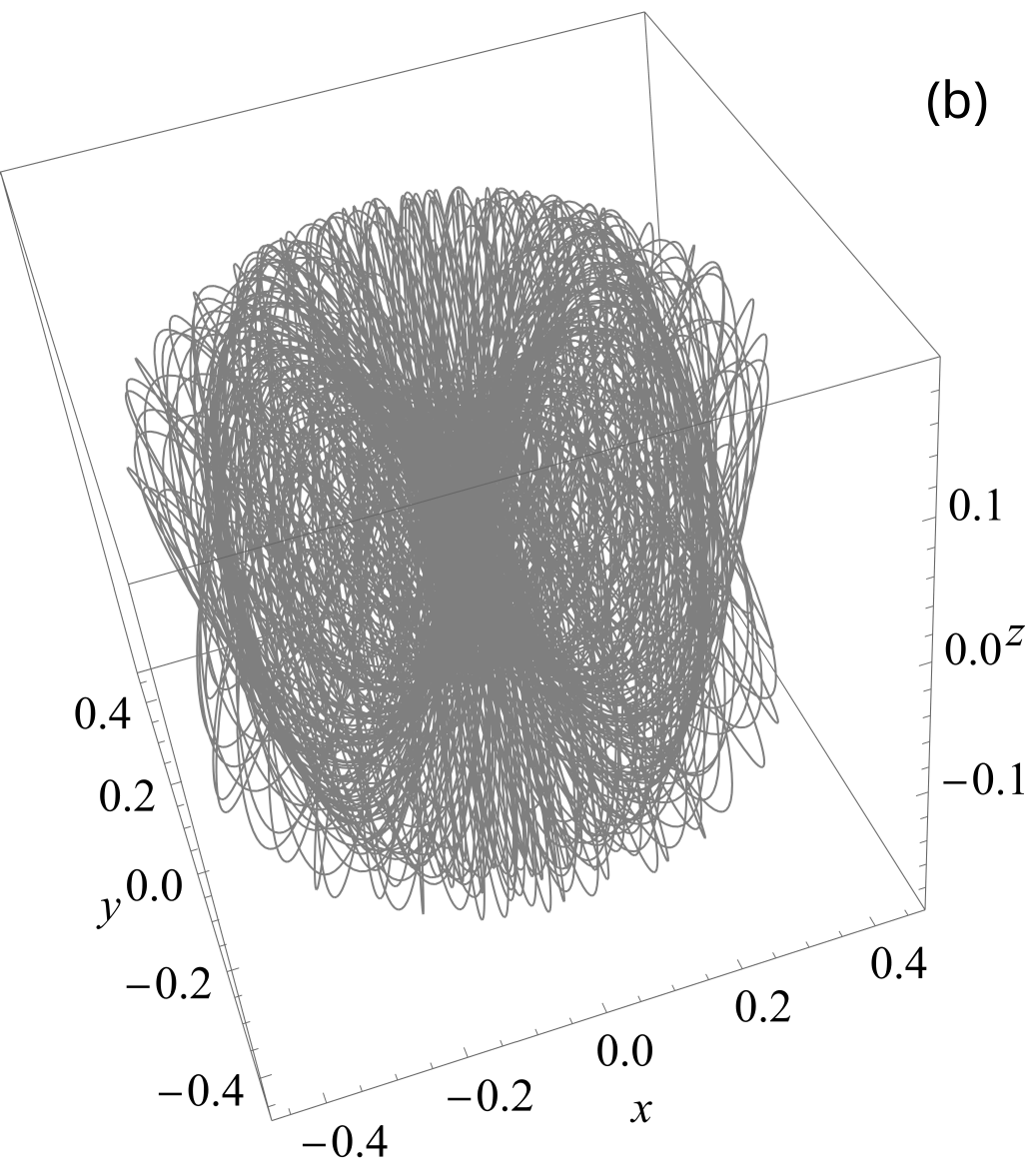}
  \end{subfigure}
    \begin{subfigure}[b]{0.49\textwidth}
    \centering \includegraphics[width=\textwidth]{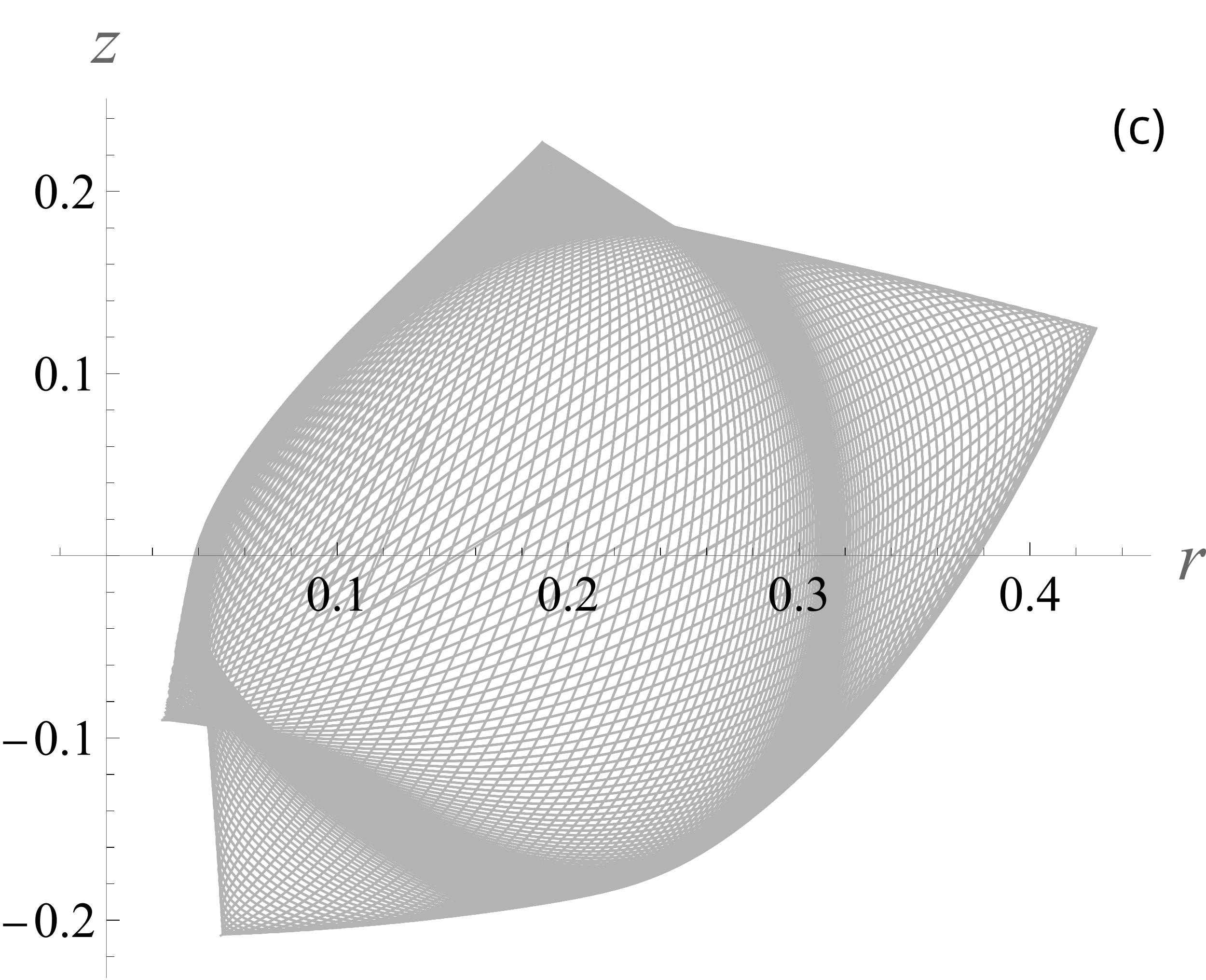}
  \end{subfigure}
   \begin{subfigure}[b]{0.40\textwidth}
    \centering \includegraphics[width=\textwidth]{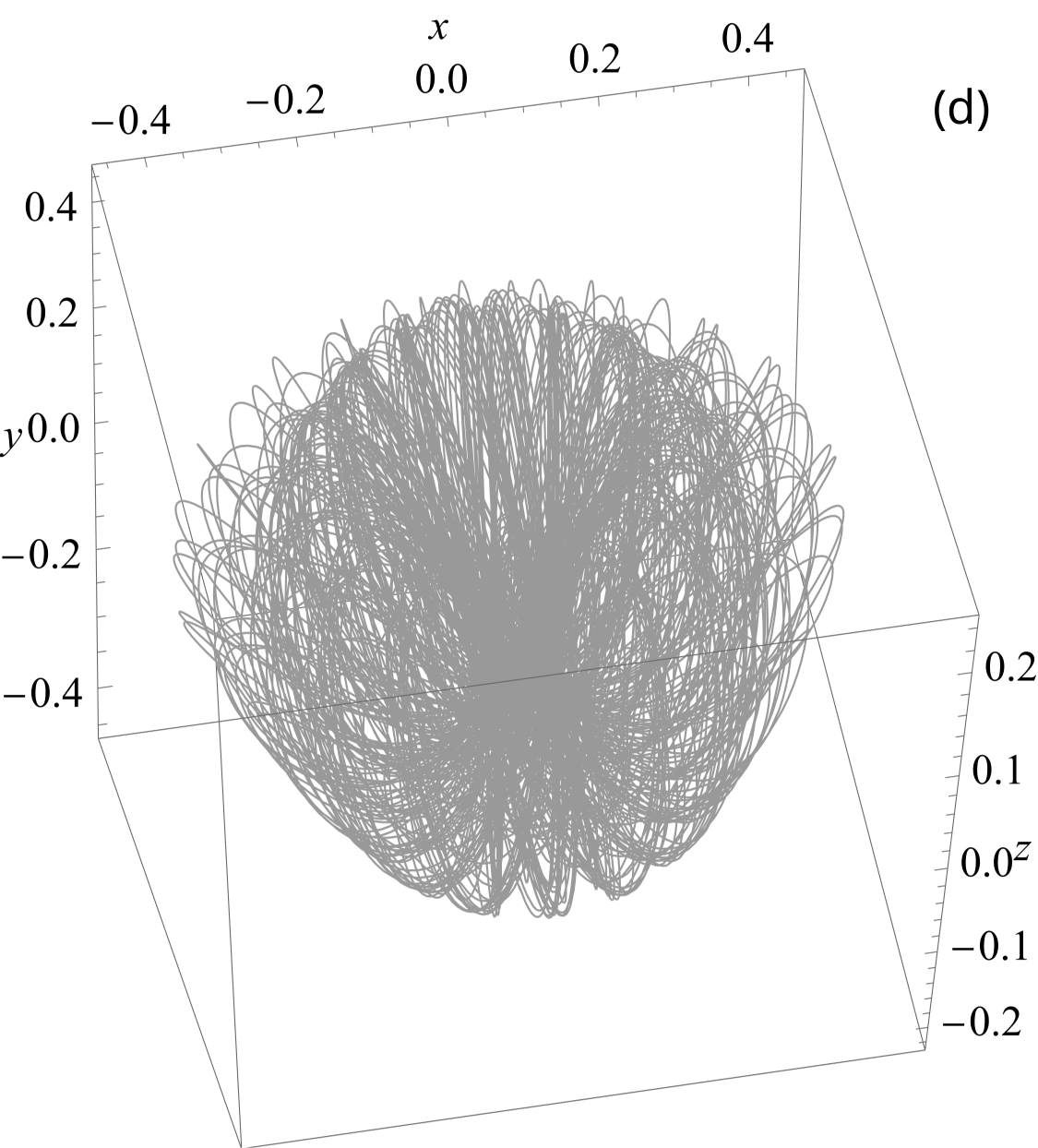}
  \end{subfigure}
   \begin{subfigure}[b]{0.47\textwidth}
    \centering \includegraphics[width=\textwidth]{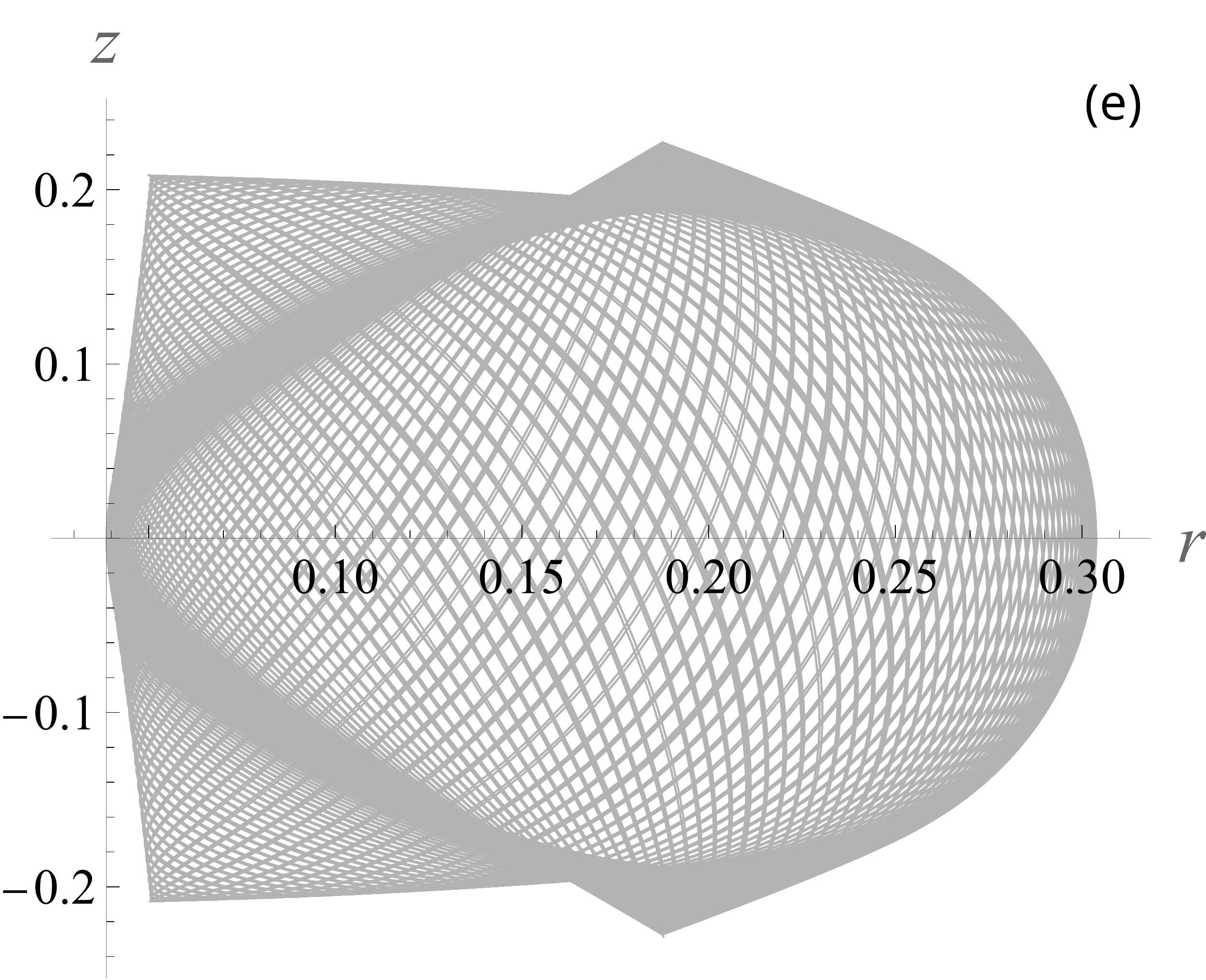}
  \end{subfigure}
   \begin{subfigure}[b]{0.39\textwidth}
    \centering \includegraphics[width=\textwidth]{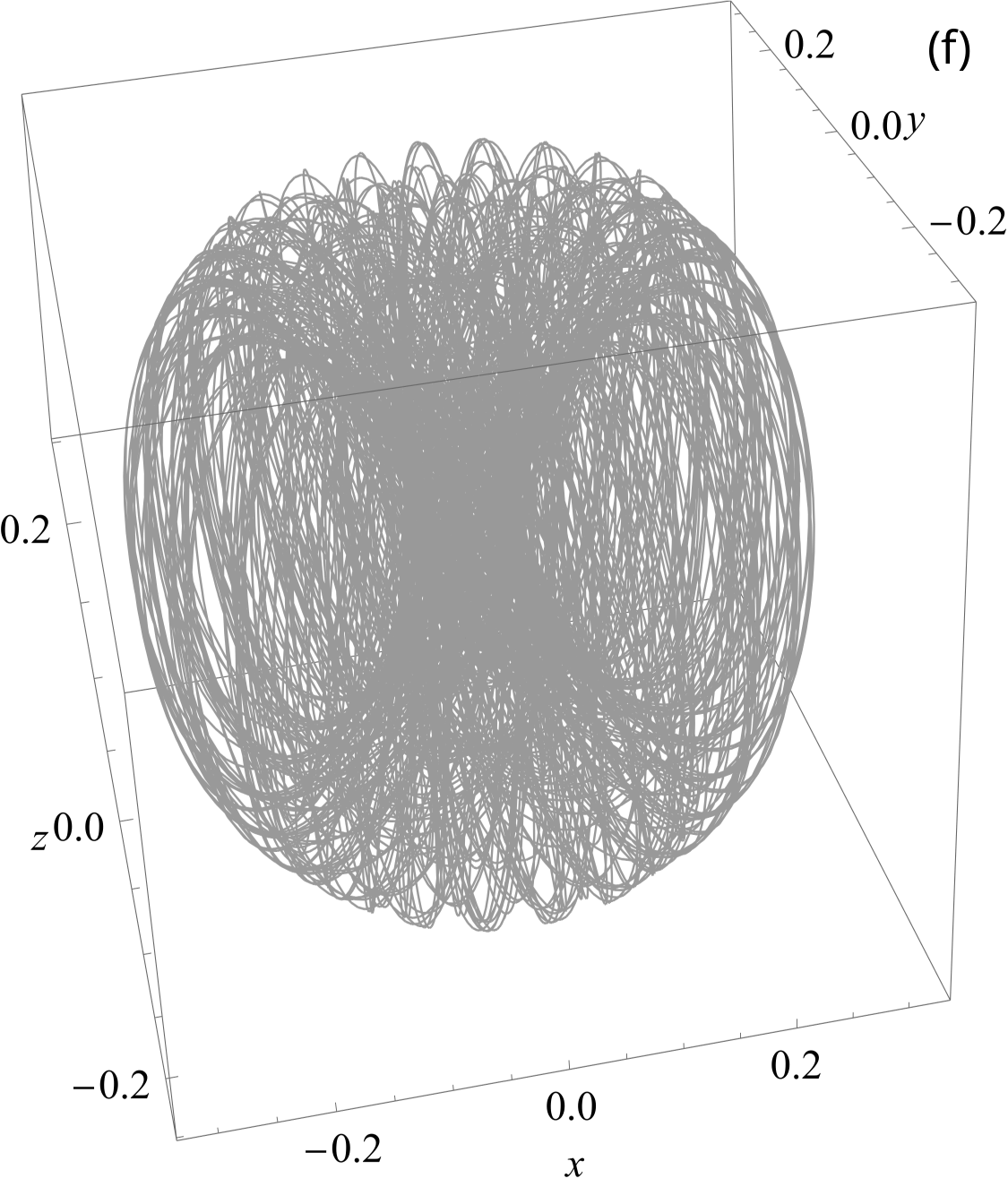}
  \end{subfigure}
  \caption{\small Time evolution of the center of mass vector $\mathbf{x}$ and the time of integration $\tau$ for quasi-periodic orbits close to the chaotic orbit $CH$ on the right, above and on the left. Subfigures: (a) "red" orbit in plane $(r,z)$, $\tau=1000$; (b) "red" orbit in $\R^3$, $\tau=1500$; (c)  "blue" orbit in plane $(r,z)$, $\tau=1000$; (d) "blue" orbit in $\R^3$, $\tau=1500$; (e) "orange" orbit in plane $(r,z),$ $\tau=1000$; (f) "orange" orbit in $\R^3$, $\tau=1500$.
  \label{fig:quasichaoticsurrounding}}
\end{figure}

In Fig~\ref{fig:J10Mm10_chaotic_captured} we show a chaotic trajectory $CH$ from Fig.~\ref{fig:hydrochaotic} (in black) surrounded by four quasi-periodic orbits on the right (in red), top (in blue), left (in orange) and bottom (in green) from four different regions. The behaviour of these surrounding trajectories to the right, above, and left of this chaotic orbit is plotted in Fig.~\ref{fig:quasichaoticsurrounding}. We do not present plots of the surrounding orbit below $CH$ coloured in green because they are reflections of plots for an orbit above $CH$ coloured in blue with respect to the axis $r$ in Fig.~\ref{fig:quasichaoticsurrounding}(c) or the plane $(x,y)$ in Fig.~\ref{fig:quasichaoticsurrounding}(d).

\begin{figure}[h!tp]
  \centering
  \includegraphics[scale=0.87]{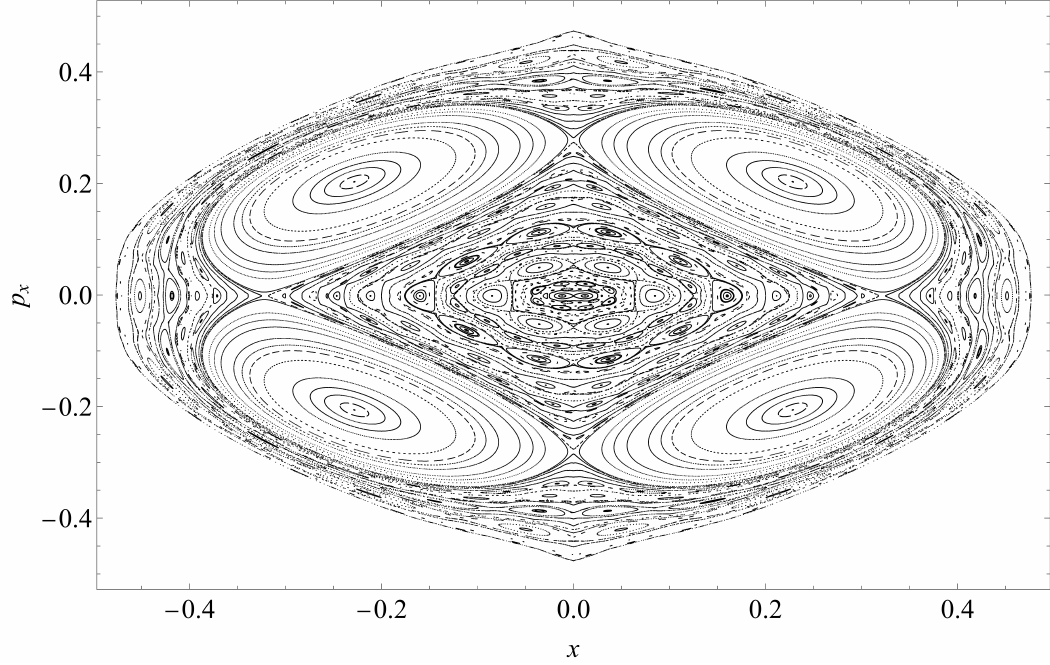}
\caption{\small The Poincar\'e section for parameters $\sigma=0.502723$ and $\delta=0.0000179305$, $p_{\varphi}=0$ and the energy value $h=0.125$ (1.7507~K). Cross-section plane $z=0$ and the orientation of the trajectories $p_z>0$.\label{fig:J10Mm10_pfeq0}}
\end{figure}

In the case where $p_{\varphi}=0$ an orbit lies in a fixed plane in space $(x,y,z)$. If we choose an invariant plane $y=p_y=0$, then the Hamiltonian $H_{\mathrm{cm}}$ in \eqref{Hcm_Edp} reduces to two degrees of freedom with canonical variables $(x,y,p_x,p_y)$. Thus, we can construct Poincar\'e sections in these variables choosing as the cross-section plane $z=0$ with variables $(x,p_x)$ in this plane and the orientation of the trajectories $p_z>0.$

The exemplary Poincar\'e section with $p_{\varphi}=0$ is shown in Fig.~\ref{fig:J10Mm10_pfeq0}. 
In the origin, three domains filled by closed phase curves are separated from each other by a figure-of-eight saddle separatrices. This central area is surrounded by subsequent quasiperiodic orbits corresponding to higher-order resonances.  An octupole resonance occupies a large part of the section with four stable fixed points surrounded by quasiperiodic orbits of stable motion (islands of stability) and the separatrix passing through four unstable fixed points. 

\begin{figure}[h!tp]
  \centering
   \includegraphics[scale=0.82]{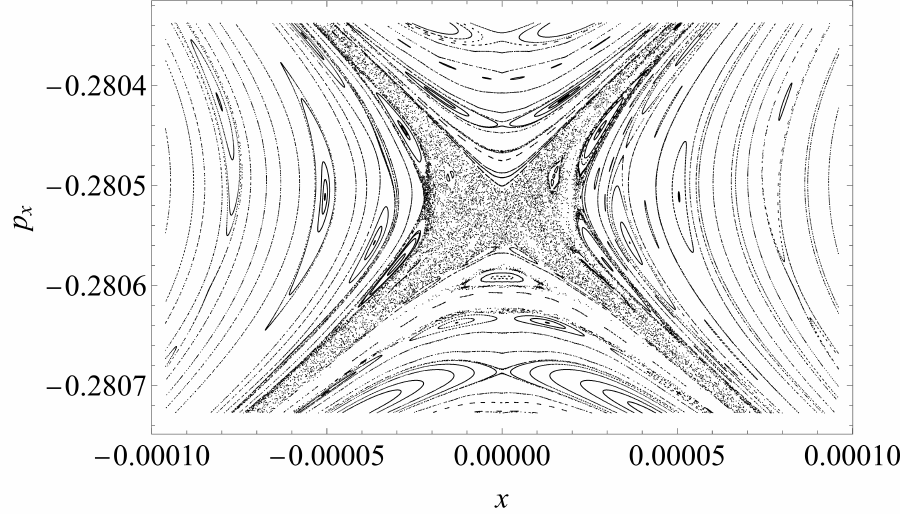}
\caption{\small Magnification of chaotic region near an unstable periodic solution in Fig.~\ref{fig:J10Mm10_pfeq0}. \label{fig:J10Mm10_pf_eq0_cross_en_1over8_magnifi}}
\end{figure}

The magnification of the vicinity of one of these points is shown in Fig.~\ref{fig:J10Mm10_pf_eq0_cross_en_1over8_magnifi}. A thin chaotic layer along the separatrix is visible, which is numerical evidence of non-integrability.

The cross-sections for Hamiltonian \eqref {eq:Hcyl} with $p_{\varphi}=0$ for different energy values are very similar. This is because $\sigma\gg\delta$ and in the potential dominates the part with a square root $\sqrt{z^2+\frac{x^2+y^2}{4}}$. Then the Hamiltonian with the standard kinetic energy and this dominating potential of degree 1 is homogeneous; thus, it has only two different energy levels: 0 and a non-zero one.

The plots for different values of $J$, $M$, $\varpi=c^*c-a^*a$, $A_1$ and $A_2$ give similar figures after a certain rescaling of coordinates and sometimes also time.  

\section{Integrability analysis and solutions on invariant submanifolds}
\label{Integrability}
\setcounter{equation}{0}

The results of numerical integration of equations of motion for particular trajectories and Poincar\'e cross-sections presented in the previous section strongly motivate a deeper qualitative study of the considered system. The most important question is whether the obtained equations of motion are solvable or integrable. In this section, we show their non-integrability. But before we proceed to prove this fact, let us make several remarks.

In general, deciding definitively whether a given system is integrable or non-integrable is non-trivial. Moreover, for non-integrable systems, investigations of partial integrability, the presence of invariant sets, or distinguished classes of particular solutions are also important. Below, we summarise the main integrability results for electromagnetic traps.

In the case of an ideal Penning trap with a quadrupolar electrostatic potential and a strong, constant magnetic field along the fixed $z$ axis, the motion of a point charge consists of bound orbits that are a superposition of three harmonic oscillations with distinct frequencies: axial, cyclotron, and magnetron; see, e.g., \cite[Sec.~3.1.3]{Ghosh}, \cite[Chap.~6]{Vogel2024}, or \cite{BG86}. Consequently, the system is integrable with three functionally independent first integrals: (i) the Hamiltonian, which separates into axial (along $z$) and radial (in the $(x,y)$ plane) parts, and (ii) the third component of the angular momentum.    

Similar solvability issues arise for the Paul trap. There, instead of a constant magnetic field, one superimposes on the static (DC) quadrupole potential an additional quadrupole potential driven by a harmonic (AC) voltage; see \cite{Paul,Dehmelt} or \cite[Chap.~2]{Ghosh}. The gradient of this potential yields the force, and each Cartesian component of Newton’s equation becomes a Mathieu equation. Thus, the model comprises three linear second-order differential equations with time-periodic coefficients. According to Floquet's theory, such a system can, in principle, be transformed into one with constant coefficients and is therefore (formally) integrable. Moreover, formal first integrals can be constructed for each Mathieu equation as described in \cite[pp.~81--82]{Contopoulos}. This integrability remains formal because the required transformation is not explicitly known, and the associated series for the integrals may not converge. Hence, stability analyses rely on numerical methods.   

However, when higher-order terms in the electric or magnetic field expansion are included to model trap imperfections, or when more complex geometries are considered, the system typically becomes non-integrable. This happens, for example, in the magnetic Ioffe–Pritchard trap: in addition to a pair of coils in a Helmholtz configuration, four straight, parallel wires (adjacent wires carrying opposite currents) are added perpendicular to the coil axis, producing magnetic-field components linear and quadratic in the spatial coordinates, see e.g. \cite{Perez-Ríos}. Other traps with higher-order multipole structure, such as the spherical hexapole “baseball” trap \cite{metcalf1987,Petrich} or mirror traps \cite{Cary}, likewise break integrability. 

Additional physical effects can also destroy integrability; for instance, \cite{Yaremko:15::} proves the non-integrability of the motion of a charged particle in an ideal Penning trap within a relativistic framework. 
 
In general, even small perturbations of an integrable model may render it non-integrable, as is explained by KAM theory, which describes qualitatively the effects of small perturbations on integrable systems (the persistence of almost all invariant tori), see \cite[Sec. 2.3]{Contopoulos} or \cite[Chap. 2]{Reichl}. Consequently, approximation techniques are employed; for electromagnetic traps with slowly varying magnetic fields, a particularly well-suited one is the guiding-center approximation, see \cite{Cary}.

The study of the dynamics of polar particles modelled by means of a dipole in an electromagnetic field is much more complex because, in addition to the translational motion of the centre of mass, it is necessary to study the rotational motion of the dipole and in general, these equations of motions are coupled, see e.g. \cite{NJPhys2020}. However, semi-classical analysis causes averaging over dipole variables, and equations of motion reduce to those describing motion for the center of mass, as for point charges in electromagnetic traps described above.

In this section, we rescale the time again so that $\sigma=1$ and introduce the parameter $\eta=\tfrac{\delta}{\sigma}$. Then the Hamiltonian \eqref{Hcm_Edp} takes the form
 \begin{equation}
H  = T+V,\quad T=\frac12\left(p_x^2+p_y^2+p_z^2\right),\,\,\, V=\sqrt{z^2+\frac{x^2+y^2}{4}}+2\eta\left(z^2+\frac{x^2+y^2}{4}\right).
 \label{eq:rescal}
\end{equation}
We will analyse the integrability of this Hamiltonian system. Let us notice that the potential $V$ consists of two homogeneous terms 
\begin{equation}
V_{1}=\sqrt{z^2+\frac{x^2+y^2}{4}},\quad  V_2=2\eta\left(z^2+\frac{x^2+y^2}{4}\right),   
\end{equation}
 which are homogeneous functions of coordinates of degrees 1 and 2, respectively. We use the implication that the integrability in the Liouville sense of the entire Hamiltonian $H=T+V$ implies the integrability in the Liouville sense of two Hamiltonians with the lowest and the highest components of the potential
 \begin{equation}
 H_1=T+V_1,\quad      H_2=T+V_2.
 \end{equation}
Hamiltonian $H_2$ is integrable because it is separable in Cartesian coordinates, but for $H_1$ we can use the necessary integrability conditions due to the differential Galois group of variational equations formulated in Theorem~\ref{thm:MoRaalg} in Appendix~\ref{App-B} for a homogeneous algebraic potential $u$
satisfying a certain polynomial equation $F(\vq,u)=0$ given in \eqref{eq:6}.  Here we have  $u=V_1$, and polynomial $F(\vq,u)=f_0(\vq)+f_1(\vq) u+f_2(\vq) u^2$ is of degree 2 with coefficients
$f_0(\vq)=-\left(z^2+\frac{x^2+y^2}{4} \right)$, $f_1(\vq)=0$ and $f_2(\vq)=1$. Potential $V_1$  has Darboux point  $\vd=(\pm1/2,0,0)$, see for definition \eqref{eq:darbi}, and $c=V_1(\vd)=\tfrac{1}{2}$ satisfying   $F(\vd,c)=0$, $\partial_u F(\vd,c)\neq0$,
$\partial_{\vq}F(\vd,c)=-\partial_uF(\vd,c) \vd$.  Hessian at this Darboux point is $ V_1''(\vd)=\operatorname{diag}(0,1,4)$. According to Theorem~\ref{thm:MoRaalg} if a Hamiltonian system defined by $H_1$ is integrable in the Liouville sense, then one of the cases 1. or 2. holds for any Darboux point satisfying the assumptions. But neither $k=1\in\scK_2$ defined in \eqref{eq:K2} nor $\lambda=4 \in \bigcup_{i=1}^6\scI_i(1),$ where the sets $\scI_i(k)$ for $j=1,\ldots,6$ are given in \eqref{eq:40}. Thus, the Hamiltonian system $H_1$ and, consequently, also $H$ in \eqref{eq:rescal} are non-integrable in the Liouville sense.

Although the Hamiltonian $H$ in \eqref{eq:rescal} is non-integrable, it has two invariant submanifolds. To analyse dynamics restricted to them, it is convenient to introduce cylindrical coordinates $(r,z,\varphi)$ and $H$ 
transforms into $\scH$ given in \eqref{eq:Hcyl} and its Hamilton equations take the form
\begin{equation}
\begin{split}
\dot r&=p_r,\quad\,\, \dot p_r= \frac{p_{\varphi}^2}{r^3}-\eta r-\frac{r}{2 \sqrt{r^2+4 z^2}},\\
\dot z&=p_z,\quad\,\, \dot p_z=-4 \eta z-\frac{2 z}{\sqrt{r^2+4 z^2}},\\
\dot \varphi&=\frac{p_{\varphi}}{r^2},\quad \dot p_{\varphi}=0.
\end{split}  
\label{eq:Hamcyle}
\end{equation}  
As mentioned above, angle $\varphi$ is a cyclic variable and the corresponding momentum $p_{\varphi}= r^2 \dot \varphi$ is a constant of motion.

\begin{figure}[h!tp]
\centering
 \includegraphics[width=0.7\textwidth]{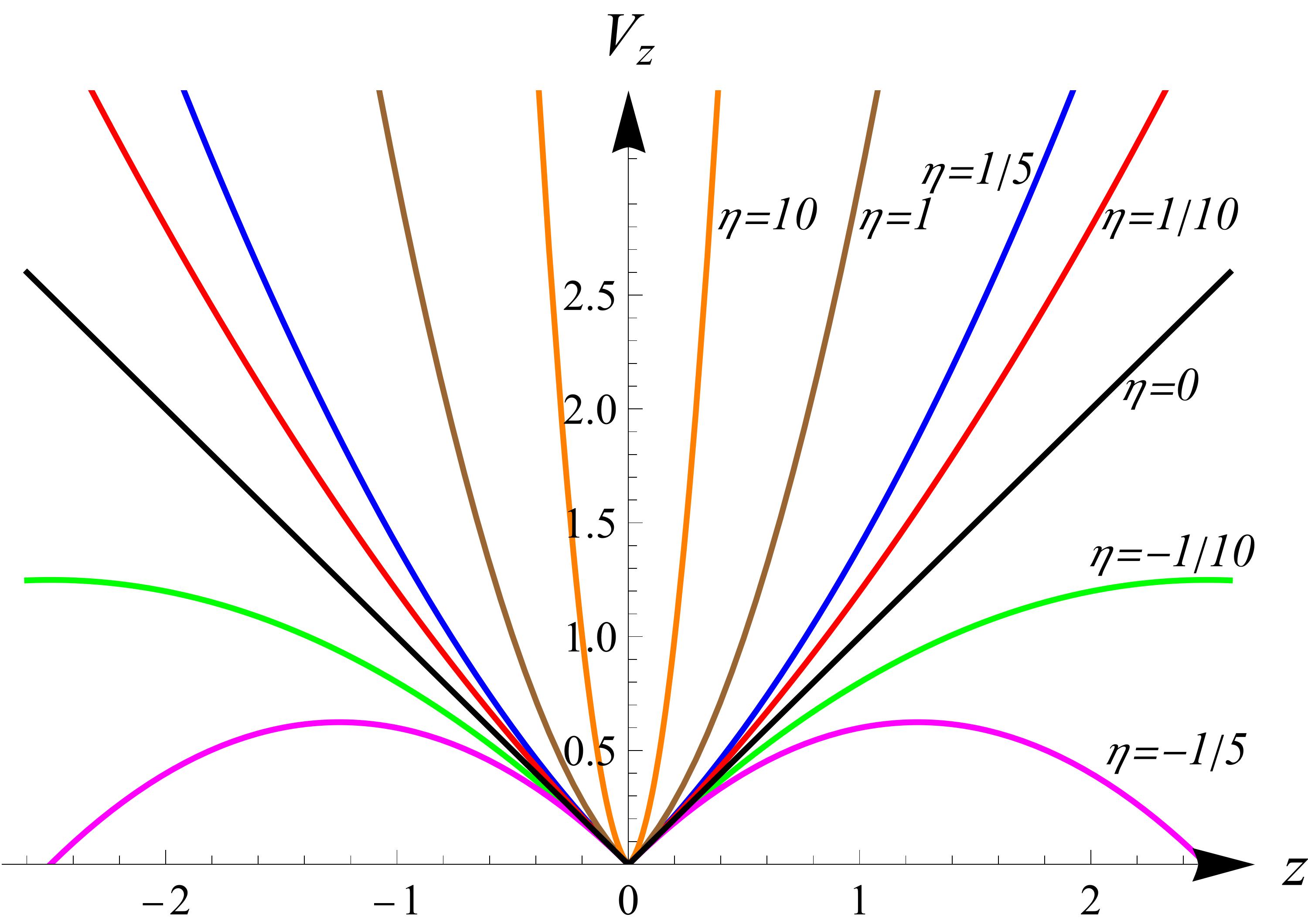}
\caption{\small Graph of  potential  $V_z(z)=2 \eta z^2+| z|$ and various values of $\eta$. \label{fig:potz}}
\end{figure}

The first invariant subspace is defined by $x=y=p_x=p_y=0$, and the particle moves along the symmetry axis $z$ according to the differential equation
\begin{equation}
  \ddot z=  -4 \eta z-\sgn z.
\end{equation}
This equation has a first integral of the form
\begin{equation}
H_z=\frac{\dot z^2}{2}+V_z,\quad V_z= 2 \eta z^2+| z|.
\end{equation}
Graphs of potential $V_z$ for various parameter values $\eta$ show Fig.~\ref{fig:potz}.

For a fixed $H_z=h_z$, this equation can be solved by quadrature
\begin{equation}\label{eq4-7}
\frac{1}{\sqrt{2}}\int \frac{1}{\sqrt{h_z-2 \eta
   z^2-| z| }}dz=t +C_z,
\end{equation}
(see eqs.~(\ref{z-pl})-(\ref{T1-2})).

\begin{figure}[h!tp]
\centering
 \includegraphics[width=0.7\textwidth]{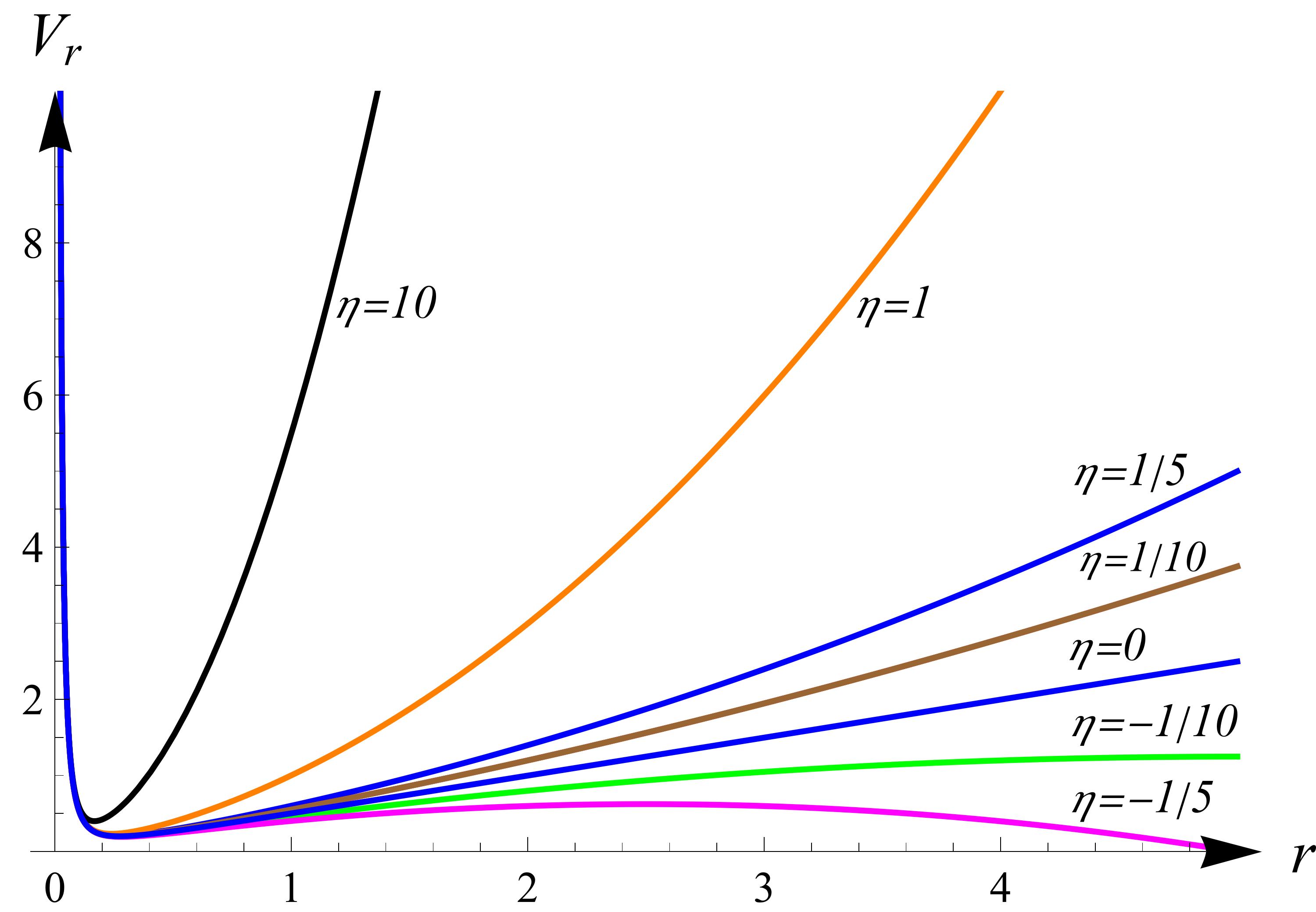}
\caption{\small Graph of the effective potential  $V_r(r)=\frac{p_{\varphi}^2}{2 r^2}+\frac{\eta
   r^2}{2}+\frac{r}{2}$ for $p_{\varphi}=\tfrac{1}{10}$ and various values of $\eta$. \label{fig:effectivpot}}
\end{figure}

The second invariant manifold is defined by equations $z=p_z=0$
and then
\[
\ddot r=\frac{p_{\varphi}^2}{r^3}-\eta r-\frac{1}{2}.
\]
It has the following first integral
\begin{equation}
H_r=\frac{1}{2}\dot r^2+V_r,\quad V_r= \frac{p_{\varphi}^2}{2 r^2}+\frac{\eta r^2}{2}+\frac{r}{2}.
\label{eq:hrr}
\end{equation}

The plots of the effective potential $V_r$ for various values of $\eta$ for a non-zero fixed value of $p_{\varphi}=\tfrac{1}{10}$ illustrate Fig.~\ref{fig:effectivpot}.

To find the minima of this potential, we analyse its derivative
\[
V_r'(r)=\frac{2 \eta r^4+r^3-2 c_z^2}{2 r^3}.
\]
The extremes of $V_r$ are among the zeros of the numerator. The discriminant of this numerator is
\[
\operatorname{disc}(2 \eta r^4+r^3-2 c_z^2,r)=-4 c_z^4 (27 + 4096 c_z^2 \eta^3),
\]
thus, for $\eta>-\frac{3}{16 c_z^{2/3}}$ is negative, which means that $V_r'(r)$ has two real different roots but only one is positive, giving the global minimum as it is visible at Fig.~\ref{fig:effectivpot}. Positive root is given by eq.~(\ref{r-min}) in Appendix \ref{App_C}.
\begin{figure}[h!tp]
\centering
 \includegraphics[width=0.7\textwidth]{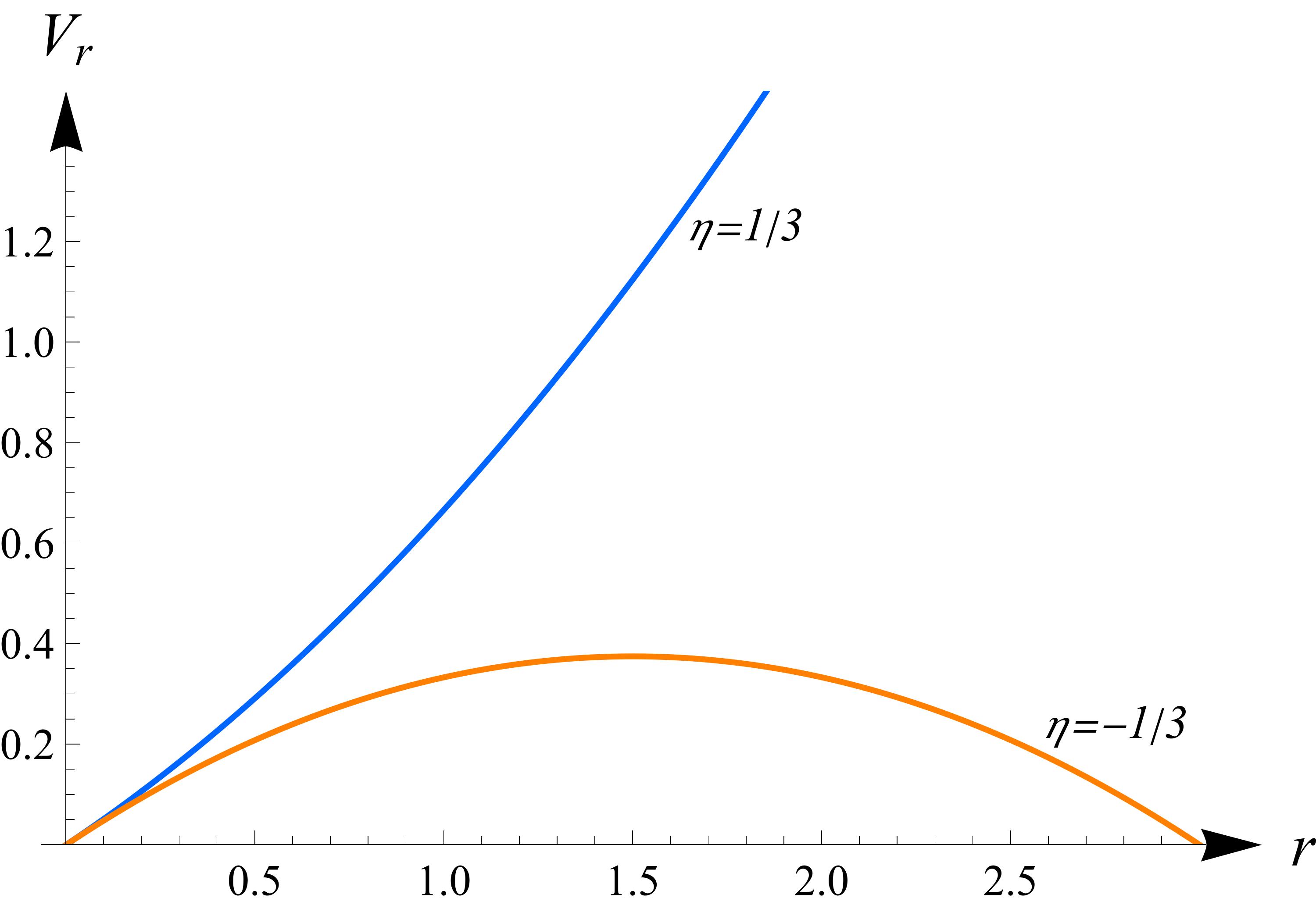}
\caption{\small Graph of the effective potential  $V_r(r)=\tfrac{r}{2}(1+\eta r)$ for $p_{\varphi}=0$ and  $\eta=\pm\tfrac{1}{3}$. \label{fig:effectivepf0}}
\end{figure}

For $p_{\varphi}=0$ effective potential factors as $V_r=\tfrac{r}{2}(1+\eta r)$. Thus, its graph is a parabola pointing upward with the vertex at $r=0$ or a parabola pointing down and passing through $r=0$ with a maximum at $r=-\tfrac{1}{2\eta}$ depending on the sign of $\eta$, see Fig.~\ref{fig:effectivepf0}.

Let us consider the case $p_{\varphi}\neq 0$. 
Looking at Fig.~\ref{fig:effectivpot}, we notice that for a fixed value $h_r>\min V_r(r)$ of first integral $H_r$ the distance $r(t)$ oscillates in a certain interval between turning points at which $\dot r=0$, that is  $V_r=h_r$ (see Appendix \ref{App_C}, eq. (\ref{r12-turn})).
From the first integral \eqref{eq:hrr} we obtain the following expression
\begin{equation}
0=\dot r^2=2(h_r-V_r(r))=2h_r-\frac{p_{\varphi}^2}{r^2}-\eta r^2-r=-\frac{1}{r^2}[\eta r^4+r^3-2h_rr^2+p_{\varphi}^2]
\label{eq:nau}
\end{equation}
and this quartic polynomial in square bracket should have two real positive roots, one of them we call $u$. Then from \eqref{eq:nau}   one can express $h_r$ through $u$ and the remaining parameters
\begin{equation}
h_r= \frac{1}{2 u^2}(c_z^2+\eta u^4+u^3),  
\end{equation}
where we chose $p_{\varphi}=c_z\neq 0$. Then substituting this expression into this on $ \dot r^2$ in \eqref{eq:nau} gives
\begin{equation}
    \dot r^2=2(h_r-V_r(r))=\frac{(r - u)}{u^2}[c_z^2 (r + u) - r^2 u^2 (1 + \eta (r + u))].
\end{equation}
One can find a solution of this differential equation in the form
\begin{equation}
\begin{split}
&r(\tau)=\frac{A_1+B_1\cn(\omega \tau)}{A_2+B_2\cn(\omega \tau)},\\
&A_1=5 c_z^2 u-u^3 \left((\eta u+1)u+(4 m-5) \omega^2\right),\quad B_1=u^3 \left((\eta u+1)u+(4 m+1) \omega^2\right)-5 c_z^2 u,\\
&A_2=u^2 \left(u (5 \eta u+2)+(5-4 m) \omega^2\right)-c_z^2,\quad B_2=  c_z^2+u^2 \left((4 m+1) \omega^2-u (5 \eta u+2)\right),
\end{split}
\end{equation}
where $\cn(x,m)$ is the Jacobi elliptic cosine with modulus $m$, and the parameters $m$ and $\omega$ satisfy the equations
\begin{equation}
\begin{split}
 p_1=&c_z^4+2 c_z^2 u^3 (7 \eta u+1)+u^6 (\eta
   u+1)^2-(16 (m-1) m+1) u^4 \omega^4=0,\\
p_2=&96 c_z^6+96 c_z^4 u^3 (13 \eta u+2)+2 (16
   (m-1) m+1) u^4 \omega^4 \left(\eta u^4+u^3-47
   c_z^2\right)\\
   &+3 c_z^2 u^6 (32 \eta
   u+23)+2 (2 m (16 m (2 m-3)+15)+1)
   u^6 \omega^6=0.   
\end{split}    
\end{equation}
The resultant of these two equations with respect to $m$ gives an equation for $\omega$
\begin{equation}
 \begin{split}
& \operatorname{res}(p_1,p_2,m)= 110592 u^{16} \omega^{12}\Big[u^8 \omega^{12}-u^4 \omega ^8 \left(c_z^4+2 c_z^2 u^3 (7 \eta
   u+1)+u^6 (\eta u+1)^2\right) \\
& +c_z^2 \left(2 \eta u^4+u^3-2 c_z^2\right)^2
   \left(4 c_z^4 \eta+c_z^2 u^2 (4 \eta u (5-2 \eta
   u)+1)+4 u^5 (\eta u+1)^3\right)  \Big] =0.
 \end{split}   
\end{equation}

\section{Conclusions}\label{Concl} 
\setcounter{equation}{0}

In this article, we study the translational motions of a spin-1 diatomic molecule in a magnetic quadrupole trap.
The molecule is prepared in its ${}^3\Sigma$ electronic states, deeply bound vibrational states, and rotational state (\ref{Psi_00rot}) of well-defined parity. The trapping potential arises from the rotational energy caused by the interaction of the total spin angular momentum of electrons and the orbital angular momentum of nuclei with the non-homogeneous magnetic field (\ref{B1}). We also calculate the second-order shifts of the rotational energy levels. They contribute additional harmonic terms
$z^2+\rho^2/4$ to the well-known square root potential $\sqrt{z^2+\rho^2/4}$ (see eq.~(\ref{V_CM})). The rough estimation of the trap's depth due to the spin Zeeman effect is $\sim 6.7~\mathrm{K}$ for $B_{\rm trap}=5~{\rm T}$ \cite{Harris2004}. If the quasi-stable molecule returns to its spinless state, the trap depth decreases to hundreds of microkelvins (see Table 1).

 We add the kinetic energy to the potential derived from the above spectral problem with linear and quadratic
Zeeman shifts. This gives us classical Hamiltonian that governs the molecule’s translational motion.

We performed a detailed analysis of this Hamiltonian system. It appears that it is not integrable. To prove this fact, we use differential Galois techniques, among other things. As the system is not integrable, its dynamics can be investigated with the help of approximated or numerical methods. We chose the second possibility and used the Poincar\'e section method to get a global overview of the system's dynamics. Using direct numerical integration of the equations of motion, we overviewed possible trajectories of the molecule in the trap.

The general conclusion of these investigations is that, for small energy values, the molecule is always trapped in a region of the size of the trap. Moreover, although the system governing the dynamics is not integrable, the chaotic behaviour is weak and does not influence the stability of the trap.    

\section*{Acknowledgements}
This research was partially funded by the Minister of Science of Poland under the “Regional Excellence Initiative” program, Project No. RID/SP/0050/2024/1, and by the National Science Center of Poland under Grant No. 2020/39/D/ST1/01632.

\appendix
\section{ Hydrogen molecule in ${}^3\Sigma$ electronic state}\label{App-A}
\setcounter{equation}{0}
\setcounter{section}{1}
\renewcommand{\theequation}{\Alph{section}.\arabic{equation}}

\begin{figure}[h!tp]
\centering
\includegraphics[width=0.45\textwidth]{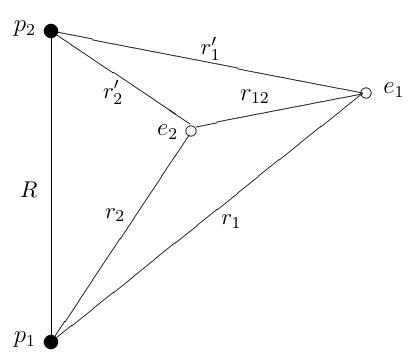}
\caption{\small Hydrogen molecule scheme showing the distances on which the electrons' potential energy $U_{ee}(R)$ depends. Protons $p_1$ and $p_2$ are ``clamped'' within the Born-Oppenheimer approximation.
Symbols $r_a$ and $r_a'$ denote the distances of the $a$-th electron from the first and the second proton, respectively. Distance between electrons is $r_{12}$. \label{H2Mpl}
}
\end{figure}

The hydrogen molecule scheme is pictured in Fig.~\ref{H2Mpl}. We use the molecule fixed frame. Applying the Born-Oppenheimer approximation, we ``clamp'' protons such that the separating vector is directed along the $z$-axis.
Picking coordinates so that proton $1$ is at the origin and proton $2$ is on the $z$ axis at the distance $R$ we put the following coordinates of particles: $\mathbf{x}_{1p}=(0,0,0)$, $\mathbf{x}_{2p}=(0,0,R)$, $\mathbf{x}_{1e}=(r_1\sin\vartheta_1\cos\phi_1,r_1\sin\vartheta_1\sin\phi_1,r_1\cos\vartheta_1)$, and $\mathbf{x}_{2e}=(r_2\sin\vartheta_2\cos\phi_2,$ $r_2\sin\vartheta_2\sin\phi_2,r_2\cos\vartheta_2)$. The vectors shown in Fig.~\ref{H2Mpl} have the following components:
\begin{eqnarray}\label{CM-H2}
\mathbf{R}&=&\mathbf{x}_{2p}-\mathbf{x}_{1p}=(0,0,R),\label{Rr1r2}\\
\mathbf{r}_1&=&\mathbf{x}_{1e}-\mathbf{x}_{1p}=(r_1\sin\vartheta_1\cos\phi_1,r_1\sin\vartheta_1\sin\phi_1,
r_1\cos\vartheta_1),\nonumber\\
\mathbf{r}_2&=&\mathbf{x}_{2e}-\mathbf{x}_{1p}=(r_2\sin\vartheta_2\cos\phi_2,r_2\sin\vartheta_2\sin\phi_2,
r_2\cos\vartheta_2),\nonumber\\
\mathbf{r}_1'&=&\mathbf{x}_{1e}-\mathbf{x}_{2p}=(r_1\sin\vartheta_1\cos\phi_1,r_1\sin\vartheta_1\sin\phi_1,
r_1\cos\vartheta_1-R),\nonumber\\
\mathbf{r}_2'&=&\mathbf{x}_{2e}-\mathbf{x}_{2p}=(r_2\sin\vartheta_2\cos\phi_2,r_2\sin\vartheta_2\sin\phi_2,
r_2\cos\vartheta_2-R).\nonumber
\end{eqnarray}
The ``primed'' arguments of  wave functions (\ref{psi0}) contain $R$:
\begin{equation}\label{r_i0}
r_a'=\sqrt{R^2+r_a^2-2r_aR\cos\vartheta_a}\,.
\end{equation}
We introduce dimensionless variables based on the atomic unit of energy.
We scale the coordinate variables by the Bohr radius, i.e. $R\to R/a_0$, $r_i\to r_i/a_0$.

We use the variational wave functions that constitute the Heitler-London approximation
for ${}^3\Sigma$ electronic state of a Hydrogen molecule:
\begin{equation}\label{Psi_minTot}
\Phi_-(\mathbf{r}_1,\mathbf{r}_2;\mathbf{s}_1,\mathbf{s}_2)=\psi_-(\mathbf{r}_1,\mathbf{r}_2)
\mathbf{S}(\mathbf{s}_1.\mathbf{s}_2)
\end{equation}
This means that we base our consideration on the variational wave function being the product of the space asymmetric wave function of the position variables
\begin{equation}\label{psi_minus}
\psi_-(\mathbf{r}_1,\mathbf{r}_2)=A_-\left(\psi_{100}(r_1)\psi_{100}(r_2')-\psi_{100}(r_1')\psi_{100}(r_2)\right),
\end{equation}
and spin-symmetric (triplet) state $\mathbf{S}(\mathbf{s}_1,\mathbf{s}_2)$
\begin{eqnarray}
|1,+1\rangle&=&|1/2,+1/2\rangle|1/2,+1/2\rangle=|\uparrow\rangle|\uparrow\rangle;\label{triplet}\\
|1,0\rangle&=&\frac{1}{\sqrt{2}}\left[\phantom{\frac11}\!\!\!\!|1/2,+1/2\rangle|1/2,-1/2\rangle
+|1/2,-1/2\rangle|1/2,+1/2\rangle\right]\nonumber\\
&=&\frac{1}{\sqrt{2}}\left(|\uparrow\rangle|\downarrow\rangle+|\downarrow\rangle|\uparrow\rangle\right);\nonumber\\
|1,-1\rangle&=&|1/2,-1/2\rangle|1/2,-1/2\rangle=|\downarrow\rangle|\downarrow\rangle.\nonumber
\end{eqnarray}
We accept the molecular orbital (MO) theory, which posits that each electron's motion can be described by single-particle functions
\begin{equation}\label{psi0}
\psi_{100}(r_a)=\frac{1}{\sqrt{\pi}}\mathrm{e}^{-r_a},\qquad \psi_{100}(r_a')=\frac{1}{\sqrt{\pi}}\mathrm{e}^{-r_a'}.
\end{equation}
The trial function (\ref{Psi_minTot}) describes the anti-bonding molecular orbital.
The normalisation function in eq.~(\ref{psi_minus}) is given by \cite[eq.~(8.60)]{Grif}, namely $A_-=\left[2(1-I^2)\right]^{-1/2}$. The overlap integral $I(R)={\rm e}^{-R}\left(1+R+R^2/3\right)$ is also calculated in this handbook (see \cite[eq.~(8.43)]{Grif}).

The components of $3$-vectors in Hamilton's function are in the coordinate system defined by the direction of the external field. When dealing with a diatomic molecule, it is convenient to use the body-fixed frame where the $z$-axis is directed along the interatomic axis (see Fig.~\ref{H2Mpl}). The relation between the components of the vector in the space-fixed frame and those in the body-fixed frame is given by the direction cosine matrix $\hat{C}$ \cite[eq.~(C.1)]{Krems}. In this paper, we use the modified matrix in which the azimuthal angle $\varphi$ is replaced by $\varphi+\pi$ and the negative polar angle $-\theta$ is substituted for $\theta$:
\begin{equation}\label{Cmatr}
\hat{C}=\left[
\begin{array}{ccc}
 -\sin\varphi & \cos\varphi\cos\theta & \cos\varphi\sin\theta \\
 \cos\varphi  & \sin\varphi\cos\theta & \sin\varphi\sin\theta \\
 0            &-\sin\theta            & \cos\theta
\end{array}\right].
\end{equation}

We calculate the energy of electrons acted upon the fields of elementary particles in a molecule as well as the external magnetic field (\ref{B1}) generated by the trap's coils. We apply the Hamiltonian \cite[eq.~(A.13)]{JPhysB2025} to the specific case of a hydrogen molecule
\begin{equation}\label{HmolH2}
H=\frac{1}{2m}\left(\mathbf{P}-\mathbf{b}_X\right)^2+H_{\rm free}+H^{(1)}_{\rm Zeeman}+H^{(2)}_{\rm Zeeman}.
\end{equation}
The functions $H^{(1)}_{\rm Zeeman}$ and $H^{(2)}_{\rm Zeeman}$ contain contributions from the interaction of the trap's magnetic field with the orbital angular momenta of electrons and nuclei. We scale the center-of-mass coordinated $(-X/2,-Y/2,Z)$ by the trap's characteristic size $D$. We scale the coordinate variables of nuclei and electrons by the Bohr radius and measure the energy in Hartree unit. We calculate the expectation values of the perturbation caused by the terms of the function (\ref{HmolH2}) step by step.

\subsubsection*{A.1 First order Zeeman shift}\label{Zee-1st}

In the specific case of two protons and two electrons, the first-order Zeeman perturbation takes the form
\begin{eqnarray}
H_{\rm Zeeman}^{(1)}&=&-\frac{1}{2}\beta_L\frac{m_e}{m_p}\left(\mathbf{B}\cdot[\mathbf{R}\times\mathbf{\Pi}]\right)\label{HmolH2-B}\\
&-&\frac{1}{4}\beta_L\left(\mathbf{B}\cdot\left\{[\mathbf{R}\times\mathbf{p}_1]
-\left(2-\frac{m_e}{m_p}\right)[\mathbf{r}_1\times\mathbf{p}_1]+\frac{m_e}{m_p}[\mathbf{r}_2\times\mathbf{p}_1]\right\}\right)\nonumber\\
&-&\frac{1}{4}\beta_L\left(\mathbf{B}\cdot\left\{[\mathbf{R}\times\mathbf{p}_2]
+\frac{m_e}{m_p}[\mathbf{r}_1\times\mathbf{p}_2]
-\left(2-\frac{m_e}{m_p}\right)[\mathbf{r}_2\times\mathbf{p}_2]\right\}\right).\nonumber
\end{eqnarray}
The first term in eq.~(\ref{HmolH2-B}) describes the interaction of the trap's magnetic field and the angular momentum of protons. In terms of dimensionless variables, the magnetic field $\mathbf{B}=(-x/2,-y/2,z)$ and the angular momentum of nuclei $\mathbf{L}=[\mathbf{R}\times\mathbf{\Pi}]$ has components (\ref{L1})--(\ref{L3}). The others are the mass polarisation
terms that have appeared because of the transformation from laboratory to nuclear center-of-mass coordinates. The constant $\beta_L$ is given by eq.~(\ref{beta_L}).

The operator $\mathbf{L}=[\mathbf{R}\times\mathbf{\Pi}]$ does not act
on the trial wave function (\ref{Psi_minTot}) as well as on the states of nuclear vibrations.
For the initial state (\ref{Psi_00rot}), the rotational energy
shifts are given by eq.~(\ref{E_1M}).

The other terms in eq.~(\ref{HmolH2-B}) define the interaction of the magnetic field
and the angular momentum of the electron cloud. The electronic state (\ref{Psi_minTot}) is composed of the atom's ground state functions (\ref{psi0}). The electrons' angular  momenta in eqs.~(\ref{HmolH2-B}) are as follows
\begin{eqnarray}
L_a^1&=&-\rmi\left(-\sin\phi_a\frac{\partial}{\partial\vartheta_a}-\cos\phi_a\cot\vartheta_a\frac{\partial}{\partial\phi_a}\right),
\nonumber\\
L_a^2&=&-\rmi\left(\cos\phi_a\frac{\partial}{\partial\vartheta_a}-\sin\phi_a\cot\vartheta_a\frac{\partial}{\partial\phi_a}\right),
\nonumber\\
L_a^3&=&-\rmi\frac{\partial}{\partial\phi_a}.\nonumber
\end{eqnarray}
Due to the cylindrical symmetry of the wave function $\Phi_-(\mathbf{r}_1,\mathbf{r}_2;\mathbf{s}_1,\mathbf{s}_2)$, the expectation values of these terms are equal to zero.

\subsubsection*{A.2 Second order Zeeman shift}\label{Zee-2nd}

The second-order Zeeman perturbation is as follows
\begin{eqnarray}
H_{\rm Zeeman }^{(2)}&=&+\frac{\beta_L^2}{16}\left(1+\frac{2m_e}{m_p}-\frac{4m_e}{m_p+m_e}\right)[\mathbf{B}\times\mathbf{R}]^2\label{HmolH2-B2}\\
&-&\frac{\beta_L^2}{8}\left(1+\frac{m_e}{m_p}-\frac{4m_e}{m_p+m_e}\right)\left\{\phantom{\frac11}\!\!\!\!\!
\left([\mathbf{B}\times\mathbf{R}]\cdot[\mathbf{B}\times\mathbf{r}_1]\right)
+\left([\mathbf{B}\times\mathbf{R}]\cdot[\mathbf{B}\times\mathbf{r}_2]\right)\right\}\nonumber\\
&+&\frac{\beta_L^2}{16}\left(2+\frac{m_e}{m_p}-\frac{4m_e}{m_p+m_e}\right)\left\{\phantom{\frac11}\!\!\!\!\!
[\mathbf{B}\times\mathbf{r}_1]^2+[\mathbf{B}\times\mathbf{r}_2]^2\right\}\nonumber\\
&+&\frac{\beta_L^2}{8}\left(\frac{m_e}{m_p}-\frac{4m_e}{m_p+m_e}\right)\left([\mathbf{B}\times\mathbf{r}_1]\cdot[\mathbf{B}\times\mathbf{r}_2]\right).
\nonumber
\end{eqnarray}
The term in the first line does not act on the electronic state function $\Phi_-(\mathbf{r}_1,\mathbf{r}_2;\mathbf{s}_1,\mathbf{s}_2)$. 
It governs the interaction of protons'
angular momentum and the trap's magnetic field. In the space fixed frame,
the first term has the form (\ref{Zee2R-H2}). For the initial state (\ref{Psi_00rot}), the correction to the rotation energy is given by eq.~(\ref{E_2M}).

Let us calculate the expectation values of the terms in the second line of eq.~(\ref{HmolH2-B2}).
We ``sandwich'' the operator between $\Phi^*_-(\mathbf{r}_1,\mathbf{r}_2;\mathbf{s}_1,\mathbf{s}_2)$ and $\Phi_-(\mathbf{r}_1,\mathbf{r}_2;\mathbf{s}_1,\mathbf{s}_2)$, and integrate. In the molecule fixed frame, the scalar product of two cross products in the second line looks as follows
\begin{eqnarray}
&&\left([\mathbf{B}\times\mathbf{R}]\cdot[\mathbf{B}\times\mathbf{r}_a]\right)=
\mathbf{B}^2Rr_a\cos\vartheta_a\\
&-&
\left(B_x'r_a\cos\phi_a\sin\vartheta_a+B_y'r_a\sin\phi_a\sin\vartheta_a
+B_z'r_a\cos\vartheta_a\right)(\mathbf{B}\cdot\mathbf{R}).\nonumber
\end{eqnarray}
The primed uppercase letters $B$ denote the components of the trap's magnetic field in the body-fixed frame pictured in Fig.~\ref{H2Mpl}.
The expectation values of terms which are proportional to the sine and cosine of the azimuthal angles vanish due to the integral $\int_0^{2\pi}{\rm d}\phi_a$ of the spherically symmetric atomic wave functions (\ref{psi0}). Integration of terms which are proportional to the cosine of the polar angle can be handled via the relations
\begin{eqnarray}
\langle\psi_{100}(r_a)|r_a\cos\vartheta_a|\psi_{100}(r_a)\rangle&=&0,\label{1st}\\
\langle\psi_{100}(r_a)|r_a\cos\vartheta_a|\psi_{100}(r_a')\rangle&=&
\frac{1}{2}R I(R),\label{1stprime}\\
\langle\psi_{100}(r_a')|r_a\cos\vartheta_a|\psi_{100}(r_a')\rangle&=&R.\label{1stprimeprime}
\end{eqnarray}
Taking into account that the trial wave function is normalised and the normalisation factor is $2(1-I^2(R))$, we finally obtain the following
\begin{eqnarray}\label{BRra}
\langle\psi_-(\mathbf{r}_1,\mathbf{r}_2)|\sum_a([\mathbf{B}\times\mathbf{R}]\cdot[\mathbf{B}\times\mathbf{r}_a])|
\psi_-(\mathbf{r}_1,\mathbf{r}_2)\rangle&=&\frac12R^2
\left[\mathbf{B}^2-B_z'(\mathbf{n}_R\mathbf{B})\right].\nonumber
\end{eqnarray}
According to the direction cosine matrix (\ref{Cmatr}), in the laboratory frame the primed $z$-component becomes
\begin{eqnarray}
B_z'&=&(\mathbf{n}_R\mathbf{B})\\
&=&-\frac{x}{2}\cos\varphi\sin\theta-\frac{y}{2}\sin\varphi\sin\theta+z\cos\theta.\nonumber
\end{eqnarray}
So, the expression between the square brackets in eq.~(\ref{BRra}) coincides with that in eq.~(\ref{Zee2R-H2}).
Collecting the contributions due to the first and second terms, we obtain the intermediate result
$$
-\frac{\beta_L^2}{16}\left(1-\frac{4m_e}{m_p+m_e}\right)\left[\mathbf{B}^2-(\mathbf{n}_R\mathbf{B})^2\right]R^2.
$$

Let us consider the matrix element of the operator in the fourth line of eq.~(\ref{HmolH2-B2}) with ${}^3\Sigma$ electronic state. Most terms of the scalar product of two cross products vanish due to the integration over the azimuthal angle. The following term survives only
\begin{eqnarray}\label{Rr1Rr2}
&&\langle\psi_-(\mathbf{r}_1,\mathbf{r}_2)|([\mathbf{B}\times\mathbf{r}_1]\cdot[\mathbf{B}\times\mathbf{r}_2])|
\psi_-(\mathbf{r}_1,\mathbf{r}_2)\rangle=\nonumber\\
&=&\left[\mathbf{B}^2-(\mathbf{n}_R\mathbf{B})^2\right]
\langle\psi_-(\mathbf{r}_1,\mathbf{r}_2)|r_1\cos\vartheta_1r_2\cos\vartheta_2|\psi_-(\mathbf{r}_1,\mathbf{r}_2)\rangle .
\end{eqnarray}
Using eqs.~(\ref{1st})--(\ref{1stprimeprime}) we calculate the matrix element
$$
\langle\psi_-(\mathbf{r}_1,\mathbf{r}_2)|r_1\cos\vartheta_1r_2\cos\vartheta_2|\psi_-(\mathbf{r}_1,\mathbf{r}_2)\rangle
=-\frac{R^2I^2(R)}{4(1-I^2(R))},
$$
where $I(R)$ is the overlap integral \cite[eq.~(8.43)]{Grif}. The energy level shift (\ref{Rr1Rr2}) is still proportional to the expression
(\ref{Zee2R-H2}).

Let us evaluate the corrections to the energy level that originated from the terms in the third line of eq.~(\ref{HmolH2-B2})
\begin{eqnarray}
[\mathbf{B}\times\mathbf{r}_a]^2&=&\mathbf{B}^2r_a^2-
(B_x'\cos\phi_a\sin\vartheta_a+B_y'\sin\phi_a\sin\vartheta_a+B_z'\cos\vartheta_a)^2r_a^2.\nonumber
\end{eqnarray}
The expectation values of the mixed terms $B_x'B_y'\cos\phi_a\sin\phi_a\sin^2\vartheta_a$, $B_x'B_z'\cos\phi_a\sin\vartheta_a\cos\vartheta_a$, and $B_y'B_z'\sin\phi_a\sin\vartheta_a\cos\vartheta_a$
disappear due to the integration of atomic wave functions (\ref{psi0}) over the azimuthal angle $\int_0^{2\pi}{\rm d}\phi_a$.

The expectation values associated with the squared in magnetic field components are as follows:
\begin{eqnarray}
\langle\psi_{100}(r_a)|r_a^2|\psi_{100}(r_a)\rangle&=&3,\nonumber\\
\langle\psi_{100}(r_a)|r_a^2\cos^2\phi_a\sin^2\vartheta_a|\psi_{100}(r_a)\rangle&=&1,\nonumber\\
\langle\psi_{100}(r_a)|r_a^2\sin^2\phi_a\sin^2\vartheta_a|\psi_{100}(r_a)\rangle&=&1,\nonumber\\
\langle\psi_{100}(r_a)|r_a^2\cos^2\vartheta_a|\psi_{100}(r_a)\rangle&=&1;
\end{eqnarray}
\begin{eqnarray}
\langle\psi_{100}(r_a)|r_a^2|\psi_{100}(r_a')\rangle&=&3I(R)+\frac15R^2X(R)+\frac{3}{10}R^2I(R),\nonumber\\
\langle\psi_{100}(r_a)|r_a^2\cos^2\phi_a\sin^2\vartheta_a|\psi_{100}(r_a')\rangle&=&
I(R)+\frac{1}{15}R^2X(R),\nonumber\\
\langle\psi_{100}(r_a)|r_a^2\sin^2\phi_a\sin^2\vartheta_a|\psi_{100}(r_a')\rangle&=&
I(R)+\frac{1}{15}R^2X(R),\nonumber\\
\langle\psi_{100}(r_a)|r_a^2\cos^2\vartheta_a|\psi_{100}(r_a')\rangle&=&
I(R)+\frac{1}{15}R^2X(R)+\frac{3}{10}R^2I(R);\nonumber\\
\\
\langle\psi_{100}(r_a')|r_a^2|\psi_{100}(r_a')\rangle&=&3+R^2,\nonumber\\
\langle\psi_{100}(r_a')|r_a^2\cos^2\phi_a\sin^2\vartheta_a|\psi_{100}(r_a')\rangle&=&1,\nonumber\\
\langle\psi_{100}(r_a')|r_a^2\sin^2\phi_a\sin^2\vartheta_a|\psi_{100}(r_a')\rangle&=&1,\nonumber\\
\langle\psi_{100}(r_a')|r_a^2\cos^2\vartheta_a|\psi_{100}(r_a')\rangle&=&1+R^2;
\end{eqnarray}
After some algebra, we obtain
\begin{eqnarray}\label{BRra2}
&&\langle\psi_-(\mathbf{r}_1,\mathbf{r}_2)|\sum_a[\mathbf{B}\times\mathbf{r}_a]^2|
\psi_-(\mathbf{r}_1,\mathbf{r}_2)\rangle=\\
&=&\mathbf{B}^2\left(
4+\frac35R^2+\frac{2}{15}R^2\frac{3-2X(R)I(R)}{1-I^2(R)}
\right)
-(\mathbf{n}_R\mathbf{B})^2\frac{R^2}{5}\left(3+\frac{2}{1-I^2(R)}\right).\nonumber
\end{eqnarray}

Collecting the contributions, we finally obtain
\begin{eqnarray}\label{psi-Hz-psi}
\langle\psi_+|H_{\rm Zeeman}^{(2)}|\psi_+\rangle&=&
-\frac{\beta_L^2}{16}\left(1-\frac{4m_e}{m_p+m_e}\right)\left[\mathbf{B}^2-(\mathbf{n}_R\mathbf{B})^2\right]R^2\\
&-&\frac{\beta_L^2}{32}\left(\frac{m_e}{m_p}-\frac{4m_e}{m_p+m_e}\right)
\left[\mathbf{B}^2-(\mathbf{n}_R\mathbf{B})^2\right]R^2\frac{I^2(R)}{1-I^2(R)}\nonumber\\
&+&\frac{\beta_L^2}{16}\left(2+\frac{m_e}{m_p}-\frac{4m_e}{m_p+m_e}\right)
\left\{\mathbf{B}^2\left(
4+\frac35R^2+\frac{2}{15}R^2\frac{3-2X(R)I(R)}{1-I^2(R)}
\right)\right.\nonumber\\
&-&\left.(\mathbf{n}_R\mathbf{B})^2\frac{R^2}{5}\left(3+\frac{2}{1-I^2(R)}\right)\right\}.
\nonumber
\end{eqnarray}
Unfreezing the nuclei gives rise to the relative motion of two atoms, which can be separated into vibrational motion and orbital motion. To uncouple them, we adopt the rigid-rotor approximation.
We expand the electronic potential in a Taylor series and restrict ourselves to the first-order term. It describes simple harmonic oscillations. To design the potential well, we use experimental data for $f^3\Sigma_{u}^+4p\sigma$ electronic state (see \cite{NIST}, constants of hydrogen molecules). Namely, the minimum electronic energy $T_e$ and the vibrational constant $\omega_e$ as the point of equilibrium and the effective spring constant, respectively.

To calculate the numerical constants $A_1$ and $A_2$ in the center-of-mass Hamiltonian (\ref{Hvbr_rot_CM}) we ``sandwich'' the $R$-dependent factors before $\mathbf{B}^2$ and $(\mathbf{n}_R\mathbf{B})^2$ between a vibrational wane function and its complex conjugate one and integrate the resulting expressions over the interval $]0,\infty[$. We approximate the vibrational excitations by Hermite polynomials. After a little algebra, we obtain
\begin{itemize}
\item ground vibrational state:
\begin{equation}\label{A1A2_gr}
A_1=0.5691906099701544,\quad A_2=0.1665675408030196\,;
\end{equation}
\item 1-st excited vibrational state:
\begin{equation}\label{A1A2_1st}
A_1=0.5369997783542894,\quad A_2=0.1613113921392951\,.
\end{equation}
\end{itemize}

\section{Differential Galois obstructions for integrability of homogeneous potentials}
\label{App-B}
\setcounter{equation}{0}
\setcounter{section}{2}
\renewcommand{\theequation}{\Alph{section}.\arabic{equation}}
Let us consider natural Hamiltonian systems with $n$ degrees of freedom governed by the following Hamilton function 
\begin{equation}
 H=\frac{1}{2}\sum_{i=1}^np_i^2+V(\vq),
 \label{eq:hamm}
\end{equation}
where $\vq=(q_1,\ldots,q_n)$ and $\vp=(p_1,\ldots,p_n)$ are canonical coordinates and momenta with potential $V(\vq)$ which is a homogeneous function of non-zero integer degree $k$ i.e. 
\[
V(\lambda q_1,\ldots,\lambda q_n)=\lambda^kV(q_1,\ldots,q_n),\quad \lambda>0.
\]

In generic cases, Hamilton equations with homogeneous potentials 
\begin{equation}
 \Dt \vq=\vp,\qquad \Dt \vp=-  \nabla V(\vq)  
 \label{eq:hamy}
\end{equation}
have particular straight-line solutions defined by non-zero solutions $\vd$ of algebraic equations
\begin{equation}
  \nabla V(\vd)=\vd,
  \label{eq:darbi}
\end{equation}
called Darboux points.  Here, $\nabla V(\vq)$ denotes the gradient of potential $V(\vq)$.  The presence of particular solutions made it possible to apply obstructions arising from the analysis of properties of differential Galois groups of the variational equations for the Hamilton equations \eqref{eq:hamy}. The presence of first integrals of Hamilton equations \eqref{eq:hamy} implies the presence of invariants of the differential Galois group, and the integrability in the Liouville sense of these equations causes Abelianity of the identity component of the differential Galois group, for details see \cite{Morales:99::c}.  Applications of these conditions to variational equations along particular solutions built by means of Darboux points gave a very famous theorem with the necessary conditions for integrability of Hamiltonian systems with homogeneous potentials formulated in \cite{Morales:99::c,Morales:01::c}. It gives restrictions on the admissible values of the Hessian eigenvalues of the potential at Darboux points $V''(\vd)$. There are many applications of this theorem.

However, the framework for applying the differential Galois theory to integrability analysis encodes the restriction of the first integrals considered to the class of meromorphic functions of coordinates and momenta.
But in applications, there appear potentials that are algebraic functions of the coordinates, i.e. they satisfy a certain polynomial equation
\begin{equation}
\label{eq:6}
F(\vq,V(\vq)):=\sum_{i=0}^mf_i(\vq) V(\vq)^i=0,
\end{equation}
where $f_i(\vq) $ are complex polynomials, and $m>0$. Then we cannot analyse meromorphic integrability because the Hamiltonian is already not meromorphic, as it is not even single-valued. The subtleties of the integrability analysis of such systems are described in the paper \cite{Combot:13::}. Here, we will only recall the necessary conditions for integrability obtained in \cite{mp:16::a}. To formulate the main theorem of this article, we first define certain sets of rational numbers
\begin{equation}
\label{eq:40}
\begin{split}
\scI_1(k):=&\defset{p + \dfrac{k}{2}p(p-1)}{p\in \Z},\\
\scI_2(k):=&\defset{ \dfrac{1}{8k}  \left[ 4  k^2 \left( p
             +\dfrac{1}{2}\right)^2 -(k-2)^{2}\right]}{p\in \Z},\\
\scI_3(k):=&\defset{ \dfrac{1}{8k}  \left[  4 k^2 \left(  p
             +\dfrac{1}{3}\right)^2 -(k-2)^{2}\right]}{p\in \Z},\\
\scI_4(k):=&\defset{ \dfrac{1}{8k}  \left[  4 k^2 \left(  p
             +\dfrac{1}{4}\right)^2 -(k-2)^{2}\right]}{p\in \Z},\\
\scI_5(k):=&\defset{ \dfrac{1}{8k}  \left[  4 k^2 \left(  p
             +\dfrac{1}{5}\right)^2 -(k-2)^{2}\right]}{p\in \Z},\\
\scI_6(k):=&\defset{ \dfrac{1}{8k}  \left[  4 k^2 \left(  p
             +\dfrac{2}{5}\right)^2 -(k-2)^{2}\right]}{p\in \Z}.
             \end{split}
\end{equation}
\begin{theorem}
  \label{thm:MoRaalg}
  Assume that an algebraic homogeneous potenial $V(\vq)$, $\vq$ of degree
  $k\in\Q\setminus\{0\}$ defined by minimal polynomial $F(\vq,u)$ satisfies the following conditions:
  \begin{enumerate}
  \item there exists a non-zero $\vd\in\C^n$  and $c\in\C$ such that
    $F(\vd,c)=0$, $\partial_u F(\vd,c)\neq0$,
    $\partial_{\vq}F(\vd,c)=-\partial_uF(\vd,c) \vd$,
    and
  \item the Hessian matrix $V''(\vd)$ has eigenvalues
    $\lambda_1, \ldots, \lambda_n$;
  \item it is integrable with  first integrals, which are complex
    meromorphic functions on $M_0$. 
  \end{enumerate}
Then 
\begin{enumerate}
\item 
either 
 \begin{equation}
k\in 
\scK_2:=\defset{\dfrac{2}{2 l +1}}{l\in\Z}, 
\label{eq:K2}
\end{equation}
\item or   $k\in\Q^{\times}\setminus \scK_2$, and for these values of $k$
   each  eigenvalue $\lambda$  of $V''(\vd)$ belongs either to the set
 $\scI(k):=\scI_1(k)\cup\scI_2(k)$, or
\begin{enumerate}
\item if $k=\pm\tfrac{3}{3l +1}$, for some $l\in\Z$, then  
\begin{equation}
\label{eq:41}
\lambda \in \bigcup_{i=1}^6\scI_i(k);
\end{equation}
\item if $k=\pm\tfrac{4}{4l +1}$, for some $l\in\Z$, then 
\begin{equation}
\label{eq:42}
\lambda \in \scI_1(k)\cup\scI_2(k) \cup \scI_3(k) ;
\end{equation}
\item if $k=\pm\tfrac{5}{5l +1}$, for some $l\in\Z$, then  
\begin{equation}
\label{eq:43}
\lambda \in \scI_1(k)\cup\scI_2(k) \cup \scI_3(k) \cup \scI_5(k) ;
\end{equation}
\item if $k=\pm\tfrac{5}{5l +2}$, for some $l\in\Z$, then  
\begin{equation}
\label{eq:44}
\lambda \in \scI_1(k)\cup\scI_2(k) \cup \scI_3(k) \cup \scI_6(k) .
\end{equation}
\end{enumerate}
\end{enumerate}
\end{theorem}
For the proof and applications of this theorem, see \cite{mp:16::a}.

\section{Turning points}\label{App_C}
\setcounter{equation}{0}

Let us focus our task on the turning points of regular motions governed by the central returning force of the trapping potential. To simplify the consideration, we rescale the dimensionless coordinates
\begin{equation}
\mathrm{r}=\eta r,\qquad\mathrm{z}=\eta z,
\end{equation}
and the evolution parameter
\begin{equation}
\mathrm{t}=\sqrt{\eta}\tau.
\end{equation}
These imply the rescaling of dimensionless momenta
\begin{equation}
\mathrm{p}_r=\sqrt{\eta} p_r,\qquad \mathrm{p}_z=\sqrt{\eta} p_z,
\end{equation}
as well as the momentum $p_\varphi$ corresponding to the cyclic angular variable
\begin{equation}
c_z^2=\eta^3 p_\varphi^2.
\end{equation}

\begin{figure}[h!tp]
\centering
\includegraphics[width=0.65\textwidth]{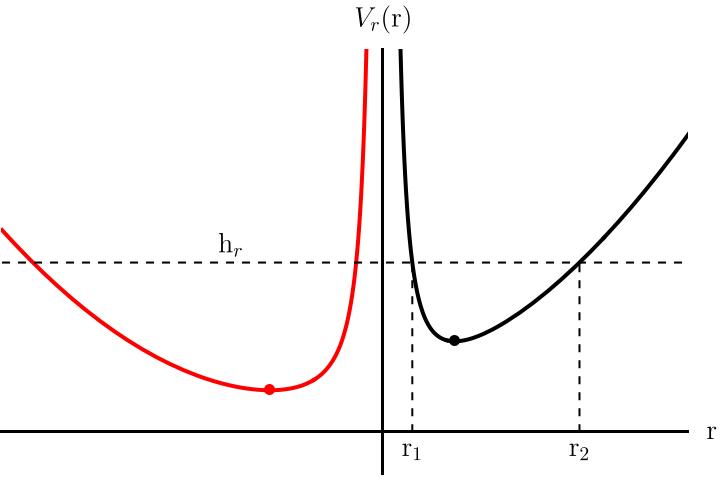}
\caption{\small Graph of the potential $V_r(\mathrm{r})$ defined in \eqref{eq:potVrrescal}. The non-physical branch is coloured in red.
\label{Graph-Vr}
}
\end{figure}

\subsection{Turning points in the manifold $z=p_z=0$}\label{Turn-r}

In the plane $z=0$, the molecule moves backward and forward in the radial potential (\ref{eq:hrr})
\begin{equation}
V_r(\mathrm{r})=\frac{c_z^2}{2\mathrm{r}^2}+\frac{\mathrm{r}^2}{2}+\frac{\mathrm{r}}{2}.
\label{eq:potVrrescal}
\end{equation}
Its graph is sketched in Fig.~\ref{Graph-Vr}.

At turning points, the radial velocity $\dot{\mathrm{r}}=0$. For a given radial energy $h_r$, they are the roots of the quartic polynomial
\begin{equation}
P_4=\mathrm{r}^4+\mathrm{r}^3-2\mathrm{h}_r\mathrm{r}^2+c_z^2,
\end{equation}
where $\mathrm{h}_r=\eta h_r$ (cf.~(\ref{eq:nau})).  We are interested in positive roots that belong to the blue branch of the graph in Fig.~\ref{Graph-Vr}. 
For a fixed energy level $\mathrm{h}_r$ and an angular momentum $c_z$ such that
\begin{equation}
c_z^2< \frac{27}{512}+\frac{9}{16}\mathrm{h}_r+\mathrm{h}_r^2-\left(\frac{9}{64}+\mathrm{h}_r\right)^{3/2},
\end{equation}
the turning points $\mathrm{r}_1<\mathrm{r}_2$ are
\begin{equation}\label{r12-turn}
\mathrm{r}_1=-\frac{1}{4}+\frac{1}{2}\left(\sqrt{2m_0}-\sqrt{D_2}\right),\quad 
\mathrm{r}_2=-\frac{1}{4}+\frac{1}{2}\left(\sqrt{2m_0}+\sqrt{D_2}\right),
\end{equation}
where 
\begin{equation}
D_2=\frac34+4\mathrm{h}_r-2m_0-\left(\frac18+\mathrm{h}_r\right)\sqrt{\frac{2}{m_0}}
\end{equation}
is the appropriate Ferrari's discriminant.
The real root of the auxiliary resolvent cubic is
\begin{equation}
m_0(\mathrm{h}_r,c_z)=
\frac18+\frac{2}{3}\mathrm{h}_r\left\{1+\sqrt{1+3\frac{c_z^2}{\mathrm{h}_r^2}}\cos\left[\alpha(\mathrm{h}_r,c_z)\right]\right\}
\end{equation}
where the trigonometric angle is
\begin{equation}
\alpha(\mathrm{h}_r,c_z)=\frac13{\arccos} \frac{\displaystyle -1+\frac{27}{16}\frac{c_z^2}{\mathrm{h}_r^3}+9\frac{c_z^2}{\mathrm{h}_r^2}}{\displaystyle \left(1+3\frac{c_z^2}{\mathrm{h}_r^2}\right)^{3/2}}.
\end{equation}

The energy level $\mathrm{h}_r$ should be greater than or equal to the extremal value of the radial potential $V_r(\mathrm{r}_{\rm min})$, where the minimum point is
\begin{equation}\label{r-min}
\mathrm{r}_{\rm min}(c_z)=-\frac18-\frac12\sqrt{2m_{00}(c_z)}+\frac12\sqrt{D_1(c_z)}.
\end{equation}
The appropriate Ferrari's discriminant is 
\begin{equation}
D_1(c_z)=\frac{3}{16}-2m_{00}(c_z)+\frac{1}{32\sqrt{2m_{00}(c_z)}}
\end{equation}
The real root of the resolvent cubic is
\begin{equation}
m_{00}(c_z)=\frac{1}{32}+2\sqrt{\frac{c_z^2}{3}}\sinh[\beta(c_z)],
\end{equation}
where the hyperbolic angle
\begin{equation}
\beta(c_z)=\frac13{\rm arcsinh}\left(-\frac{3}{64}\sqrt{\frac{3}{c_z^2}}\right)
\end{equation}

\subsection{Turning points in the manifold $x=y=p_x=p_y=0$}\label{Turn_z}

On the vertical line, the molecule moves backward and forward in the radial potential
\begin{equation}
V_z=2\mathrm{z}^2+|\mathrm{z}|.
\end{equation}
For a given azimuthal energy level $\mathrm{h}_z=\eta h_z$ we have the following turning points $\mathrm{z}_1<\mathrm{z}_2$:
\begin{equation}
\mathrm{z}_1=\frac14-\frac14\sqrt{1+8\mathrm{h}_z}\,, \qquad \mathrm{z}_2=-\frac14+\frac14\sqrt{1+8\mathrm{h}_z}\,.
\end{equation}
 We obtain the periodic orbit by calculating the quadrature (\ref{eq4-7}). In terms of the scaled variable $\mathrm{z}=\eta z$ and the time $\mathrm{t}=\sqrt{\eta} \tau$ we have
\begin{itemize}
\item[$\mathrm{z}\geq 0$]:$\,\,\,\mathrm{ t}\in[0,\mathrm{T}]$
\begin{equation}\label{z-pl}
\mathrm{z}(\mathrm{t})=-\frac14+\frac14\sqrt{1+8\mathrm{h}_z}\cos\left(2\mathrm{t}-\mathrm{T}\right);
\end{equation}
\item[$\mathrm{z}\leq 0$]:$\,\,\,\mathrm{t}\in[\mathrm{T},2\mathrm{T}]$
\begin{equation}\label{z-mn}
\mathrm{z}(\mathrm{t})=\frac14-\frac14\sqrt{1+8\mathrm{h_z}}\cos\left(2\mathrm{t}-3\mathrm{T}\right).
\end{equation}
\end{itemize}
By the capital letter $\mathrm{T}$ we denote the one-half period
\begin{equation}\label{T1-2}
\mathrm{T}=\arccos\left(\frac{1}{\sqrt{1+8\mathrm{h}_z}}\right).
\end{equation}

\end{document}